\RequirePackage{fix-cm}
\documentclass[natbib,smallextended]{svjour3}       
\smartqed  
\usepackage{graphicx}
\usepackage{natbib}
\usepackage{color}
\usepackage{amssymb}
\usepackage{lineno}
\usepackage{txfonts}
\usepackage{amstext}
\usepackage{mathabx}
\usepackage{multirow}
\usepackage{ulem}
\usepackage{lscape}
\usepackage{txfonts}
\usepackage{subfigure}
\usepackage{float}
\usepackage{color,soul}
\usepackage{bm}
\usepackage{lscape}

\makeatletter 
\@mparswitchfalse%
\makeatother
\usepackage[colorinlistoftodos]{todonotes}
\normalmarginpar 

\usepackage[bookmarks = true, bookmarksnumbered = true, pdfstartview = FitH, pdfpagelayout = SinglePage, colorlinks = true, urlcolor = blue, citecolor = blue]{hyperref}

 \journalname{Space Science Reviews}

\newcommand{\degre}{$^{\circ}$}
\newcommand{\Rj}{\,R$_\mathrm{J}$}

\begin{document}

\title{Jupiter Science Enabled by ESA's Jupiter Icy Moons Explorer}

\titlerunning{Jupiter Science from JUICE}        

\author{Leigh N. Fletcher \and
        Thibault Cavali\'e \and
        Davide Grassi \and
        Ricardo Hueso \and
        Luisa M. Lara \and
        Yohai Kaspi \and
        Eli Galanti  \and
        Thomas K. Greathouse  \and
        Philippa M. Molyneux \and
        Marina Galand  \and
        Claire Vallat \and
        Olivier Witasse \and
        Rosario Lorente \and
        Paul Hartogh \and
        Fran\c{c}ois Poulet \and
        Yves Langevin \and
        Pasquale Palumbo \and
        G. Randall Gladstone \and
        Kurt D. Retherford \and
        Michele K. Dougherty \and
        Jan-Erik Wahlund \and
        Stas Barabash \and
        Luciano Iess \and
        Lorenzo Bruzzone \and
        Hauke Hussmann \and
        Leonid I. Gurvits  \and
        Ondřej Santolik \and
        Ivana Kolmasova \and
        Georg Fischer \and
        Ingo M\"uller-Wodarg \and
        Giuseppe Piccioni \and
        Thierry Fouchet  \and
        Jean-Claude G\'erard \and
        Agustin S\'anchez-Lavega \and
        Patrick G. J. Irwin \and
        Denis Grodent  \and
        Francesca Altieri \and
        Alessandro Mura \and
        Pierre Drossart \and
        Josh Kammer \and
        Rohini Giles \and
        St\'ephanie Cazaux \and
        Geraint Jones \and
        Maria Smirnova \and 
        Emmanuel Lellouch \and 
        Alexander S. Medvedev \and
        Raphael Moreno \and
        Ladislav Rezac \and 
        Athena Coustenis \and
        Marc Costa}


\institute{Leigh N. Fletcher \at
              School of Physics and Astronomy, University of Leicester, University Road, Leicester, LE1 7RH, UK \\
              \email{leigh.fletcher@leicester.ac.uk} 
           \and
              Thibault Cavali\'e \at
              Laboratoire d'Astrophysique de Bordeaux, Univ. Bordeaux, CNRS, B18N, all\'ee Geoffroy Saint-Hilaire, 33615 Pessac, France
           \and
              Davide Grassi, Pasquale Palumbo, Giuseppe Piccioni, Francesca Altieri, Alessandro Mura \at
              Istituto di Astrofisica e Planetologia Spaziali - Istituto Nazionale di Astrofisica, Via del Fosso del Cavaliere, 100, I-00133, Roma, Italy
           \and
              Ricardo Hueso, Agustin S\'anchez-Lavega  \at
              F\'isica Aplicada, Escuela de Ingenier\'ia de Bilbao Universidad del Pa\'is Vasco UPV/EHU, Plaza Ingeniero Torres Quevedo, 1, 48013 Bilbao, Spain
           \and
              Luisa M. Lara \at
              Instituto de Astrof\'isica de Andaluc\'ia-CSIC, c/Glorieta de la Astronom\'ia 3, 18008 Granada, Spain
           \and
              Yohai Kaspi, Eli Galanti, Maria Smirnova  \at
              Dept. of Earth and Planetray Science, Weizmann Institute of Science, Rehovot, Israel, 76100
           \and
              Thomas K. Greathouse, Philippa Molyneux, G. Randall Gladstone, Kurt D. Retherford, Josh Kammer, Rohini Giles  \at
              Southwest Research Institute, San Antonio, TX 78228, United States 
           \and
              Marina Galand \at
              Department of Physics, Imperial College London, Prince Consort Road, London SW7 2AZ, UK
           \and
              Claire Vallat, Rosario Lorente \at
              European Space Agency (ESA) - ESAC Camino Bajo del Castillo s/n Villafranca del Castillo, 28692, Villanueva de la Ca\~nada (Madrid), Spain
           \and
              Olivier Witasse \at
              European Space Agency (ESA), European Space Research and Technology Centre (ESTEC), Noordwijk, Netherlands
           \and
              Marc Costa \at
              Rhea Group, for European Space Agency, ESAC, Madrid, Spain
           \and
              Paul Hartogh, Alexander S. Medvedev, Ladislav Rezac \at
              Max-Planck-Institut f\"ur Sonnensystemforschung, 37077 G\"ottingen, Germany
           \and
              Fran\c{c}ois Poulet, Yves Langevin \at
              Institut d'Astrophysique Spatiale, CNRS/Universit\'e Paris-Sud, 91405 Orsay Cedex, France
           \and
              Michele K. Dougherty, Ingo M\"uller-Wodarg \at
              Blackett Laboratory, Imperial College London, London, UK
           \and
              Jan-Erik Wahlund \at
              Swedish Institute of Space Physics (IRF), Uppsala, Sweden
           \and
              Stas Barabash \at
              Swedish Institute of Space Physics (IRF), Kiruna, Sweden
           \and
              Luciano Iess \at
              Dipartimento di ingegneria meccanica e aerospaziale, Universit \'a La Sapienza, Roma, Italy
           \and
              Lorenzo Bruzzone \at
              University of Trento, Department of Information Engineering and Computer Science, Remote Sensing Laboratory, Via Sommarive 14, Trento, I-38123, Italy
           \and
              Hauke Hussmann \at
              Deutsches Zentrum f\"ur Luft- und Raumfahrt (DLR), Berlin, Germany
           \and
              Leonid I. Gurvits \at
              Joint Institute for VLBI ERIC, Oude Hoogeveensedijk 4, 7991 PD Dwingeloo, The Netherlands
           \and
              Leonid I. Gurvits \at
              Aerospace Faculty, Delft University of Technology, Kluyverweg 1, 2629 HS Delft, The Netherlands
           \and
              O. Santolik, I. Komalsova \at
              Department of Space Physics, Institute of Atmospheric Physics of the Czech Academy of Sciences, Prague, Czechia;  Faculty of Mathematics and Physics, Charles University, Prague, Czechia
           \and
              Georg Fischer \at
              Space Research Institute, Austrian Academy of Sciences, Graz, Austria
           \and
              Thibault Cavali\'e, Thierry Fouchet, Pierre Drossart, Emmanuel Lellouch, Raphael Moreno, Athena Coustenis \at
              LESIA, Observatoire de Paris, Universit\'e PSL, Sorbonne Universit\'e, Universit\'e Paris Cit\'e, CNRS, 5 place Jules Janssen, 92195 Meudon, France
           \and
              Jean-claude G\'erard, Denis Grodent \at
              LPAP, STAR Institute, Universit\'e de Li\`ege, Belgium
           \and
              Patrick G. J. Irwin \at
              Atmospheric, Oceanic and Planetary Physics, Department of Physics, University of Oxford, Parks Rd, Oxford OX1 3PU
           \and
              St\'ephanie Cazaux \at
              Faculty of Aerospace Engineering, Delft University of Technology, Delft, The Netherlands
           \and
              Geraint Jones \at
              UCL Mullard Space Science Laboratory, Hombury St. Mary, Dorking RH5 6NT, UK and The Centre for Planetary Sciences at UCL/Birkbeck, London WC1E 6BT, UK
           \and
              G. Randall Gladstone, Kurt D. Retherford \at
              University of Texas at San Antonio, San Antonio, TX, United States
           \and
              Pierre Drossart \at
              Institut d'Astrophysique de Paris, CNRS, Sorbonne Université, 98bis Boulevard Arago, 75014 Paris
}

\date{Received: date / Accepted: date}

\maketitle
\tableofcontents

\begin{abstract}
ESA's Jupiter Icy Moons Explorer (JUICE) will provide a detailed investigation of the Jovian system in the 2030s, combining a suite of state-of-the-art instruments with an orbital tour tailored to maximise observing opportunities.  We review the Jupiter science enabled by the JUICE mission, building on the legacy of discoveries from the Galileo, Cassini, and Juno missions, alongside ground- and space-based observatories.  We focus on remote sensing of the climate, meteorology, and chemistry of the atmosphere and auroras from the cloud-forming weather layer, through the upper troposphere, into the stratosphere and ionosphere.  The Jupiter orbital tour provides a wealth of opportunities for atmospheric and auroral science: global perspectives with its near-equatorial and inclined phases, sampling all phase angles from dayside to nightside, and investigating phenomena evolving on timescales from minutes to months.  The remote sensing payload spans far-UV spectroscopy (50-210 nm), visible imaging (340-1080 nm), visible/near-infrared spectroscopy (0.49-5.56 $\mu$m), and sub-millimetre sounding (near 530-625\,GHz and 1067-1275\,GHz).  This is coupled to radio, stellar, and solar occultation opportunities to explore the atmosphere at high vertical resolution; and radio and plasma wave measurements of electric discharges in the Jovian atmosphere and auroras.  Cross-disciplinary scientific investigations enable JUICE to explore coupling processes in giant planet atmospheres, to show how the atmosphere is connected to (i) the deep circulation and composition of the hydrogen-dominated interior; and (ii) to the currents and charged particle environments of the external magnetosphere.  JUICE will provide a comprehensive characterisation of the atmosphere and auroras of this archetypal giant planet.

\end{abstract}

\section{Introduction}
\label{intro}


Jupiter is our closest and best example of a hydrogen-dominated gas giant planet, representing a class of objects $\sim10\times$ the size of Earth that may be commonplace across our galaxy. The formation and migration of such a large planet (317.8 Earth masses, approximately 0.1\% of the mass of the Sun) shaped the architecture of our Solar System, such that the origin of Jupiter is an essential piece of the puzzle of planetary system evolution, providing a window on the epoch of planet formation.  Jupiter provides a perfect planetary-scale laboratory for the exploration of atmospheric physics and chemistry (e.g., climate, meteorology, and convective processes on a rapidly-rotating hydrogen-rich world), without the complicating influences of terrestrial topography or large seasonal changes\footnote{Jupiter's axial tilt is $3^\circ$ and its orbital eccentricity introduces a 14\% maximum change in insolation at Equator \citep{Levine1977}, providing limited seasonal forcing during its 11.9-year orbit}.  Jupiter also provides a means to explore how the layers within a giant planet are coupled, from the interior to the external plasma environment, and vice versa.  For example, the interaction between the upper atmosphere and the plasma environment of the magnetosphere creates an auroral lightshow that is unrivalled in the Solar System. The influence of solar ultraviolet light on the chemicals in Jupiter's stratosphere generates a rich atmospheric chemistry.  And motions within the interior, from the metallic hydrogen to the deep atmosphere, influence the ever-shifting clouds and colours in the visible atmosphere.

For all these reasons and more, a comprehensive investigation of Jupiter as the archetypal giant planet is one of the two primary goals of ESA's Jupiter Icy Moons Explorer (JUICE), Europe's first mission to the Jupiter system \citep{23witasse}.  The emergence of habitable worlds within Gas Giant systems is explored by \citet{23tobie} and \citet{23tosi}, focussing on Ganymede, Europa and Callisto.  The wider Jovian system, and the magnetosphere, are covered by \citet{23masters}, \citet{23schmidt} and \citet{23denk}, here we focus on the Jupiter scientific investigations enabled by the JUICE orbital tour and its suite of state-of-the-art instruments.  Jupiter science, particularly atmospheric, magnetospheric, and auroral science and how they connect to the wider system of potentially-habitable satellites, formed a key component of ESA's Jupiter mission from the outset, when it was first formulated as the multi-spacecraft Laplace mission in 2007 \citep{09blanc} for ESA's Cosmic Vision.  The science case evolved as it became the Jupiter Ganymede Orbiter \citep[JGO,][]{10blanc}, ESA's contribution to the Europa-Jupiter System Mission (EJSM) between 2008 and 2011.  Finally, Jupiter exploration was a cornerstone in the science case for JUICE \citep{13grasset}, which was selected (2012) and adopted (2014) as ESA's first `L-class' mission, and which launched on April 14th, 2023.

The science case presented by \citet{13grasset} built on the discoveries of the Galileo orbiter (1995-2003) and \textit{in situ} probe (1995); the flybys of Pioneer 10 and 11 (1973, 1974), Voyager 1 and 2 (1979), Cassini (2000) and New Horizons (2007); and the wealth of remote sensing investigations from ground-based and earth-orbiting observatories.  These previous missions had provided snapshots of Jupiter at specific times, often lacking adequate sampling of Jovian variability over minutes (e.g., auroras, lightning), days (e.g., storm plumes, impacts), months (belt/zone changes), and years (vortices) to determine the mean atmospheric state and the drivers of variability.  JUICE would provide a continuity of data coverage over long temporal baselines to address the shortcomings of the previous snapshots, particularly the challenging Galileo observations due to the failed deployment of its high-gain antenna.  JUICE would also use broad and quasi-simultaneous spectral coverage from the UV to the sub-millimetre to probe different atmospheric layers.  At the time of mission adoption, the Jupiter science case \citep{13grasset} aimed to provide \textit{`the first four-dimensional climate database for the study of Jovian meteorology and chemistry,'} creating a global picture of the processes shaping the Jovian atmosphere \textit{`from the thermosphere down to the lower troposphere.'}  This led to three science objectives to characterise atmospheric (a) dynamics and circulation; (b) composition and chemistry; and (c) vertical structure and clouds.  It also determined a series of mission requirements that would contribute to the design of the JUICE orbital tour, sampling both low- and high-latitude domains over a long span of time.

Since the JUICE Jupiter science case was developed, both the orbital tour \citep{23boutonnet} and the payload capabilities have been fully specified.  Furthermore, NASA's Juno mission has been providing new discoveries and insights into the planet's interior, atmosphere, and magnetosphere since its arrival at Jupiter in 2016 \citep[e.g.,][]{17bolton}.  Juno's elliptical polar orbit brought the spacecraft close to Jupiter every $\sim53$ days (reducing to $\sim40$ days during the extended mission) to provide high-resolution regional views, whereas JUICE will have a near-equatorial orbit that provides opportunities for longer-term monitoring and global views.  The tour strategies for Juno and JUICE are therefore different and complementary.  Furthermore, Juno's exploration of the deep interior via gravity sounding and microwave remote sensing complements the JUICE observations at lower pressures.  Given the wealth of new discoveries from Juno and supporting Earth-based observations since the original JUICE objectives were developed, and new insights gained from the culmination of the Cassini mission at Saturn between 2004-2017, this paper revisits and significantly updates the JUICE Jupiter science case.

This paper is organised as follows.  Section \ref{science_case} provides a brief review of the Jupiter science case for JUICE, in light of the latest discoveries, and focusing on key questions and objectives that must be addressed by the tour and payload.  Requirements for the tour, and the observation opportunities needed to address the science objectives, are discussed in Section \ref{opportunities}.  Details of the payload relevant to Jupiter science, and how the instruments will operate both independently and synergistically to achieve the science goals, are provided in Sections \ref{instruments}-\ref{synergies}.  We place the JUICE science case into broader context of other missions and astronomical facilities operating in the 2030s in Section \ref{support}, and emphasise the need for Earth-based support from amateur and professional observers.  Finally, Section \ref{summary} confirms how the instruments and tour achieves closure of the science requirements.

\section{Jupiter Scientific Objectives}
\label{science_case}

\begin{figure*}[ht]
\begin{centering}
\centerline{\includegraphics[angle=0,width=\textwidth]{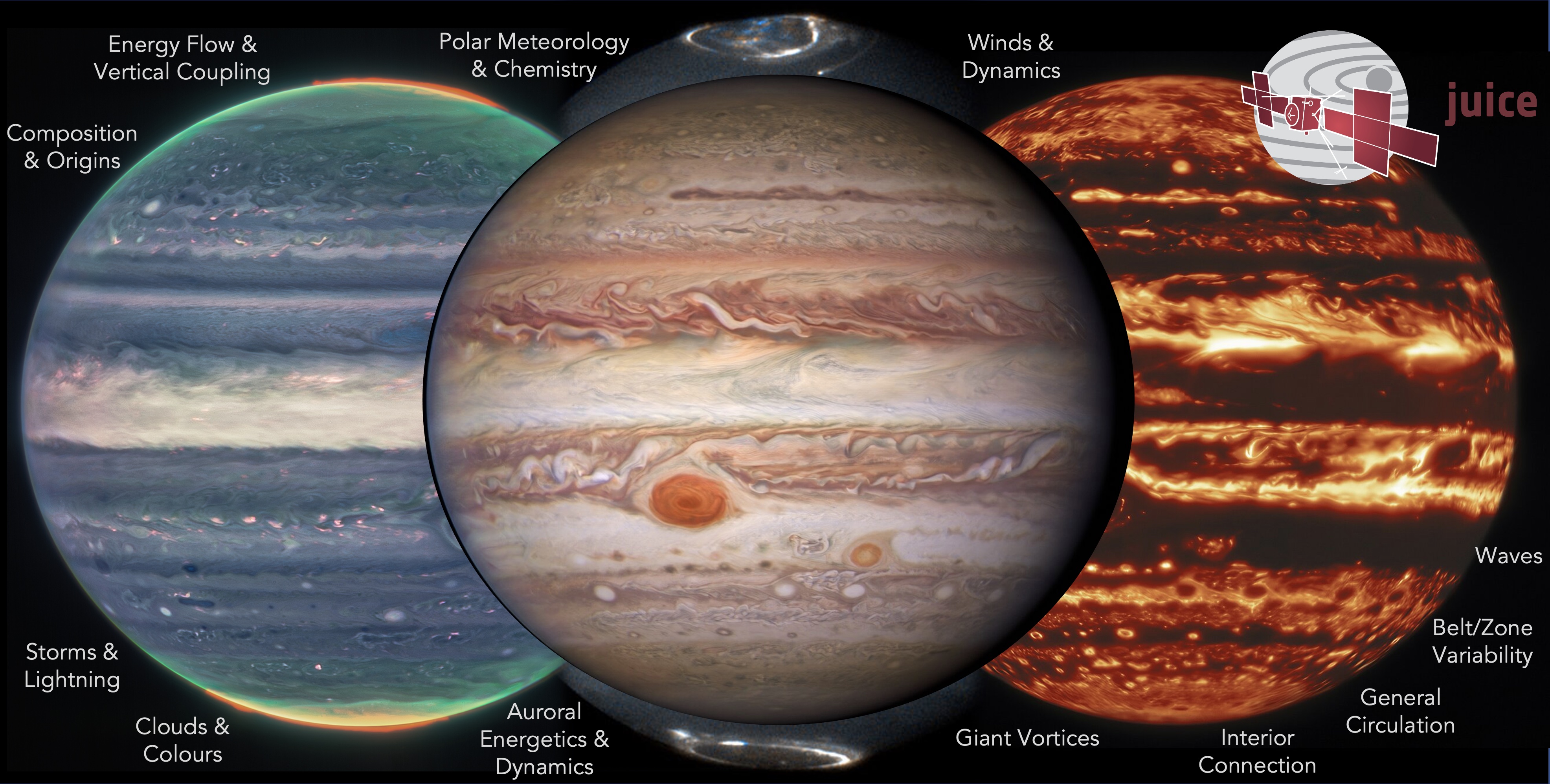}}
\caption{Summary of the Jupiter science enabled by the JUICE mission.  Images show Jupiter in visible light from Hubble (centre, Credit: NASA, ESA, NOIRLab, NSF, AURA, M.H. Wong and I. de Pater \textit{et al.}), near-infrared from JWST (left, using 3.6 $\mu$m (red), 2.12 $\mu$m (yellow-green), and 1.5 $\mu$m (cyan); credit: NASA, ESA, CSA, Jupiter ERS Team; image processing by Judy Schmidt), and in the 5-$\mu$m window (right, credit:  Gemini Observatory, NOIRLab, NSF, AURA, M.H. Wong \textit{et al.}), where clouds from the visible/near-IR images (sensing $\sim0.1-3$ bars) appear in silhouette against the thermal background from the 4-6 bar region.  Auroral emissions from H$_3^+$ can be seen in the JWST image, and in the UV in Hubble observations (centre-top and centre-bottom, credit: NASA, ESA. J. Clarke).}
\label{science_summary}
\end{centering}
\end{figure*}

Jupiter's atmosphere, and its connections to both the deep interior and external plasma environment, are to be explored via a carefully-designed remote sensing investigation (Section \ref{instruments}) across a $\sim4$-year orbital tour that samples a range of illumination conditions, geometries, and orbital inclinations (see Section \ref{opportunities}).  The original JUICE Jupiter science case was subdivided into three scientific objectives that sought to `characterise atmospheric (a)  dynamics and circulation; (b) composition and chemistry; and (c) vertical structure and clouds.' This led to a traceability matrix with 11 specific science investigations (Table \ref{tab:objectives}, and summarised in Figure \ref{science_summary}), and 12 level-one science requirements (based on Science Requirements Document JUI-EST-SGS-RS-001).  These requirements on the spacecraft capabilities and tour are emphasised in \textbf{boldface font} and discussed in detail in this Section, but first we briefly introduce the payload elements that will be crucial to achieving the JUICE Jupiter-science objectives:  an ultraviolet spectrograph \citep[UVS,][and Section \ref{uvs}]{23gladstone}; visible-light camera \citep[JANUS,][and Section \ref{janus}]{23palumbo}; near-infrared mapping spectrometer \citep[MAJIS,][and Section \ref{majis}]{23poulet}; sub-millimetre wave instrument \citep[SWI,][and Section \ref{swi}]{23hartogh}; a radio science experiment for atmospheric occultations (3GM, \citealt{23iess} and Section \ref{3gm}; PRIDE, \citealt{23gurvits}); and a radio and plasma wave instrument \citep[RPWI,][and Section \ref{rpwi}]{23wahlund}.  These studies, conducted synergistically by six onboard instruments and an Earth-based experiment (PRIDE), will be used to achieve the scientific goals described in the following sections.

\begin{table}[ht]
    \centering
    \begin{tabular}{l|l}
    \hline

\textbf{JA} & \textbf{Characterise the atmospheric dynamics and circulation} \\
JA.1 & Investigate the dynamics and variability of Jupiter's weather layer.\\
JA.2 & Determine the thermodynamics of atmospheric meteorology.\\
JA.3 & Quantify the roles of wave propagation and atmospheric coupling on energy and material \\
& transport.\\
JA.4 & Investigate auroral structure and energy transport mechanisms at high latitudes.\\
JA.5 & Understand the interrelationships between the ionosphere and thermosphere.\\
\hline
\textbf{JB} & \textbf{Characterise the atmospheric composition and chemistry} \\
JB.1 & Determine Jupiter's bulk elemental composition to constrain formation and evolution. \\
JB.2 & Investigate upper atmospheric chemistry and exogenic inputs from the stratosphere to the \\
& thermosphere. \\
JB.3 & Study spatial variation in composition associated with discrete phenomena and polar vortices. \\
JB.4 & Determine the importance of moist convection in meteorology, cloud formation, and chemistry. \\
\hline
\textbf{JC} & \textbf{Characterise the atmospheric vertical structure and clouds} \\
JC.1 & Determine the three-dimensional temperature, cloud and aerosols structure from Jupiter’s upper \\
 & troposphere to the lower thermosphere. \\
JC.2 & Study coupling by waves, eddy mixing and global circulation across atmospheric layers.   \\
\hline
    \end{tabular}
    \caption{JUICE Jupiter Scientific Objectives}
    \label{tab:objectives}
\end{table}

\subsection{Jupiter's Dynamic Weather Layer}
\label{dynamics}

Investigations of Jovian dynamics and meteorology are naturally biased to the dayside top-most clouds, where contrasts in colours, and rapid motions of small-scale meteorological phenomena, reveal the banded structure of winds, aerosols, temperatures, and gaseous composition.  This two-dimensional perspective samples a relatively unique interface, where the condensate cloud decks start to mingle with sunlight; where adiabatic lapse rates (both dry and saturated) become influenced by radiative heating to produce the statically-stable upper troposphere; and where photolysis of gaseous compounds by ultraviolet light can produce colourful hazes.  Although visible-light images inform much of what we know today about atmospheric dynamics on Jupiter-sized worlds \citep{04ingersoll, 05vasavada,19sanchez_jets}, spectroscopy is needed to sample all the different layers within a planetary atmosphere.  Spectroscopy from the UV to the sub-millimetre (Fig. \ref{spectra}) provides an invaluable tool to access the vertical dimension, probing the depths below the visible cloud tops ($\sim500-1000$ mbars), and extending measurements through the cold-trap of the tropopause ($\sim100$ mbar), the radiatively-controlled stratosphere (mbar-$\mu$bar pressures), and into the ionosphere and thermosphere (nbar pressures).  Spectroscopy from JUICE will exploit reflected sunlight, i.e., the solar spectrum with significant absorption from methane and other species to sound the vertical distribution of aerosols, and thermal emission, i.e., hydrogen-helium collision-induced opacity overlain by tropospheric absorption and stratospheric/ionospheric emission bands.  JUICE will use nadir views, limb views, and solar/stellar occultations, to probe the vertical domain and the transfer of energy, momentum and material between adjacent atmospheric layers.

\begin{figure*}[ht]
\begin{centering}
\centerline{\includegraphics[angle=0,width=\textwidth]{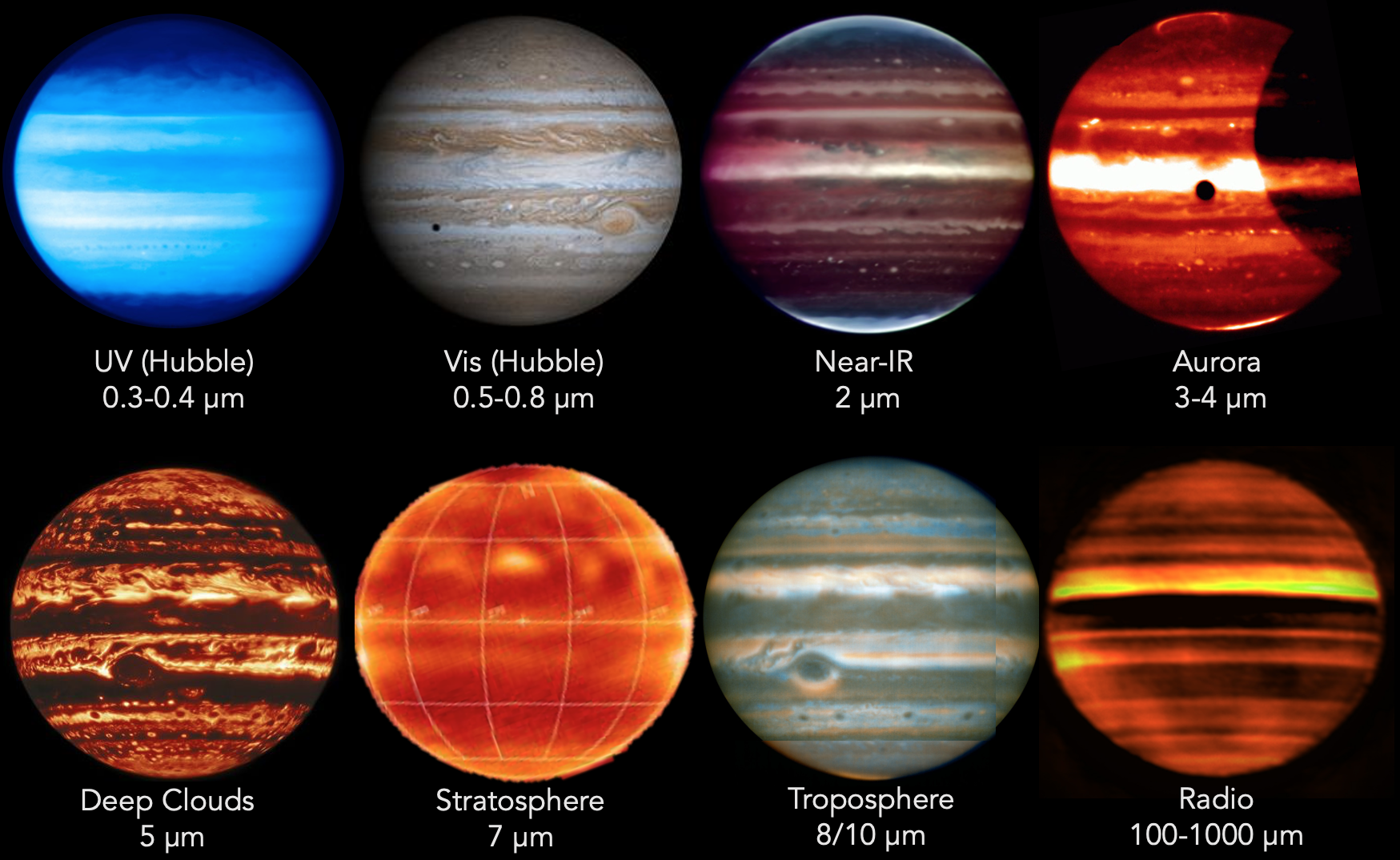}}
\caption{Multi-wavelength remote sensing of Jupiter provides access to both reflected sunlight (UV to near-IR) and thermal emission (mid-IR to radio).  These false-colour images demonstrate the appearance of the atmosphere at different wavelengths.  JUICE UVS will measure scattered sunlight from upper-tropospheric aerosols.  JANUS and MAJIS observations (below approximately 3 $\mu$m) sense clouds, chromophores and winds in the cloud decks using both the continuum and strong CH$_4$ absorption bands \citep{17hueso_jup, 20grassi}.  MAJIS will be able to observe H$_3^+$ emission from Jupiter's ionosphere and auroras between 3-4 $\mu$m (VLT/ISAAC observations, Credit: ESO), as well as thermal emission from the deep cloud-forming layers (4-6 bars) near 5 $\mu$m \citep[Gemini/NIRI observation,][]{20wong}.  Although JUICE lacks mid-IR capabilities \citep[VLT/VISIR observations sensing the upper troposphere at 0.1-0.5 bars, and stratosphere at 1-10 mbar,][]{17fletcher_neb} and radio-wavelength capabilities \citep[VLA observations,][]{16depater}, sub-millimetre sounding by SWI will probe the stratospheric temperatures and winds.  The approximate sensitivity of the JUICE instruments to different altitudes is shown in Fig. \ref{synergies_fig}.}
\label{em_spec}
\end{centering}
\end{figure*}

\begin{figure*}[p]
\begin{centering}
\centerline{\includegraphics[angle=0,width=\textwidth]{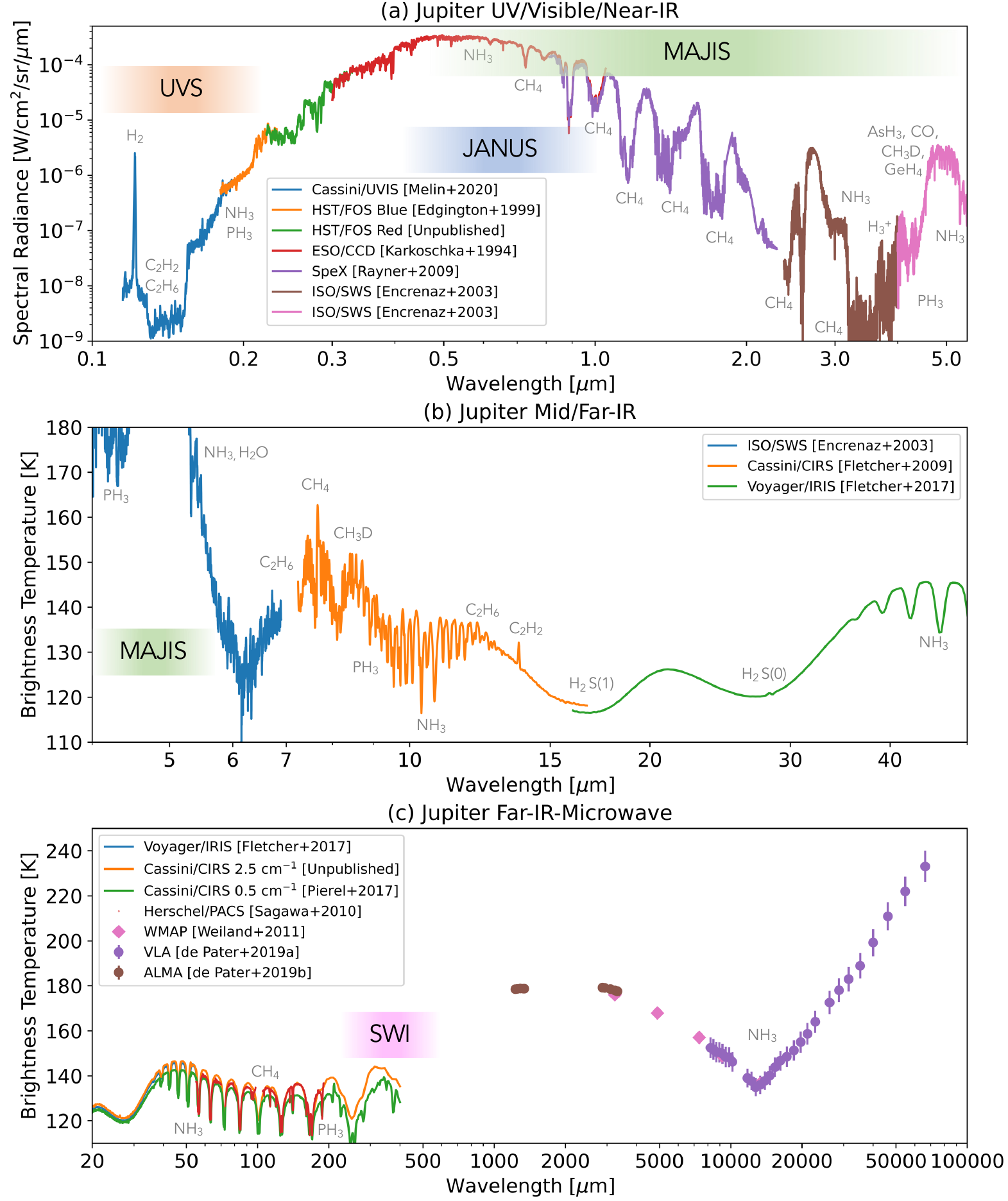}}
\caption{Overview of Jupiter's reflected ($\lambda<4$ $\mu$m) and thermal emission ($\lambda>4$ $\mu$m) spectra, with key molecular features labelled, and the approximate ranges covered by UVS, JANUS, MAJIS and SWI. UV, visible, and near-IR spectra in (a) were created from a low-latitude spectrum from Cassini/UVIS \citep{20melin}; Hubble FOS spectra at $6-20^\circ$N acquired in November 1992 with the blue and red detectors \citep{99edgington}; disc-averaged measurements from the European Southern Observatory \citep{94karkoschka} converted from albedo to spectral radiance assuming the solar spectrum of \citep{18meftah}; disc-averaged measurements from IRTF SpeX instrument \citep{09rayner} approximately scaled to match adjacent datasets; and disc-averaged ISO/SWS measurements from \citet{03encrenaz}. Mid- and far-IR spectra in (b) were from ISO/SWS, plus low-latitudes averages from Cassini/CIRS \citep{09fletcher_ph3} and Voyager-1/IRIS \citep{17fletcher_sofia}. Far-IR to microwave spectra in (c) were from averaged Cassini/CIRS spectra \citep{17pierel}, Herschel/PACS observations \citep{10sagawa}; and disc-averaged brightnesses from WMAP and ALMA in the millimetre \citep{11weiland, 19depater_alma} and VLA in the centimetre \citep{19depater_vla}.  }
\label{spectra}
\end{centering}
\end{figure*}

This capability to view the Jovian atmosphere in three dimensions will be exploited over a variety of spatial scales, from the largest circulation patterns, to the smallest storm systems and waves. 

\subsubsection{Belt/Zone Circulation}
\label{trop_dynamics}

\begin{figure*}[ht]
\begin{centering}
\centerline{\includegraphics[angle=0,width=\textwidth]{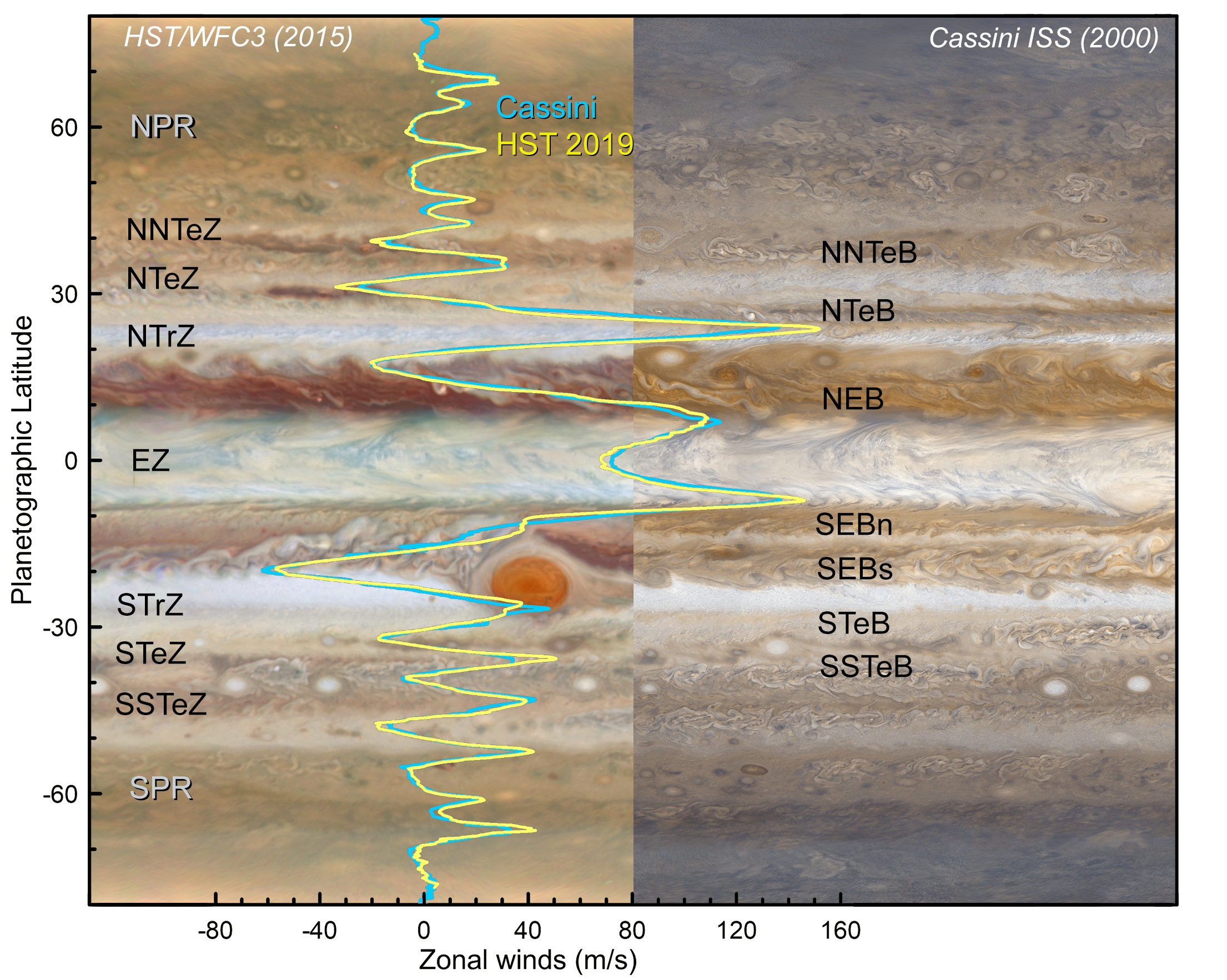}}
\caption{Jupiter belts and zones, defined by the zonal winds, compared to contrasts in colour and reflectivity. White zones and reddish belts alternate in latitude following the anticyclonic and cyclonic shear of the zonal jets. The locations and overall characteristics of the jets and the bands are stable in time, but the magnitude of the winds and the intensity of the belt/zone colors are variable. Zonal winds in this figure come from Cassini in 2000 \citep{03porco} and from Hubble images from 2019 following an equivalent analysis to that presented in \citet{17hueso_jup}. The conventional names of zones (left) and belts (right) are given. The HST background on the left comes from the HST/OPAL program and is available at \url{http://dx.doi.org/10.17909/T9G593}. The Cassini map is available at NASA photojournal as image PIA02864.}
\label{fig:Jupiter_bands}
\end{centering}
\end{figure*}

The dominance of the Coriolis force in the momentum balance on a rapidly rotating planet leads to the generation of a system of planetary bands.  Jupiter's system of zonal (east-west) jets has been remarkably stable over multiple years \citep{01garcia,03porco,17hueso_jup, 17tollefson}, despite significant variability in cloud coverage and aerosols.  The jets themselves appear to be maintained by an upscale flow of energy, from the smallest scales to the largest scales, with eddies and storms feeding momentum into the zonal flows \citep{19sanchez_jets}.  Given Jupiter's rapid rotation, the jets are in geostrophic balance, as the Coriolis force is in balance with the forces exerted by the pressure gradient, and the thermal wind equation \citep{04holton} relates the vertical shear on the winds to latitudinal temperature contrasts in the upper troposphere.  At altitudes above the top-most clouds, the tropospheric winds are found to decay with increasing height \citep{81pirraglia, 04flasar_jup, 16fletcher_texes}, and the temperature gradients and zonal jets are so well co-aligned that both are used to define the latitudes of Jupiter's canonical warm, cyclonic `belts' and cool, anticyclonic `zones.'  These axisymmetric bands sometimes (but not always) exhibit contrasts in aerosol properties - zones are often considered to be bright and reflective, as volatile species like ammonia become saturated and condense to NH$_3$ ice at the cool temperatures of the zones.  Conversely, aerosols evaporate/sublime in the warmer and typically cloud-free belts.  But this correspondence between aerosols and the belt/zone boundaries is only well defined around the equator, with its typically-white Equatorial Zone (EZ) bordered by the typically-brown North and South Equatorial Belts (NEB and SEB). At mid-latitudes, the banding becomes more tightly packed, with Tropical Zones giving way to a series of Temperate Belts in each hemisphere, each bordered by prograde (eastward) jets on the equatorward edge, and retrograde (westward) jets on their poleward edge.  Here the correspondence between the thermal/wind banding and the aerosol properties begins to break down \citep{20fletcher_beltzone}.  The last detectable zonal jets, around $65-70^\circ$ in each hemisphere, give way to a polar region dominated by smaller-scale vortices and large cyclones \citep{17orton_juno,Mura2022}, albeit still with some form of latitudinal organisation (see Section \ref{vortices}).  A diagram presenting the zone/belt structure is shown in Figure \ref{fig:Jupiter_bands}.

Jupiter's belts and zones therefore appear to differ as a function of latitude, and their appearance at least at wavelengths sensitive to aerosols appears to change over poorly understood timescales \citep{17fletcher_cycles, 18antunano_EZ, 19antunano}.  To better understand the circulation patterns associated with the planetary banding, JUICE will probe their vertical aerosol and gaseous structures via spectroscopy, and characterise the fluxes of momentum and energy into the zonal jets.  Crucially, the JUICE orbital tour enables long-term monitoring of the winds, clouds, and composition, to see how they change along with the axisymmetric `upheavals' to their appearance.  For example, the North Equatorial Belt undergoes periods of northward expansion and contraction with a 4-5 year period \citep{17fletcher_neb}; the Equatorial Zone exhibits periodic clearings of clouds with a 6-7 year period \citep{18antunano_EZ}; the North Temperate Belt exhibits spectacular plume activity on a 4-5 year period \citep{16sanchez}; and the South Equatorial Belt displays disturbances, fades (whitening) and revivals with periods of 3-7 years \citep{96sanchez, 17fletcher_seb}.  These timescales, or at least their half-cycles, are within reach of the JUICE mission.  

\textbf{Measuring Winds:}. Determination of wind speed and direction requires the monitoring of cloud tracers \citep{19sanchez_jets}, usually over one Jupiter rotation (10 hours), but sometimes over smaller time-scales (0.5-2.0 hr) on particularly active regions in convective storms, turbulent regions or inside vortices.  Continuum-band imaging, i.e., away from strong methane absorption, where the atmosphere is relatively transparent down to the NH$_3$-ice cloud tops, and methane-band imaging (i.e., sensing the upper tropospheric hazes) can be used to determine how winds vary with altitude, a direct measure of the vertical wind shear that can be compared to maps of tropospheric temperatures derived from continuum spectra measured in the sub-millimetre.  The resultant wind maps can reveal zonal and (weak) meridional motions, and resolve the motions of individual eddies to understand momentum convergence on the zonal jets \citep{06salyk}, as well as the kinetic-energy and turbulence spectra at the cloud tops.  By observing how this changes with time, such maps will allow JUICE to explore the variability of the energetics of the jets, particularly in relation to discrete storm activity and planet-wide changes.  JUICE will this be able to \textbf{`globally determine the vertical structure of zonal, meridional and vertical winds and eddy fields to understand the mechanisms driving zonal jets and meteorological activity (R1-J-5).'}  The close-in orbit of Juno, whilst providing high-resolution regional views of atmospheric phenomena with JunoCam \citep{17hansen}, cannot provide the global temporal coverage needed to study the global windfield.

\begin{figure*}[ht]
\begin{centering}
\centerline{\includegraphics[angle=0,width=1.3\textwidth]{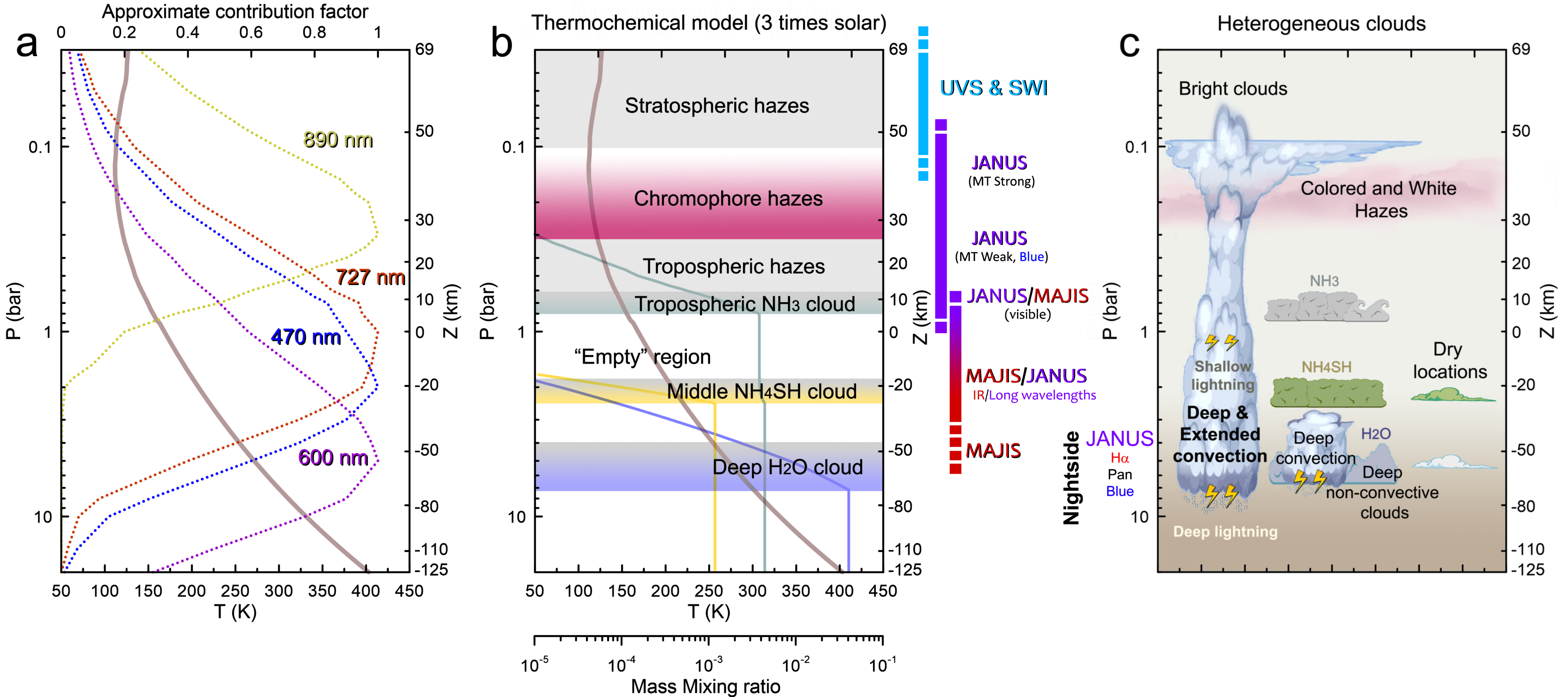}}
\caption{Vertical structure of Jupiter's troposphere and lower stratosphere. Deriving the vertical cloud structure at different locations from JANUS/MAJIS data will require the use of radiative transfer models. Reflected-sunlight observations will be sensitive to levels from $\sim100$ mbar (in the methane absorption band with JANUS) to at least 2.5 bar (in the IR images from MAJIS) with some contributions from deeper layers \citep{23wong}. (a) In a cloudless atmosphere imaging filters from the blue to red wavelengths can penetrate deep in the atmosphere limited by Rayleigh scattering and methane absorptions at specific wavelengths such as in 727 and 890 nm (contribution functions for single wavelengths from \citealt{21dahl}). (b) The nominal cloud structure in Jupiter consists of layers of ammonia, ammonia hydrosulphide and water clouds with approximate cloud bases at around 0.7, 2.5, 5-7 bar, respectively, depending on the local abundance of condensables.  The thermochemical calculation shown here assumes 3 times solar abundance of condensables \citep[see][for details]{99atreya}, but the NH$_3$ abundance below the condensation level is not well mixed, and is now known to vary substantially as a function of altitude and latitude \citep{17li}. The upper ammonia cloud limits the penetration depth of visible light. Above the condensate clouds are higher-altitude hazes with varied properties in different Jupiter regions that can be sampled with a combination of methane band images and observations in near-IR wavelengths. (c) The real cloud structure is probably very heterogeneous with locations of deep convection, dry areas and intermediate cloud systems.}
\label{fig:Jupiter_clouds}
\end{centering}
\end{figure*}

This discussion naturally raises the question of how the windspeeds change as a function of depth.  Infrared imaging, particularly in the 4.5-5.7 $\mu$m range, senses thermal emission from the 2-6 bar region, with clouds in silhouette (i.e., absorbing) against the bright background.  Tracking of cloud tracers at these longer `M-band' wavelengths may enable JUICE to measure windshear immediately below the NH$_3$ ice clouds, down to the levels where the NH$_4$SH cloud is expected to form via combination of NH$_3$ and H$_2$S, and possibly the condensation levels of H$_2$O clouds (see Figure \ref{fig:Jupiter_clouds}).  The Galileo probe revealed that winds appeared to strengthen from the cloud tops to the 5-bar level \citep{98atkinson} for a single location (the jet stream separating the NEB and EZ), whilst microwave contrasts measured by Juno \citep{20oyafuso} were suggestive of the same strengthening of zonal winds at all latitudes down to the $\sim6$ bar level of the H$_2$O cloud \citep{21fletcher}.  However, degeneracies between ammonia absorption and physical temperatures prevent a unique interpretation of microwave data, so JUICE will attempt to use visible and near-infrared observations to directly determine windshear and atmospheric stability across all of Jupiter's belts and zones down to approximately $\sim5$ bars.

\textbf{Clouds and Hazes:} JUICE has two further techniques to determine the properties of the belts and zones - by mapping the distributions of aerosols and gases.  The vertical distribution of aerosols - both condensed volatiles like NH$_3$ ice, and photochemical hazes like, possibly, hydrazine N$_2$H$_4$ - can be derived by modelling reflected-sunlight spectra in the near-infrared, as the differing strengths of gaseous CH$_4$ absorption provide sensitivity across a range of altitudes.  The phase function of aerosol scattering can be used to investigate the size, shape, and possible chemical composition of the aerosols. This remains a considerable unknown - the clouds are certainly not pure condensates \citep[e.g.,][]{10sromovsky_iso,Perez-Hoyos2020}, but could be aggregates of multiple compounds, seeded around a cloud-condensation nucleus that could be photochemical in origin \citep[e.g.,][]{04west}.  However, breaking the degeneracies between the optical properties, composition, and vertical structure requires sampling the aerosol population under a range of illumination conditions and viewing geometries, from nadir low-phase imaging in noon sunlight, to observations of the dawn and dusk terminator regions.  Darkening as observations approach the planetary limb and terminator can provide invaluable constraint on the aerosol properties.  Diagnostic spectral signatures of pure NH$_3$ ices, H$_2$O ice and NH$_4$SH all exist in the near-infrared accessible to MAJIS, particularly near 2-3 $\mu$m \citep{02baines, 10sromovsky_iso, 18sromovsky_nh3}, which can be used to understand the existence of fresh ices in regions of strong convective activity.  

The JUICE orbital tour was required to \textbf{`provide sufficient spectral, latitudinal, illumination and phase angle coverage to investigate Jovian aerosols from the condensation clouds to upper tropospheric and stratospheric hazes (R1-J-10).'}  The nature of these Jovian aerosols is expected to be tied to their formation environments, such as fresh white condensates in cold zones, or stagnant UV-photolysed hazes in quiescent vortices, so JUICE will \textbf{`relate the global temperature and wind structure to visible properties (albedo, winds, clouds) and atmospheric chemistry (R1-J-3).'}  Finally, the ability to view the limb of Jupiter at high spatial resolution, using visible images and infrared spectral maps, could also provide access to thin, tenuous haze layers in the lower stratosphere, which in turn can be tied to the vertical thermal structure derived via sub-millimetre sounding the solar/stellar occultations. Figure \ref{fig:Jupiter_clouds} shows the various altitudes and cloud layers probed by the remote sensing instruments. 

\textbf{Volatiles and Disequilibrium Species:}. Clouds are intricately linked to the supply of volatile species (NH$_3$, H$_2$S , H$_2$O) to condense on condensation nuclei, and colourful aerosols are linked to the supply of chemicals (e.g., PH$_3$, NH$_3$, sulphur-bearing species) that can be photolysed by UV irradiation above the clouds. The strength of vertical mixing within the belts and zones is a crucial missing piece to understand this puzzle.  Upper tropospheric PH$_3$ and NH$_3$ are known to be elevated in the anticyclonic zones and depleted in the cyclonic belts \citep{86gierasch, 06achterberg, 09fletcher_ph3, 20grassi}.  The EZ is the only region with a significant deep column of NH3 gas below the ammonia clouds \citep{16depater, 17li}. The ammonia rich EZ renders this region dark at microwave wavelengths sensitive to depths from 1 bar all the way down to 100 bar.  Saturn displays a similar connection between gaseous species and the meridional circulation on the scale of the belts and zones \citep[e.g.,][]{11fletcher_vims, 13laraia}.  

The meridional circulation derived from temperature measurements and jet decay was first explored during the Voyager era \citep{81pirraglia, 83conrath}. The circulation associated with the equatorial zones and belts has been likened to the Earth's Hadley circulation \citep{70barcilon}.  On the other hand, the mid-latitude jets may be similar to Earth's Ferrel-like circulations \citep{20fletcher_beltzone} that may exhibit different directions (upwelling, downwelling) above and below the water-condensation level \citep[as revealed by contrasts in microwave brightness associated with NH$_3$ and temperatures in the 0.1-100 bar region,][]{21duer, 21fletcher}.  This hypothetical vertically-stacked series of cells with different circulation regimes is testable using measurements of gaseous species as tracers and the distribution of moist convection inferred from lighting \citep{00ingersoll, 20fletcher_beltzone}.  However, the belt/zone variability of several gaseous species accessible in the 4.0-5.7 $\mu$m range, including AsH$_3$, GeH$_4$, CO, and H$_2$O, remains unclear, largely due to the challenge of accessing the trace abundances, and degeneracies associated with the distribution of aerosols \citep{17giles, 18bjoraker, 20grassi}.  Both PH$_3$ (160-180 nm) and NH$_3$ ($>160$ nm) also provide absorption in the ultraviolet \citep{98edgington, 20melin}, sensing higher altitudes of the upper troposphere where photochemical depletion dominates.  It is possible that vertical motions, and associated transport of materials, is localised within discrete meteorological features (see Section \ref{convection}), rather than being elevated over an entire planetary band.  JUICE will map the spatial distributions of each of these species, and monitor their variation over months and years, in an effort to understand the belt/zone circulation patterns within and above the cloud-forming region of the troposphere.

At even higher pressures, below the cloud-forming layers, Juno has revealed that the cloud-level winds persist down to approximately 3000 km depth, decaying away before reaching the transition to metallic hydrogen \citep{18kaspi,20kaspi,18guillot}.  The truncation of these winds could potentially be due to stabilising compositional gradients or radiative zones at great depth \citep{20christensen}, but must occur before differential winds reach the conducting, uniformly rotating interior where the dynamo originates \citep{17cao}.  Juno's microwave radiometer can probe below the clouds to depths of $\sim300$ km, but unexpectedly found that NH$_3$ still showed spatial belt/zone variability and global depletion \citep{17bolton, 17li, 17ingersoll}.  The absence of microwave remote sensing and close-perijove gravity measurements on JUICE means that it will not directly reproduce these Juno discoveries, but it will scrutinise the interface region down to $\sim5$ bars - the weather layer sitting above the deeper troposphere - using infrared spectroscopy to understand how it couples to the deeper circulation patterns revealed by Juno.

\subsubsection{Vortices}
\label{vortices}

\begin{figure*}[ht]
\begin{centering}
\centerline{\includegraphics[angle=0,width=\textwidth]{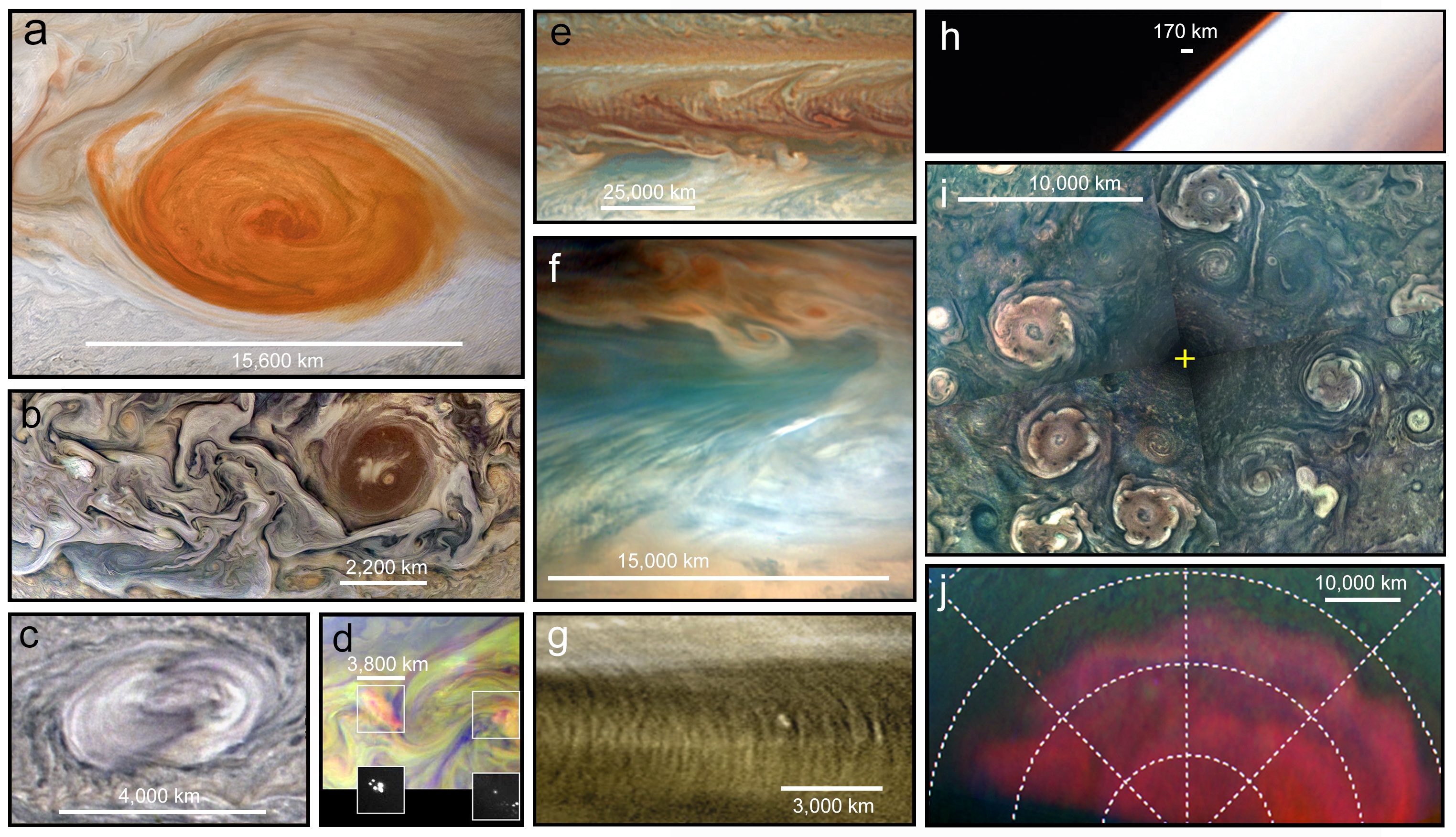}}
\caption{Morphology and variety of atmospheric features at different spatial scales (colours are approximate and contrasts have been enhanced for visibility):  (a) Jupiter's Great Red Spot. (b) Turbulent filamentary features at a 52\degre N (white structures) near a dark brown vortex, both showing cyclonic morphologies in their clouds. (c) Convective storm at 31\degre S. (d) Dayside storms with lightning observed at the same location on the nightside. (e) Series of short-scale gravity waves in Jupiter’s cloud at 17\degre N above a series of large dark features at 8\degre N in Jupiter's North Equatorial Belt. (f) One of the dark projections of the North Equatorial Belt, sometimes known as a 5-$\mu$m hotspots. (g) New Horizons observation of small-scale gravity waves in Jupiter's atmosphere. (h) Galileo observations of Jupiter's limb in violet and near infrared light at 756\,nm. (i) Composite map of Jupiter's North polar region in polar projection from Junocam observations obtained on different perijoves. (j) combination of visible and near IR observations of Jupiter's South polar region sampling polar hazes structured as a circumpolar wave. Latitudinal grid is superimposed each 10\degre.  Credits and sources: (a) and (b). Junocam images acquired on February 12, 2019 with credits: NASA / JPL-Caltech / SwRI / MSSS / Kevin M. Gill. (c) Excerpt from a Junocam observation obtained on June 2, 2020. (d) Combination of Galileo SSI images obtained on May 4, 1999. NASA / JPL-Caltech. Image (e) is an HST observation from April 1, 2017 from \citet{18simon}. Image (f) is a Junocam observation obtained on Sept. 16, 2020. (g) New Horizons views from the MVIC instrument of equatorial waves on Feb. 28, 2007 with credits from NASA/Johns Hopkins University Applied Physics Laboratory/Southwest Research Institute. (h) Galileo December 20, 1996; NASA / JPL-Caltech. (i) Image composite from Junocam images obtained on Feb. 17, April 10, June 2, and July 25 of 2020. NASA/JPL-Caltech/SwRI/MSSS / Gerald Eichstädt, John Rogers. (j) Adapted from \citet{08barrado}.}
\label{fig:montage_vortices}
\end{centering}
\end{figure*}

Jupiter's banded appearance is disrupted by the presence of a diverse collection of geostrophic vortices, both anticyclones (high-pressure centres, with anticlockwise circulation in the southern hemisphere) and cyclones (low-pressure centres, with clockwise circulation in the southern hemisphere), as displayed in Figure \ref{fig:montage_vortices}.  These vortices possess the same sign of vorticity as the environment in which they are embedded, and are prevented from migrating with latitude by the strong shears associated with the system of zonal jets. Jupiter displays a fundamental asymmetry between the two types - anticyclones appear larger and more numerous than cyclones.  Anticyclones appear to grow at the expense of other anticyclones, as was the case for Oval BA, which formed from three smaller anticyclones in the South Temperate Belt \citep{01sanchez}.  These vortices are relatively shallow `pancake-like' structures, with horizontal extents orders of magnitude greater than their depths.  They are thought to possess a midplane somewhere in the cloud-forming region \citep{14palotai, 20lemasquerier}, where tangential velocities and the pressure differences are at a maximum. Their windspeeds decay via the thermal wind equation (a `despinning') with both altitude and depth.  Thus an anticyclone will exhibit a cold anomaly in the upper troposphere, a cyclone will exhibit a warm anomaly, and such thermal contrasts have been confirmed by mid-infrared thermal imaging \citep[e.g.,][]{10fletcher_grs, 20wong}. Below the vortex midplane, there is evidence from Juno for warm cores beneath anticyclones, and cold cores beneath cyclones \citep{21bolton, 21parisi}. However, these deep levels will not be accessible to JUICE.

\textbf{Anticyclones: } The Great Red Spot (GRS) (Figure \ref{fig:montage_vortices}a) is the largest and longest-lived of all the vortices observed in planetary atmospheres \citep{95rogers}. Internally the GRS exhibits a variety of meteorological phenomena \citep{Sanchez-Lavega2018}, and the interaction of these large-scale vortices with the surrounding environment also has a significant effect on Jovian dynamics.  The Great Red Spot deflects jet streams to the north and south, which generate a `wake' of turbulent activity that promotes moist convective plumes (see Section \ref{convection}).  The peripheral winds appear to entrain material within the GRS \citep{Sanchez-Lavega2021}, such that the unidentified compounds responsible for the  orange-red haze \citep{19baines} are either irradiated for longer within the stagnant top of the anticyclone, or some unusual chromophore is supplied from below via secondary circulation with the vortex itself.  UV and infrared spectroscopy will be able to measure the aerosols and gaseous composition in the GRS compared to its surroundings, and to contrast this largest anticyclone to smaller white anticyclonic ovals \citep{Anguiano-Arteaga2021}.  Furthermore, the GRS has been steadily shrinking from $\sim$40000\,km in 1879 to its current value of $\sim$15000\,km \citep{18simon_grs}, resulting in changes to its velocity field, vorticity, and temperature structure at the upper cloud level.  Another large anticyclone, Oval BA, undergoes colour changes, from white to red and back again \citep{08cheng}.  By the early 2030s, JUICE will be able to assess any changes to velocities, vorticity, aerosol coverage and gaseous composition associated with these variable anticyclones.

\textbf{Cyclones: } Temporal variability is not just a feature of anticyclones.  Cyclonic vortices come in diverse shapes and sizes, from elongated and quiescent brown barges, to chaotic `Folded Filamentary Regions' (FFRs) at mid-to-high latitudes, to organised arrays of circumpolar (CPCs) and polar cyclones (PCs) at both poles (Figure \ref{fig:montage_vortices}i).  The connections between the dynamics of these different types of cyclones remains unclear, but cyclones do appear to promote moist convective activity (see Section \ref{convection}), which may be partially responsible for the chaotic and ever-changing appearance of the FFRs.  The CPCs and PCs revealed by Juno \citep{18adriani} challenge our understanding of Jupiter's polar domain - an octagonal arrangement at the north pole, and a pentagonal (sometimes hexagonal) arrangement at the south pole, whose long-term stability reveals the dynamics of atmospheric turbulence and the `beta-drift' of cyclones \citep{21gavriel}.  The inclined phase of JUICE, with sub-spacecraft latitude reaching up to 33$^\circ$, will provide a new glimpse of the polar domain, with its FFRs and polar cyclones, several years after the culmination of the Juno mission.  Whilst the JUICE inclination (see Section \ref{tour}) does not match the polar orbit of Juno, it does provide a long-term vantage point to observe how these polar features move and change over daily or weekly timescales, and a broader infrared spectral range to study their aerosols and composition.  Thus the JUICE mission will be able to \textbf{`determine the three-dimensional properties of discrete atmospheric features at high spatial resolution and track them over time (e.g., Great Red Spot, vortices, atmospheric plumes, 'brown barges’) (R1-J-6)'. }

\subsubsection{Convective Storms and Lightning}
\label{convection}

On the smallest scales, the high-resolution imaging and spectroscopy of JUICE will be able study individual storm cloud features as a window onto moist convection in hydrogen-rich atmospheres.  Although latent heat release at phase transitions can drive heat transport in atmospheres, the high molecular weight of condensates (compared to the hydrogen atmosphere) can have an inhibiting effect on convection, which must be overcome \citep{95guillot, 02hueso, 17leconte} to generate the storms that we see.  Such storms may help to regulate heat flux through the tropospheric layers, and as such may play a role in the thermal evolution and banded structure of Jupiter.

Storm plumes are observed as small white spots in dynamically-active domains (Figure \ref{fig:montage_vortices}c), such as the centres of cyclonic vortices \citep{Hueso2022}, or the wake of the GRS \citep{02baines}.  Individual cumulus-like clouds can be seen in Juno high-resolution imaging, often adding texture to larger-scale stratiform clouds.  These cumulus clouds are most likely powered by the latent heat release of water condensation in the $\sim6$-bar region, providing enough buoyancy to rise through the hydrogen-rich air.  Shallow convection, at altitudes too cold for liquid water and potentially associated with latent heat release in NH$_3$-ice cloud layers, may also be occurring \citep{20becker,Hueso2022}, and the complex blend of water and ammonia ice may be forming slushy `mushballs,' \citep{20guillot_mushball}, which trap NH$_3$ gas, precipitate, and then release their payload at several tens of bars \citep{20guillot_ammonia}.  Thus convective motions, and associated precipitation, play a vital role in shaping the vertical structure of aerosols and gaseous composition, both on the largest scales (belts and zones) and smallest scales (surrounding individual storm plumes).

Remote sensing from JUICE will examine these thunderstorms, determining the vertical aerosol structure in the upper troposphere, and the spatial distribution of volatiles (e.g., NH$_3$ and H$_2$O) and disequilibrium species (e.g., PH$_3$, AsH$_3$, GeH$_4$, CO) as tracers of vertical motions.  Spectroscopic maps in the UV and infrared will be compared to the morphology of the cumulus clouds at the highest spatial resolutions. 

JUICE will also examine the distribution and energetics of Jovian lightning, using nightside imaging to detect flashes \citep[e.g.,][]{86borucki, 99little, 04dyudina, 07baines}, and listening for radio emissions generated by electrical discharges in the Jovian atmosphere and propagating through the plasma environment. Using this technique during the first quarter of the Juno mission, the Waves instrument has made about two thousand lightning detections. This represents the largest data set on Jovian lightning processes collected to date. Close to Jupiter, low dispersion rapid whistlers occurred at frequencies from 50\,Hz to 20\,kHz \citep{18kolmasova}, and the so-called Jupiter dispersed pulses (JDPs) were recorded at frequencies between 10\,kHz and 150\,kHz \citep{Imai2019}. The rapid whistlers have dispersion from units of milliseconds to a few tens of milliseconds, and the dispersion of JDPs is even lower. The latter might propagate in the free space ordinary mode through low density regions in Jupiter’s ionosphere \citep{Imai2019}. The third kind of Jovian lightning radio bursts are long dispersion whistlers lasting for several seconds, which were previously detected at frequencies of several kHz in different regions of the Io torus between 5 and 6 Jovian radii using Voyager 2 measurements \citep{79gurnett,Kurth1985}. Due to the larger periapsis distances of JUICE ($\sim$9 to 20 Jovian radii, see Section \ref{opportunities}) compared to Juno or the Voyagers, it will be more difficult for JUICE RPWI to detect lightning radio pulses. However, lightning whistlers from very high latitudes as well as JDPs going through the ionospheric low-density patches might propagate that far and be recorded. 

In Jupiter's atmosphere, lightning activity is predicted to show the largest emission in the H$\alpha$ line at 656-nm \citep{Borucki1996}. However, nightside emissions (Figure \ref{fig:montage_vortices}d) of lightning observed by Galileo at 656-nm were ten times weaker than expected \citep{13dyudina}. This can be a consequence of deep lightning at pressures larger than a few bar. Shallow lightning on Jupiter at pressures near 1 bar were discovered by Juno at high latitudes \citep{20becker} and might be brighter at 656-nm than in clear filters. Dayside lightning in Saturn was observed in blue wavelengths during a large-scale storm \citep{13dyudina}.  Thus, searches of lightning in Jupiter's atmosphere by JUICE will have to use a combination of filters to test different scenarios of depth and intensity of the lightning.  Furthermore, JUICE will search for Transient Luminous Events (TLEs) in the ultraviolet from the night side in Jupiter's upper atmosphere \citep{20giles_tle}. Although lightning statistics from Galileo had suggested that lightning was predominantly found within the Jovian belts \citep{00gierasch}, Juno observations of microwave sferics and rapid whistlers indicated increased lightning activity at the middle and higher latitudes \citep{18brown,18kolmasova}. These distributions of lightning activity provides constraints on moist convection and the deep abundance of water \citep{98yair, 14sugiyama, 15li}.  JUICE will therefore re-examine the relationships between the distribution of lightning and the distribution of cumulus-like clouds, and \textbf{`determine the influence of moist convective processes by mapping the frequency, distribution and depth of tropospheric lightning (R1-J-7).'}

\subsubsection{Tropospheric Waves}
\label{waves}

Jupiter's atmosphere exhibits wave phenomena at a variety of scales, each providing a means of characterising the background atmosphere through which they propagate.  Longitudinal waves and curvilinear structures have been observed at the smallest scales of Galileo orbiter \citep{09arregi} and Juno imaging \citep{20orton}, and interpreted to be gravity (i.e., buoyancy waves in a stably-stratified atmosphere) or inertia-gravity waves (i.e., sensing the Coriolis effect).  Mesoscale waves modulating cloud opacity and reflectivity, often found in regions of cyclogenesis, have been observed by the Hubble Space Telescope (HST, Figure \ref{fig:montage_vortices}e), Juno, and ground-based telescopes \citep{18fletcher_waves, 18simon}.  And larger planetary-scale waves have been observed at low latitudes, including the equatorially-trapped Rossby wave responsible for the chain of 5-$\mu$m hotspots (visibly-dark formations, Figure \ref{fig:montage_vortices}f) on the jet separating the NEB and EZ \citep{90allison,06arregi}, and the large-scale thermal waves that often occur over the NEB during periods of expansion and contraction \citep{17fletcher_neb}.  Large-scale wave motions were recorded as movies by Cassini over two months in late 2000 \citep[e.g.,][]{13choi}.  The polar regions are covered by high hazes that stand out in images taken in methane absorption filters and in the ultraviolet, with their edge at about 65 deg latitude undulating with wavenumber 12 \citep[Figure \ref{fig:montage_vortices}j,][]{Sanchez-Lavega1998,Barrado-Izagirre2008}.  These Rossby waves, which require a gradient of the Coriolis parameter as the restoring force, are known to modulate aerosol reflectivity and upper tropospheric temperatures, but as yet the distribution of gaseous species and their phase speeds remains unclear.   

Understanding the origins of the waves (e.g., from instabilities, or arising from convection, or some other means) and the deposition of energy during wave breaking (providing or removing momentum from the zonal flows) requires assessment of their motions.  Measuring the phase velocities of waves requires long-term imaging and cloud tracking, with timescales tuned to the phenomenon of interest.  A JUICE requirement was therefore to \textbf{`classify the wave activity in the Jovian atmosphere, both horizontally (multi-spectral imaging) and vertically (R1-J-11).'}  We will return to the influence of wave phenomena on the stratosphere and upper atmosphere in the following sections, as a means to couple the meteorology of the troposphere with the circulations at higher altitudes.

\subsection{Chemistry and Circulation in the Middle Atmosphere}
\label{strat_dynamics}

The previous section described JUICE requirements for remote sensing of the troposphere, at the interface between the deep interior and the cloud-forming regions accessible to multi-wavelength observations.  However, a key strength of JUICE is its ability to probe the radiatively-controlled middle atmosphere, namely the stratosphere above the tropopause.  Here the stratified thermal structure is determined by a radiative balance \citep[e.g.,][]{20guerlet} between heating (absorption by CH$_4$ gas and aerosols) and cooling (thermal emission from ethane, acetylene, and to a lesser extent CH$_4$), and UV photolysis of methane generates a complex network of hydrocarbon species \citep{05moses} with emission features throughout the mid-infrared.  The JUICE remote sensing payload is required to \textbf{`characterise the three-dimensional temperature structure of the upper troposphere, stratosphere, and thermosphere (R1-J-9.5),'} enabling a comprehensive study of the Jovian middle atmosphere.

\begin{figure*}[ht]
\begin{centering}
\centerline{\includegraphics[angle=0,width=1.1\textwidth]{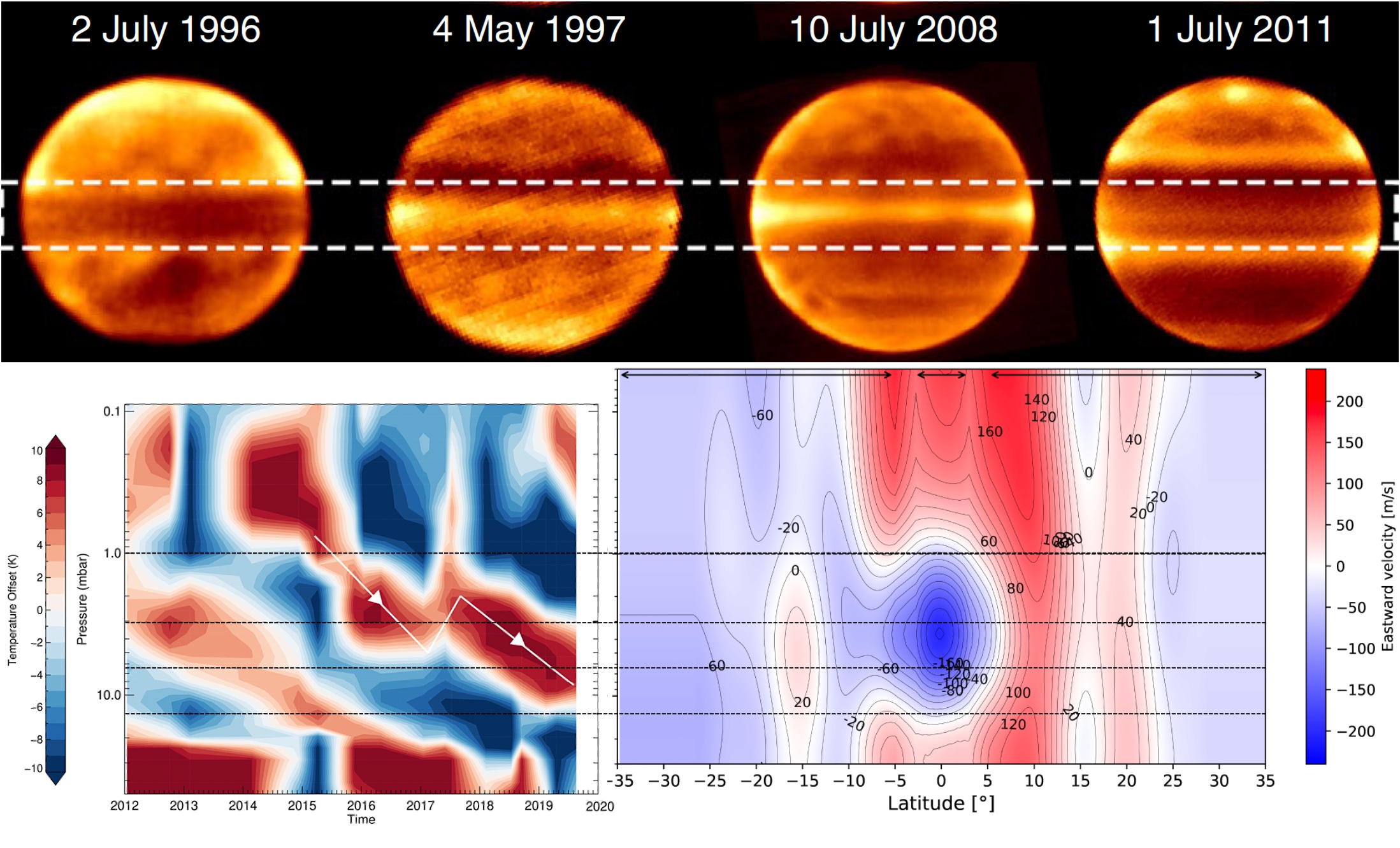}}
\caption{Jupiter's stratospheric equatorial oscillation was initially discovered from thermal infrared observations similar to those shown in the upper panel (adapted from \citealt{21antunano}). It is characterised by vertically alternating temperature extrema that descend with time. At a given pressure level (the 3, 6.4, and 13.5\,mbar levels are indicated with horizontal dashed lines), positive maxima occur approximately every 4.5 years. Over the years, this oscillation has shown some variability in its periodicity as demonstrated by \citet{20giles_qqo} (left panel). The oscillation is not only temporal, but also spatial, as demonstrated on the right panel by the vertically stacked prograde and retrograde jets at the equator (adapted from \citealt{21benmahi}).}
\label{Stratospheric_circulation}
\end{centering}
\end{figure*}

\textbf{Stratospheric Temperatures:} Jupiter's stratospheric circulation exhibits similar zonal organisation as the troposphere, with bands of warmer and cooler regions revealed via thermal imaging (Figure \ref{Stratospheric_circulation}).  The zonal organisation is strongest at low-latitudes, where Jupiter's equatorial stratospheric oscillation  \citep[often referred to as the Quasi-Quadrennial Oscillation, or QQO,][]{91leovy,91orton} modulates the 10-mbar thermal contrasts at the equator on a $\sim4$-year timescale.  Mid-infrared spectroscopy revealed that this is associated with a downward-moving chain of warm and cool anomalies \citep[and associated changes in stratospheric zonal jets,][]{04flasar_jup, 17cosentino, 20giles_qqo}, but long-term monitoring reveals that its period and phase can be substantially perturbed by tropospheric upheavals \citep{21antunano}.  The vertical structure of this pattern, which is thought to be driven by waves emanating from the troposphere and interacting with the mean flow \citep{99friedson}, can be sounded via emission in the sub-millimetre \citep{21cavalie,21benmahi}, particularly with CH$_4$ and H$_2$O from the lower stratosphere up to 10 $\mu$bar, but also via occultations - radio occultations as JUICE passes behind Jupiter as seen from Earth, and stellar/solar occultations as a star or the Sun sets behind the Jovian limb \citep[e.g.,][]{10greathouse}.  These techniques will reveal the vertical structure of the QQO.  Given the expected longevity of JUICE, the pattern of stacked wind/temperature anomalies will be monitored during their descent, sampling a full $\sim4$-year period over the duration of the mission.

\textbf{Gaseous Tracers:} Moving to mid-latitudes, the stratosphere again demonstrates coupling to the underlying troposphere.  The spatial distribution of stratospheric acetylene is asymmetric between the northern and southern hemisphere \citep{10nixon, 16fletcher_texes, 18melin}, possibly due to a difference in the strength of vertical mixing, rather than differences in the efficiency of photochemical production.  Furthermore, significant stratospheric thermal wave activity, likely to be near-stationary Rossby waves, are often reported in the northern hemisphere \citep{17fletcher_neb} where the acetylene is at its maximum.  Ultraviolet spectra from Juno, which exhibit strong C$_2$H$_2$ absorption, reveal that this abundance declines strongly towards the poles, as expected from the annually-averaged solar insolation, also indicating that horizontal equator-to-pole mixing is not strong enough to distribute C$_2$H$_2$ uniformly with latitude \citep{07nixon,21giles_cxhy}. On the other hand, the meridional distribution of C$_2$H$_6$ does not follow the mean solar insolation and increases towards the poles \citep{07nixon}, and altitude-latitude advective transport models cannot reproduce these C$_2$H$_2$ and C$_2$H$_6$ distributions \citep{18hue}.  

Recent observations of the polar regions have revealed an even more complex situation, in which the distributions of C$_2$H$_x$ species are not zonal. Local enhancements in abundances are correlated with the location of the diffuse auroras \citep{Sinclair2017b,Sinclair2018,Sinclair2019}. Other species, like HCN and CO$_2$, also show enhancements/depletion in the polar and auroral regions \citep{06lellouch,Cavalie2022b}. This is a strong indication that auroral chemistry plays an important role in these regions.  Thermal-infrared observations of the polar stratosphere and upper troposphere reveal cold polar vortices \citep{06simon, 16fletcher_texes}, possibly as a consequence of efficient radiative cooling from polar aerosols \citep{20guerlet}, for which C$_2$ species are probable precursors.  However, tying the aerosol distribution observed in reflected sunlight directly to the thermal structure will remain a challenge, particularly as a consequence of auroral heating discussed in Section \ref{aurora}. JUICE will investigate the thermal structure of the stratosphere via sub-millimetre sounding, as well as the distribution of tracers like C$_2$H$_2$ and C$_2$H$_6$ via UV spectroscopy \citep[these hydrocarbons have signatures in the FUV below 180 nm,][]{83gladstone}. It will also study the distribution of species deposited by Comet Shoemaker-Levy 9 (SL9) like HCN, CO, CS and H$_2$O in the sub-millimetre \citep[e.g.,][]{95lellouch_CO,97feuchtgruber}, to understand the middle-atmospheric circulation, from the equatorial QQO, to the mid-latitude waves, to the polar domain.  

\textbf{Stratospheric Winds:} The global stratospheric temperature field, and the distributions of hydrocarbons, are just two ingredients required for understanding the global circulation. Measurements of winds in the Jovian stratosphere have presented a challenge, given the absence of identifiable cloud tracers to monitor the flows at millibar pressures. Indirect determinations via stratospheric temperature gradients and the thermal wind equation \citep{04flasar_jup, 06read_jup} are subject to substantial uncertainties, because there is an altitude gap of two scale heights between the level where the cloud-top winds are used as initial condition (in the upper troposphere), and the levels where the winds are derived (in the stratosphere).  Direct and absolute measurement of stratospheric winds relies on high-resolution spectroscopy to reveal the Doppler shifts of individual sub-millimetre emission lines, recently demonstrated using the Atacama Large Millimeter/submillimeter Array (ALMA) \citep{21cavalie,22benmahi}.  The winds are measured at the levels probed by the spectral lines, usually in the middle stratosphere with the sensitivity of ALMA.  The derivation of the full stratospheric wind field then requires near-simultaneous temperature measurements to be combined with the wind observations using the thermal wind equation, as shown in \citet{21benmahi}.  Although it is an essential piece to constrain general circulation models and quasi-periodic stratospheric oscillations like Jupiter's QQO (see Figure \ref{Stratospheric_circulation}), such a combination has only been obtained once in ten years of ALMA operations.  The JUICE payload will enable the first maps of stratospheric winds, using emission from CH$_4$, H$_2$O, CO, HCN and CS lines.  Here we take advantage of an influx of chemical species from exogenic sources, such as long-lived oxygen-bearing species (H$_2$O, CO$_2$, and CO), HCN, and CS, many of which were deposited by the impact of the SL9 comet in 1994 \citep{95lellouch, 04harrington} and have been slowly diffusing with latitude ever since \citep{Moreno2003,06lellouch}.  High-resolution and high-sensitivity spectroscopy of these emission lines enables simultaneous sounding of stratospheric temperatures and direct wind measurements in the 10 $\mu$bar to 100 mbar range, with scale-height resolution, a first for an orbiting spacecraft.  Combining sub-millimetre sounding, performed every orbit for the entire JUICE tour, with UV, IR and radio occultations will provide significant opportunities for synergistic science for stratospheric circulation (see Section \ref{synergies}), with the goal to \textbf{`globally determine the vertical structure of zonal, meridional and vertical winds (R1-J-5).'}

\subsection{Global Composition and Origins}
\label{origins}

The origin and migration of Jupiter played a central role in shaping the present-day configuration of our Solar System.  Its bulk chemical composition (in comparison to that of the Sun) provides a window on the composition of the protosolar nebula at the time of planet formation \citep{03atreya, 04lunine, 20venturini}.  JUICE does not aim to reproduce the Juno capabilities of probing the bulk NH$_3$ \citep{17li} and H$_2$O \citep{20li_water} content of Jupiter's deep interior, nor will it perform gravity sounding of the zonal flows and density gradients in the hot and fluid interior \citep[e.g., the diffuse core extending up to 50\% of Jupiter's radius into the overlying molecular envelope,][]{17wahl}.  Furthermore, the picture of a relatively homogeneous hydrogen-helium envelope, mixed by efficient convection associated with its high intrinsic luminosity \citep[e.g.,][]{04guillot, 18li} is no longer suitable, with a gradient in heavier elements being more likely.  However, JUICE will contribute to our understanding of Jupiter's atmospheric composition from the cloud-forming region and upwards, and is specifically required to \textbf{`provide estimates of elemental abundances and isotopic ratios in the atmospheric envelope to constrain the composition of the deep troposphere and the origin of external material (R1-J-9),'} as woll be discussed below.


\subsubsection{Tropospheric Composition and Origins}
\label{trop_comp}
Thermochemistry in a strongly-reducing environment leads to the most cosmogonically abundant elements (C, N, O, S, P) appearing in their hydrogenated forms (CH$_4$, NH$_3$, H$_2$O, H$_2$S, PH$_3$).  At the low temperatures of the upper troposphere, the volatile compounds will condense to form clouds \citep[ices of NH$_3$ and H$_2$O, and a combination reaction to form NH$_4$SH,][]{73weidenschilling}, such that the main reservoirs for these gases are hidden at depth below the clouds.  Furthermore, Juno has revealed that NH$_3$ is not well-mixed even below its $\sim700-$mbar cloud base, remaining variable and depleted down to $\sim60$ bars \citep{17ingersoll, 17li}.  Spectroscopic measurements from JUICE will sound the abundances above, within, and immediately below the cloud-formation levels.  At these altitudes, the abundances are governed by a combination of vertical mixing, saturation, and photolytic destruction (e.g., NH$_3$ is photolysed to form N$_2$H$_4$, a possible contributor to aerosols in the upper troposphere).  JUICE contributions are therefore likely to be lower limits for tropospheric volatile enrichments N/H, S/H and O/H.  Nevertheless, mapping the latitudinal variability of these species (in relation to the belt/zone structure), and observing how they vary with time, places bulk measurements of these gases into a wider context, to understand how representative they might be of the planet as a whole.

Similarly, disequilibrium species like PH$_3$, AsH$_3$, GeH$_4$ and tropospheric CO can be studied via combination of UV and 4.0-5.7 $\mu$m spectroscopy.  These species are only present in the upper troposphere as the rate of vertical mixing is faster than the rate of their thermochemical destruction, leading to `quenched' abundances that are representative of deeper, kilobar levels \citep{84kaye}.  Their spatial and temporal variability therefore provides estimates of elemental abundances of P/H, As/H, and Ge/H that would be otherwise inaccessible.  Furthermore, tropospheric PH$_3$ and CO are limited by chemical reactions with water, providing an indirect means of estimating the deep water abundance via thermochemistry \citep{Bezard2002,Visscher2010,Cavalie2022a}. Thus, JUICE can provide indirect constraints on deep elemental abundances, without actually sampling below the cloud-forming layers.

The ratios of isotopes within a particular molecule can also reveal insights into the nature of its original reservoirs, and the proportion of ices incorporated into the forming protoplanets.  In particular, Jupiter's deuterium-to-hydrogen (D/H) ratio can be measured in methane \citep{03owen}, using absorption features of CH$_3$D in the 4.0-5.7 $\mu$m range. Separating CH$_3$D abundances from the properties of aerosols is challenging, and will rely on the techniques described in Section \ref{trop_dynamics}, exploiting reflected sunlight observations under multiple illumination conditions and geometries.  Nevertheless, the JUICE estimate of D/H in methane can then be compared to estimates of the D/H in hydrogen \citep{01lellouch} to understand the fractionation of deuterium between different molecules.  It will also be compared to the direct in situ measurement from the Galileo probe \citep{00mahaffy}, to assess how well that measurement represents the global composition of Jupiter. 

\subsubsection{Stratospheric Composition and Evolution}
\label{strat_comp}
Jupiter's stratospheric composition is determined by photochemistry of methane (which does not condense at Jovian temperatures), alongside the influx of exogenic species from interplanetary dust, bolides and larger impactors (asteroids and comets) entering the upper atmosphere \citep{05moses,Hue2018}.  As described in Section \ref{strat_dynamics}, JUICE remote sensing will be able to map the spatial distribution of stratospheric hydrocarbons, primarily C$_2$H$_2$ and C$_2$H$_6$ in the UV \citep{20melin, 21giles_cxhy,23giles,23sinclair} as well as CH$_3$C$_2$H in the sub-millimetre. An additional goal is to \textbf{`constrain the origin of external material (R1-J-9)'} via measurements of abundances and isotopic ratios in externally sourced materials.  This includes species like H$_2$O, CS, CO and HCN originating from the cometary impact (SL9) in 1994.  The latitudinal distribution of these species is expected to be governed by stratospheric circulation and diffusion since the time of impact \citep{Moreno2003,06lellouch}, but additional ongoing sources of exogenic materials (e.g., interplanetary dust, connections with Jupiter's rings, or with satellites via magnetic field lines) could be determined via new spatial maps acquired by JUICE in the sub-millimetre \citep[e.g.,][]{86connerney, 17moses}.  The vertical profiles of H$_2$O and CO can be mapped with scale-height resolution, providing an indication of the origin of the external oxygen.  

JUICE has another technique to determine the origin of these exogenic species - high spectral-resolution observations in the sub-millimetre will provide estimates of isotopic ratios in those molecules, for comparison with the wider Jovian environment.  Examples include D/H, $^{16}$O/$^{18}$O, $^{16}$O/$^{17}$O in water and CO; $^{12}$C/$^{13}$C in CO; and $^{12}$C/$^{13}$C and $^{15}$N/$^{14}$N in HCN, each of which may allow us to connect Jupiter's stratospheric species back to source populations in comets and other icy bodies.  For example, if Jupiter's external water originates from interplanetary dust particles and if these particles would enter the planet with slow velocities (so as not to be dissociated by the heat generated during atmospheric entry), then we would expect to find cometary D/H ratios in H$_2$O, i.e. 1--8$\times$10$^{-4}$ \citep{Anderson2022}.  Conversely, if the water were produced by comet impacts and would thus result from the recombination of cometary oxygen with Jovian hydrogen, then the water would exhibit a Jupiter-like D/H ratio, i.e. $\sim$2$\times$10$^{-5}$ \citep{Lellouch2001}.   By comparing the potential origins of stratospheric species with what is known of Jupiter's global composition, JUICE will be able to place new constraints on the formation and subsequent evolution of the gas giant.


Finally, JUICE may be lucky and glimpse examples of ongoing evolution in stratospheric composition.  The rate of impacts is being refined by Earth-based video monitoring of flashes in the Jovian atmosphere \citep{18hueso}, revealing bolide flashes (i.e., impactors disintegrating in the upper atmosphere) at a detectable rate estimated to be 0.4-2.6 per year.  Juno UVS has detected an impactor from orbit, and considering the probabilities of capturing such an event in a UVS scan, estimated a rate that was considerably higher \citep{21giles_impact}. Larger impactors, such as the 1994 Comet Shoemaker Levy 9 \citep{04harrington}, or the 2009 `Wesley' asteroidal impactor \citep{10sanchez, 10hammel}, remain much rarer \citep{Zahnle2003}.  Each of these events may have observable consequences for atmospheric composition (injecting water and silicate-rich materials, and producing high-temperature shock chemistry within the entry point), and the likelihood of tracking impactors in advance will improve with the commissioning of the forthcoming Vera Rubin observatory (see Section \ref{support}).  The flexibility and agility of the JUICE spacecraft to react to unique (and potentially unexpected) events will be discussed in Section \ref{opportunities}.

\subsection{Energetics of the Ionosphere, Thermosphere, and Auroras}
\label{ionosphere}

Beyond the dynamics of the troposphere, and the circulation and chemistry of the stratosphere, JUICE will explore the interface between the neutral atmosphere and the external charged-particle environment. Jupiter's ionosphere and thermosphere epitomises the JUICE goal of exploring coupling between components of the Jovian system, being influenced by the circulation and wave propagation of the lower atmosphere, and the deposition of energy in the polar region via electrons propagating along magnetic field lines.  Particle and fields in situ measurements, by PEP, RPWI and J-MAG, will directly monitor the energy and momentum exchange processes in the magnetosphere responsible for these energetic auroral particles impacting the atmosphere \citep[c.f.,][]{23masters}. This gives the electromagnetic energy flux and exact energy distributions flowing downward along the field lines, and can be compared to the auroral emissions detected in the atmosphere at various wavelengths. It will also put further constraints on the ionization, related ion-molecule and thermosphere aerosol formation chemistry, and thermal altitude profiles in the Jovian ionosphere, and electrodynamic processes that contribute to the atmosphere thermal balance. The locations of the auroral electron acceleration regions and radio emissions from atmospheric lightning can furthermore be remotely monitored by radio wave detections by RPWI. 

Above the methane homopause, solar EUV and auroral precipitation shape a region comprising H$_2$, He and H.   The ionosphere is formed from layers of thermal plasma embedded in the neutral atmosphere, from EUV/XUV ionisation or impact ionisation from high-energy precipitating particles. JUICE will explore this upper atmospheric region by means of (i) both radio occultations and stellar occultations to determine electron densities (from refraction and Doppler frequency shifting) and the temperature structure of the neutral atmosphere; and via (ii) imaging spectroscopy (UV and infrared emissions from H$_2$, H, and the H$_3^+$ ion) from the auroral regions to low latitudes.  Direct measurements of stratospheric winds via sub-millimetre Doppler shifts of spectral lines will also provide constraints on the winds in this region (see Section \ref{synergies}).  

\subsubsection{Thermospheric Circulation and the Energy Crisis}
\label{thermosphere}

From stellar occultations and observations of the H$_3^+$ ion \citep[generated on the dayside via the reaction of H$_2$ with H$_2^+$ produced by the ionisation of H$_2$ under solar Extreme Ultraviolet (EUV) radiation,][]{89drossart}, it is known that Jupiter's upper atmosphere and exosphere at low to mid-latitudes are systematically far hotter than can be explained by solar heating alone \citep{04yelle}, a conundrum known as the `energy crisis.' The auroral regions are bombarded by electrons that cover a broad range of energies, with low energy electrons (eVs) mainly contributing to heating the atmosphere, and the more energetic ones (100 eV and above) producing excitation, ionization, dissociation and subsequent auroral emissions, chemistry, and heating. However, thermospheric circulation models \citep[e.g.,][]{98achilleos,20yates} suggest that the strong Coriolis forces associated with Jupiter's rapid rotation should trap this energy at high latitudes.  However, recent H$_3^+$ observations during a potential solar wind compression event have revealed a steady decrease in temperature from the auroral regions to the equator, confirming the potential redistribution of energy from high to lower latitude regions \citep{21odonoghue}. This heating possibly occurs in pulses associated with a solar-wind compression, which facilitates the transport of heat to lower latitudes \citep{Yates2014}. Additional sources of heating from below (e.g., dissipation of gravity and acoustic waves propagating from the troposphere) may contribute to low-latitude variability in H$_3^+$ emission \citep{03schubert}, such as the possible excess heating over the Great Red Spot \citep{16odonoghue}. Such sources, localised and sporadic, also contribute to the highly structured and variable H$_3^+$ density profile in altitude \citep{Matcheva2001}. Furthermore, heating not only expands the atmosphere but can also drive vertical winds and affect turbulence. The location of the homopause, below which the main atmospheric species are well mixed and above which there is a diffusive separation of species by mass, is hence a tracer of energy deposition and dynamics.   

Remote sensing by JUICE will \textbf{`characterise the three-dimensional temperature structure of the ... thermosphere (R1-J-9.5)'}.  Vertically propagating waves observed in occultations throughout the atmosphere in the IR and the UV will be directly correlated with potential sources in the lower atmosphere (e.g., moist convective events and plumes) to understand their contribution to the energy budget.  The combination of UV and IR occultations is a powerful means to characterise Jupiter’s atmospheric vertical structure and composition, the dynamical coupling between layers, and the source of energy sources in the upper atmosphere. They will be complemented by measurements of the stratospheric windfield by SWI, and radio occultations \citep[e.g.,][]{Gupta2022}. In addition, UV observations will be critical to derive the global variations of the homopause in terms of height and eddy diffusion coefficient, characterising the amount of mixing, based on hydrocarbon tangential column densities from multiple UV stellar occultations, complemented by dayglow HeI 58.4\,nm resonance line observations (e.g., \citealt{Parkinson2006}, \citealt{Vervack1995}). Temperature profiles associated with H$_2$ can be derived from solar and stellar occultations in the UV \citep{15koskinen}. Furthermore, UV dayglow maps could help to identify the origin of the H Lyman alpha bulge, its possible connection with the auroral activity or with thermospheric/exospheric circulation, and its relation of any possible longitudinal asymmetry in HeI 58.4\,nm. 

\subsubsection{Jupiter's Auroras}
\label{aurora}

\begin{figure*}[ht]
\begin{centering}
\centerline{\includegraphics[angle=0,width=\textwidth]{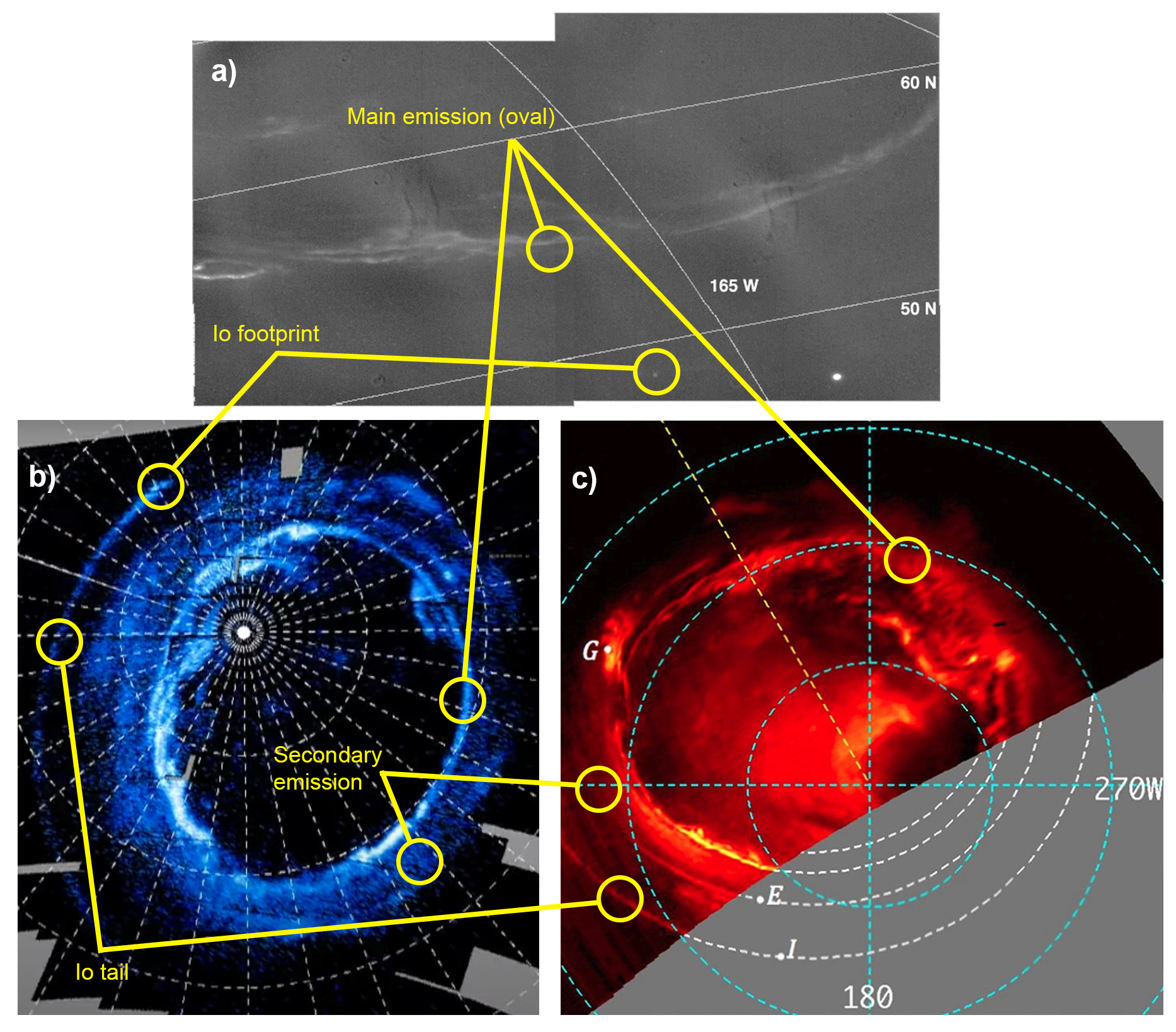}}
\caption{Example auroral images obtained from Jovian orbit at three wavelengths, at different epochs and orientations, with some key features labelled. Lines of System-III longitude and latitude are shown on each. See Fig. 2 of \citet{15grodent} for a comprehensive overview of features at UV wavelengths. a) Galileo SSI visible light image of the northern aurora, at a spatial resolution $\sim$26 km/pix, obtained in November 1997; presented in Fig. 1 of \citet{99vasavada}, and released as NASA PIA01602 (Credit: NASA/JPL-Caltech). b) Polar projection of May 2017 data from Juno UVS, here showing the southern aurora, presented in detail in Fig. 2 of \citet{21bonfond} (Credit: NASA/JPL-Caltech/SwRI/UVS/ULi\`{e}ge). c) Juno JIRAM mosaic map of the southern auroral oval, taken during August 2016. White points are predicted footprint positions indicated by letters I, E, and G for the moons Io, Europa, and Ganymede, respectively. Presented in detail by \citet{17mura}, Fig. 2 (credit: NASA/JPL-Caltech/SwRI/ASI/INAF/JIRAM).}
\label{auroras}
\end{centering}
\end{figure*}

Before the Juno mission, the main emissions (shown in Fig. \ref{auroras}) were thought to be due to co-rotation breakdown in the middle magnetosphere, with emissions related to upward currents \citep{01cowley} from the auroral ionosphere.  Juno observations have shown that particles and plasma wave phenomena are tightly linked in Jupiter's low-altitude auroral regions \citep{Kurth2018,Bonfond2021,Sulaiman2022}, where different latitudinally separated zones are linked to upward (zone I) and downward (zone II) electric currents \citep{20mauk}. \citet{Kurth2018} showed that the electron distributions and significant density depletions corresponding to zone I are coincident with brief but very intense broadband plasma waves propagating downward in the whistler mode at frequencies below ~10 kHz. \citet{Sulaiman2022} identified  H$^+$ / H$_3^+$ cyclotron waves in zone I in the presence of energetic upward H$^+$ beams and downward energetic electron beams, and large-amplitude solitary waves in zone II.  Juno has also detected evidence of large-scale electrostatic potentials above the main aurora, with broad electron energy distributions \citep{20allegrini}. Hence, we still probably lack a complete understanding of all the phenomena that produce the aurora.  

The variable magnetic footprints of Io, Europa, and Ganymede are visible equatorward of the main oval \citep[Fig. \ref{auroras},][]{08grodent}, associated with Alfv\'en, ion cyclotron, and whistler mode waves \citep{Sulaiman2020}, but the complex nature of their morphology and hence of the moon-plasma interaction, revealed by Juno close-in observations \citep{Mura2018}, is still to be fully understood.  A magnetic footprint of Callisto was tentatively observed by \citet{Bhattacharyya2018} and further observations are necessary to confirm it.  Interior to the main oval, the acceleration of charged particles may be responsible for the fluctuating polar-cap aurora \citep[e.g.,][]{03grodent}, with signatures of flares associated with magnetic reconnection on the dayside magnetopause being observed \citep[e.g.,][]{17ebert}, but the debate about whether the polar region is open or closed to the solar magnetic flux remains unresolved.  These auroral signatures, from the polar cap, to the main oval and satellite footprints, are all highly variable on timescales of minutes and hours \citep{04clarke}; signatures of possible substorm-like injections have been proposed \citep{Bonfond2021}.  \citet{Greathouse2021} reported that the bright polar emissions observed by the HST and Juno UVS on the day side dim substantially or disappear between midnight and dawn.  The JUICE UVS will have extended periods of time to scan the aurora on the night side to study this dimming in more detail, especially during the high inclination phase. Long-term HST imaging programmes \citep{09nichols, 09clarke, 17nichols}, as well as Juno IR and UV imaging \citep{Mura2017,Bonfond2017}, have provided substantial increases in our understanding of auroral morphology. In particular, the comparison between UV and IR is crucial to understand the complexity of the auroral-related energization processes \citep{Gerard2018,Gerard2023}, but a long-term programme of JUICE auroral monitoring will provide vantage points that are not possible from Earth-based facilities.

Visible light studies with JANUS will greatly enhance our understanding of the aurora in this wavelength range, which, unlike the IR and UV emission, cannot be observed on the dayside of Jupiter, thus ruling out studies from the Earth or its vicinity. Limited visible light observations by the Galileo solid-state imaging (SSI) system (e.g. Fig. \ref{auroras}) captured several classes of auroral features from a survey of the northern auroral region \citep{99vasavada}. These included a continuous primary arc a few hundred km wide, observed at around 245 km above the 1-bar level. This varied in morphology with local time between a single arc and a multiply-branched feature. A variable secondary arc, associated with the region just beyond Io's torus was also visible, plus a diffuse ``polar cap'' emission. A spot and tail associated with the magnetic footprint of Io was also observed. As the visible emission is too faint for Juno's JunoCam, JANUS will provide unprecedented data, at spatial resolutions as small as tens of km per pixel, and occasionally at high cadence, allowing the dynamics of these features to be captured.

The vertical structure of the auroral curtain, and its spectral properties, have been explored at high spatial resolution in the ultraviolet \citep{Bonfond2017,20mauk}, infrared \citep{17adriani_juno,Dinelli2017} and visible \citep{99vasavada}, which sense different altitudes and chemical compounds. With JANUS and MAJIS imagery at perijove, along with MAJIS and UVS occultations, JUICE will study the dynamic auroras in three dimensions and their temporal evolution. 

\textbf{Auroral Coupling to the Stratosphere:} Ion-neutral chemistry at high latitudes may be responsible for the unusual nature of upper tropospheric and stratospheric aerosols in the polar hazes \citep{02friedson,Wong2000}, possibly fractal aggregates \citep{13li} that are numerous and reflective in the infrared \citep{08barrado}, and dark and absorbent in the ultraviolet.  Remote sensing in the UV and infrared will characterise the composition and scattering properties of these polar hazes, to understand their asymmetric properties between the northern and southern poles, and their influence on radiative balance.  The consequences of auroral heating are visible at stratospheric altitudes \citep[e.g.,][]{18sinclair}, possibly via direct Joule heating or radiative effects in the auroral-associated hazes.  Sub-millimetre observations will enable joint measurements of stratospheric temperatures and winds within these auroral zones, to assess the penetration levels of auroral energy into the stratosphere.  Measuring the distributions of species like C$_2$H$_2$ and C$_2$H$_6$ \citep[e.g.,][]{18sinclair,23sinclair,23giles}, H$_2$O, HCN and CO$_2$ \citep{20benmahi,Cavalie2022b,06lellouch}, will also help shed light on chemistry occurring in the auroral regions, like aerosol production \citep{Perry1999,Wong2000,Wong2003,Friedson2002} and heterogeneous chemistry involving them \citep{Perrin2021}.

JUICE is required to \textbf{`investigate the unique atmospheric properties of Jupiter’s polar regions, including the influence of auroral energy deposition and ion chemistry on the atmospheric temperatures, energy budget, chemistry and cloud/ haze formation (R1-J-4)'}. As we discuss in Section \ref{opportunities}, the JUICE tour enables a long-term study of the Jovian auroras across multiple wavelengths, with observations tuned to the timescales of the various phenomena, observing variable emissions over minutes, hours and days as the conditions within the solar wind fluctuate.  The auroras will be particularly scrutinised as JUICE reaches higher orbital inclinations.  

\subsection{Summary of Objectives}

The JUICE science case is summarised by the eleven objectives listed in Table \ref{tab:objectives}, and by the boldface science requirements included in the text above.  The requirements state that the mission \textbf{`shall have the capability to investigate the spatial variability of Jovian dynamics, chemistry and atmospheric structure in three dimensions (R1-J-1)'}, and must provide \textbf{`long-term time-domain investigations of atmospheric processes over $2+$ years, with a frequency tuned to the timescales of interest (R1-J-2).'}  To do so, JUICE \textbf{`must support global and regional spectroscopic mapping of the sources and sinks of key atmospheric species tracing atmospheric circulation and chemistry with spatial resolution $<200$ km/px in the VIS-NIR range, $<1000$ km in UV and with about a scale height vertical and 2000-4000 km horizontal resolution in the sub-millimeter range, repeated over a range of timescales from days to years (R1-J-8).'}  All other science requirements (R1-J-1 to J-11) are recorded in the previous sections.  An overarching theme of the JUICE science case is that the different regimes (the interior, atmospheric layers, and magnetosphere) are not decoupled from one another, so cannot be investigated in isolation - JUICE will explore the connections between all of the components within this system.  We now turn to how the payload and Jupiter tour have been designed to meet these objectives.

\section{JUICE Tour:  Jupiter Observing Opportunities}
\label{opportunities}

To meet the Level-1 science requirements described in Section \ref{science_case}, Level-2 requirements on the mission and spacecraft design were identified (Science Requirements Document JUI-EST-SGS-RS-001).  To summarise those relevant to Jupiter science, the mission and spacecraft were designed to:

\begin{itemize}
    \item Support inertial pointing for solar and stellar occultations [...] in orbit around Jupiter.
    \item Support nadir and off-nadir pointing for imaging.
    \item Support raster pointing for Jupiter [...] when objects are larger than fields of view.
    \item Support a spot-tracking mode for Jupiter.
    \item Perform limb tracking manoeuvre during radio occultations of Jupiter.
    \item Support instrument pointing needs according to [their] requirements.
    \item Provide sufficient temporal coverage for Jovian atmospheric and magnetospheric science, [with a] Jovian tour [that] shall be at least 2.5 years.
    \item Start Jupiter atmosphere observations 6 months before the Jupiter Orbit Insertion and continue during the Ganymede phase.
    \item Provide an orbit with inclination of at least $30^\circ$ with respect to the Jupiter equatorial plane.
    \item Co-align the boresights of instruments JANUS, MAJIS, UVS, GALA and SWI.
    \item Provide repeated observations of the same latitudes and cloud features with frequencies tuned to the timescales of interest (hours to months).
    \item Enable complete latitudinal, phase angle and local solar time coverage of Jupiter by remote sensing instruments.
    \item Provide at least two opportunities for remote sensing of Jupiter during the Europa phase when the distance to Jupiter is at a minimum.
    \item React on short timescales to new and unexpected events in the Jovian atmosphere, such as major storms or impacts. JUICE pointing can be updated up to one week in advance, and commanding up to 3 days before the uplink, both in exceptional cases.
    \item Enable sounding of the Jovian atmosphere in radio-, stellar and solar occultations at all latitudes, repeating observations at regular times during the mission.
    \item Provide capabilities for (1) global mosaics/scans of Jupiter repeated once every 2.5 hours to build up 360-degree longitude coverage during a ten-hour rotation, and (2) repetitive imaging of discrete cloud features/ region with hourly frequency for cloud feature tracking while the feature transits from west to east (full rotation in 10 hours), repeated on subsequent rotations.
    \item Enable joint observation campaigns with remote sensing instruments (submm, IR, visible and UV), as well as ENA and Radio observations for the study of the auroral region of Jupiter, as well as the coupling between the Galilean satellites, the Jovian magnetosphere, and the high latitude regions of Jupiter (thermosphere, ionosphere and magnetosphere).
\end{itemize}

\begin{figure*}[ht]
\begin{centering}
\centerline{\includegraphics[angle=0,scale=.4]{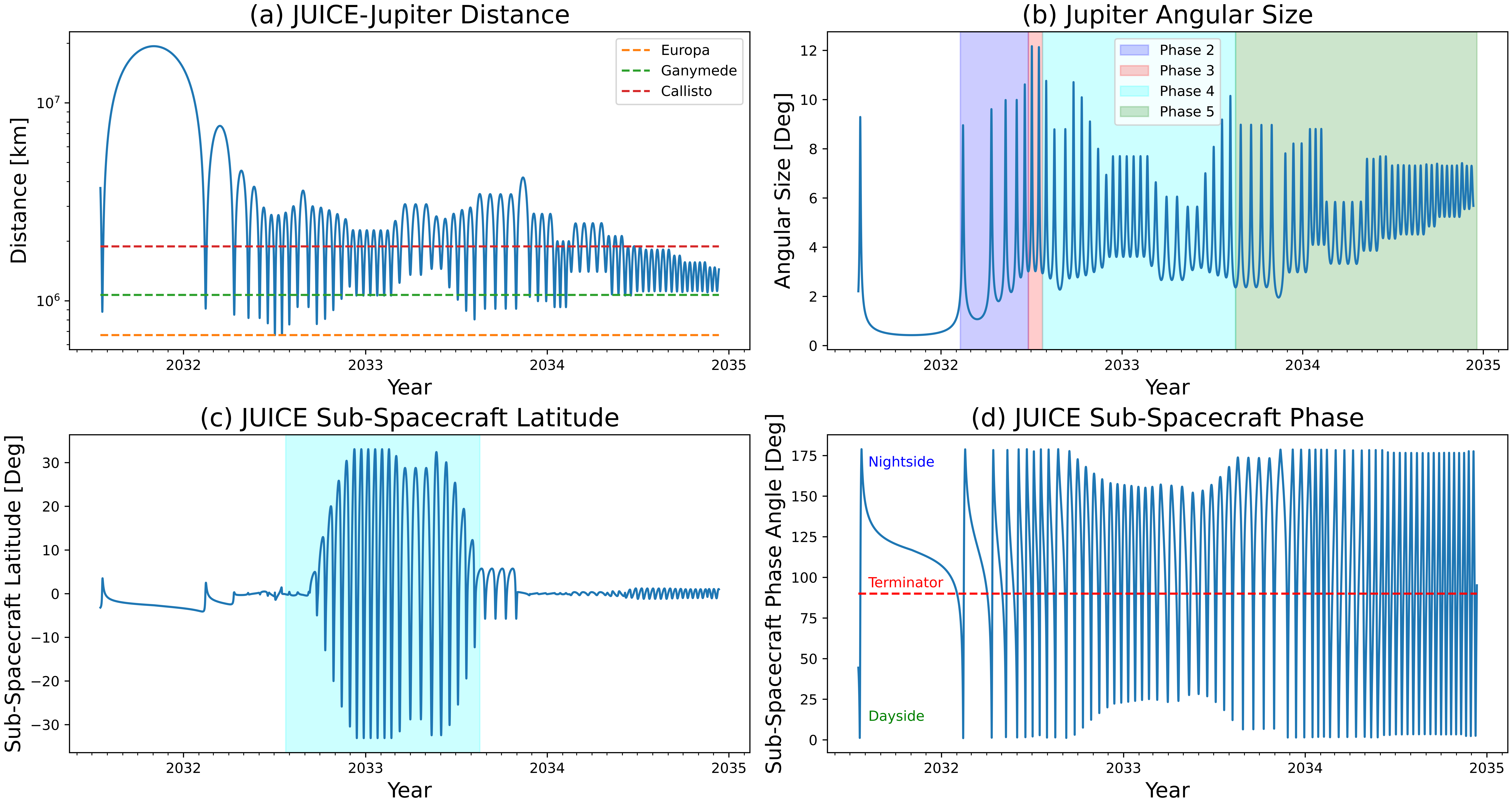}}
\caption{Overview of the JUICE orbital tour of Jupiter using Crema 5.0, showing (a) the distance to Jupiter in km; (b) the angular size of Jupiter from JUICE's vantage point; (c) the sub-spacecraft latitude; and (d) the phase angle (low phase implies dayside, high phase implies nightside).  In (a) we show the mean orbital distances of Europa, Ganymede and Callisto for comparison. In (b) we highlight phases 2, 3, 4 and 5 of the tour.}
\label{tour}
\end{centering}
\end{figure*}

The JUICE orbital tour of the Jovian system is described by \citet{23boutonnet}, and was designed to meet these requirements and enable the activities of the comprehensive payload discussed in Section \ref{instruments}.  The JUICE orbit differs substantially from that of Juno, enabling global views over longer duration from a low-inclination orbit, and with an inclined phase offering high-resolution views of the southern hemisphere.  Unlike Juno, JUICE does not spin, therefore simplifying some of the remote sensing observation sequences described in Section \ref{instruments}.  Here we briefly describe the different phases of the tour of interest to Jupiter science, which are shown in Figs. \ref{tour} and \ref{resolution}.  

\begin{figure*}[ht]
\begin{centering}
\centerline{\includegraphics[angle=0,width=\textwidth]{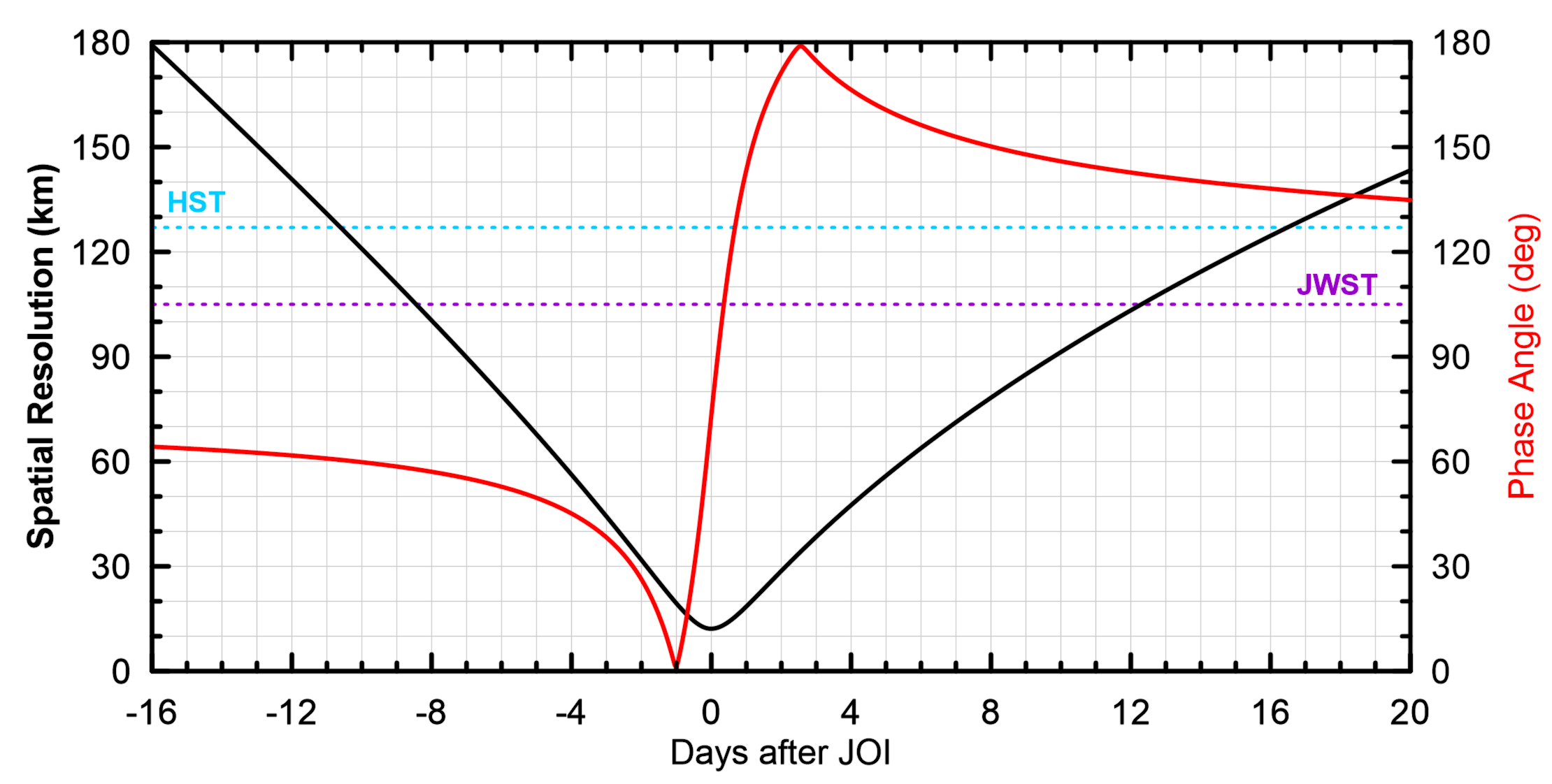}}
\caption{Spatial resolution of JANUS (black line) in the days surrounding Jupiter Orbit Insertion and compared to HST and JWST resolutions (cyan and purple lines, respectively). Phase angle is also plotted (red line).}
\label{fig:Janus_Approach}
\end{centering}
\end{figure*}

\begin{figure*}[ht]
\begin{centering}
\centerline{\includegraphics[angle=0,width=\textwidth]{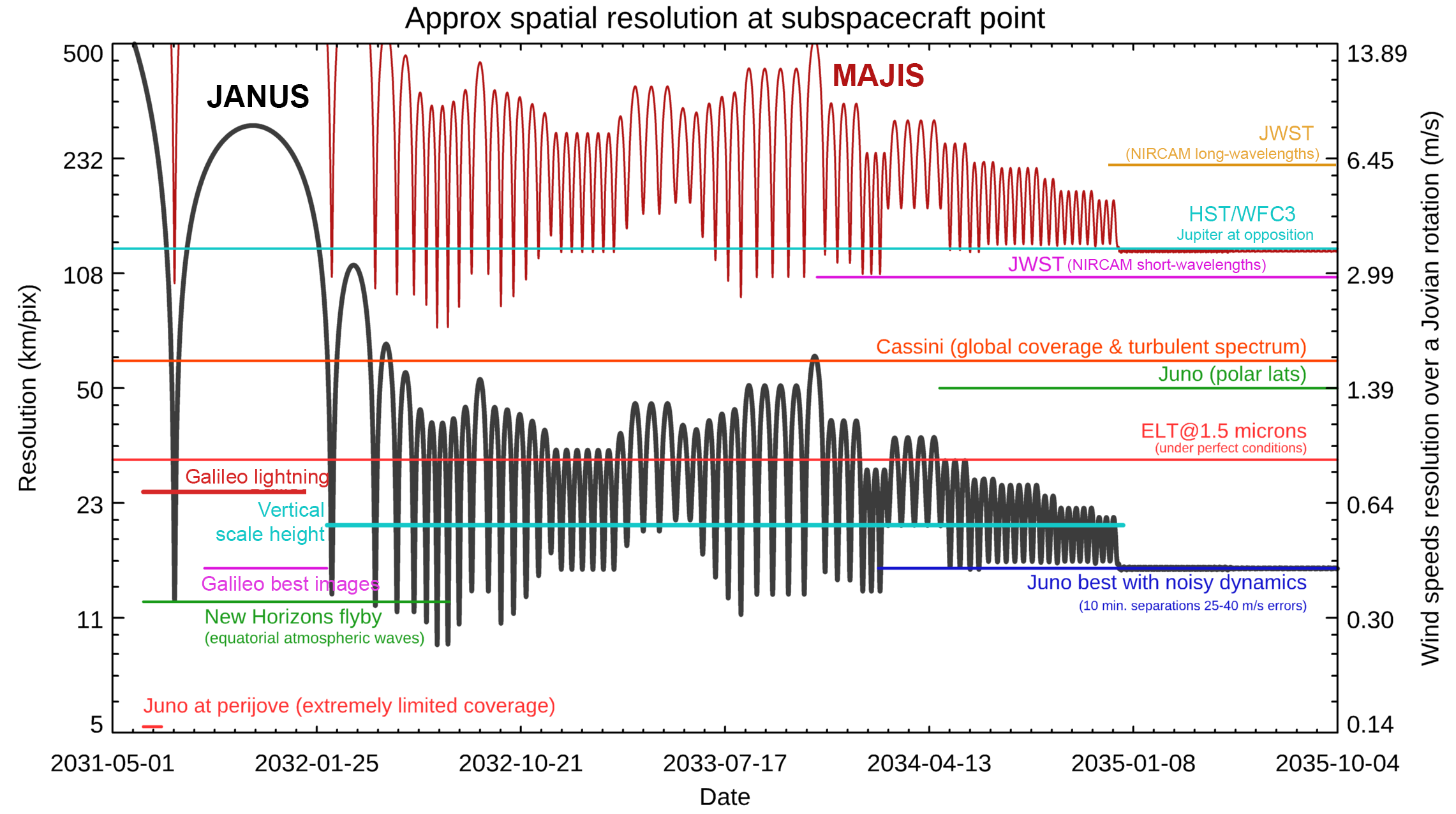}}
\caption{Spatial and wind speed resolutions of JANUS and MAJIS as a function of time, based on the Crema 5.0 trajectory and compared to  instrumentation on other facilities. }
\label{resolution}
\end{centering}
\end{figure*}

\textbf{Phase One} begins some six months before Jupiter Orbit Insertion (JOI) in July 2031 according to the Crema 5.0 (Consolidated report on mission analysis) trajectory, and encompasses the first elliptical orbit around Jupiter until February 2032.  The long approach to Jupiter will permit atmospheric monitoring and the generation of low-resolution movies, with the spatial resolution afforded by the JANUS camera exceeding the 150 km/pixel of the HST WFC3 instrument some 11.0 days before JOI and will exceed JWST spatial resolution at 8.5 days before JOI, respectively (see Figure \ref{fig:Janus_Approach}).  For much of the long first orbit, the resolution will be 200-300 km/pixel, enabling global monitoring of atmospheric phenomena.  Higher-resolution observations begin in earnest in \textbf{Phase Two,} between February and June 2032, as the JUICE orbital energy is reduced.  Five close flybys of Jupiter occur during this phase (perijoves 2-6, with perijove 1 being orbit insertion in July 2031), and provide the first real opportunity to test the JUICE remote sensing investigations with distances down to 11--13 R$_\mathrm{J}$ from Jupiter (Figure \ref{tour}).  Jupiter remote sensing will work in tandem with satellite remote sensing and magnetospheric measurements during this period, with four Ganymede flybys and one Callisto flyby during Phase 2.

\textbf{Phase Three} provides the highest spatial-resolution Jupiter observations of the entire tour, as JUICE performs two close flybys of Europa in July 2032.  Although much of the spacecraft resources and data volume will be dedicated to the Europa encounters, Jupiter science activities will aim to take advantage of these high-resolution opportunities at perijove 7 and 8 (9.4 R$_\mathrm{J}$ from Jupiter).

Following the Europa encounters, JUICE then uses multiple flybys of Callisto to increase its orbital inclination up to $\gtrsim33^\circ$ during \textbf{Phase Four}, the inclined phase between July 2032 and August 2033 (Figure \ref{tour}c).  With $\sim23$ perijoves during this period, with distances ranging from 11-20 R$_\mathrm{J}$ from Jupiter, JUICE will be afforded with improved views of the atmosphere and auroras in the polar domains. The maximum orbit inclination will be attained between December 2032 and February 2033 and the highest sub-spacecraft latitude at perijove of $\gtrsim33^\circ$ will be reached in May-June 2033 (perijoves 26 and 27). JUICE will spend more than 6 months with an orbit more inclined than $30^\circ$.  Furthermore, JUICE closest approaches and highest spatial resolutions during the inclined phase will be in the southern hemisphere, complementing the northern-hemisphere perijoves of Juno's extended mission.

As JUICE returns to the equatorial plane, it commences the low-energy \textbf{Phase Five}, circularising the orbit and providing perijoves that are much more frequent, approximately every two weeks between August 2033 (perijove 32) and December 2034 (perijove 67).  Spatial resolutions of visible images match those of the best Galileo images (10 to 40 km/pixel) throughout this phase, with closest approaches varying between 12-20 R$_\mathrm{J}$ before reaching Ganymede's orbit at 15 R$_\mathrm{J}$.  Fewer satellite encounters are envisaged during this period (four for Callisto, seven for Ganymede), leaving a number of uninterrupted orbits for Jupiter science.  Finally, in December 2034, JUICE enters orbit around Ganymede, with \textbf{Phase Six} completing its primary mission by September 2035.  Short time periods during the Ganymede orbital phase will be devoted to the monitoring of Jupiter's atmosphere, following up on dynamical, chemical, and meteorological phenomena discovered during Phases 1 through 5.

\subsection{Segmentation of the Tour}

\begin{figure*}[ht]
\begin{centering}
\centerline{\includegraphics[angle=0,width=\textwidth]{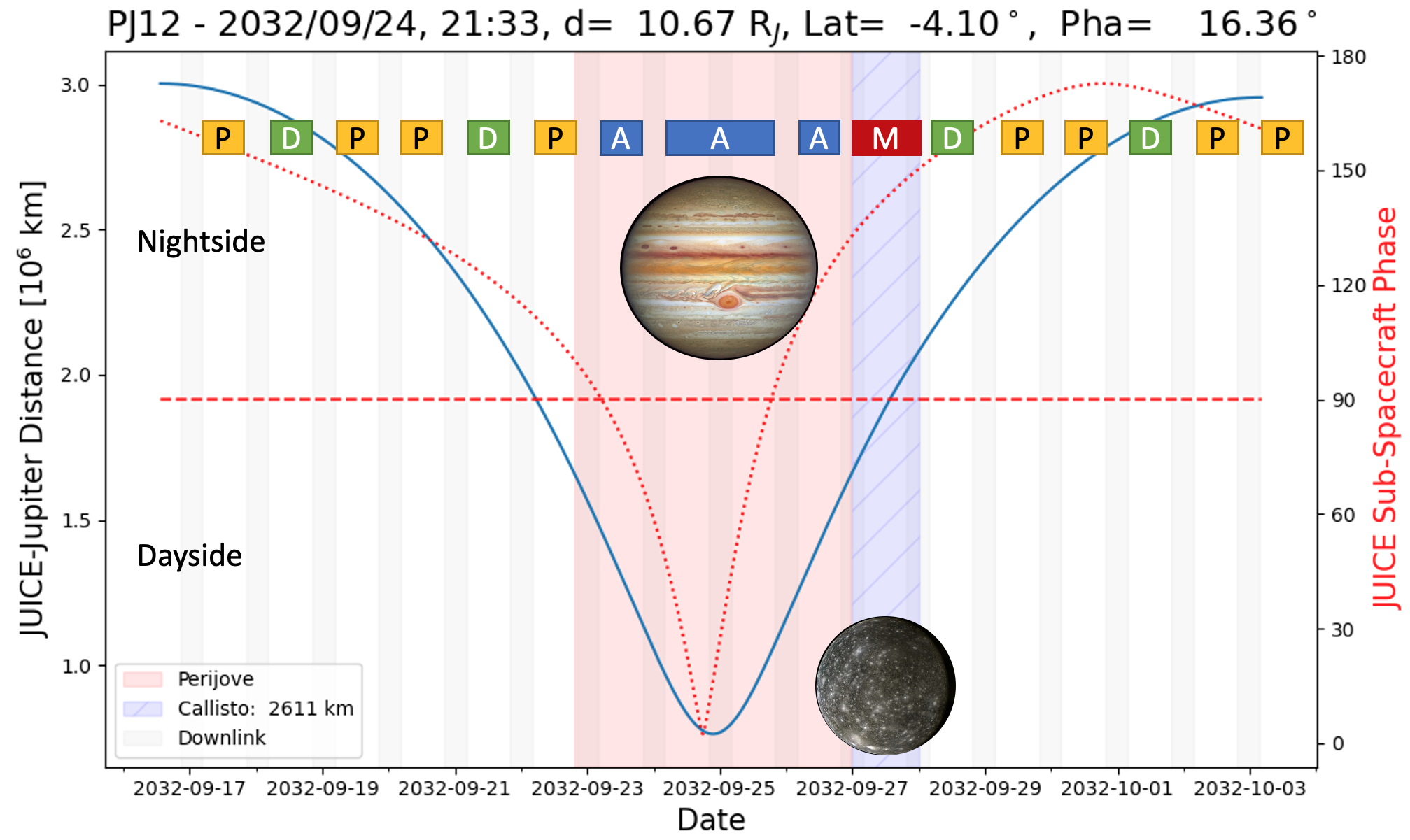}}
\caption{Perijove 12 is used as a generic orbit, from apojove to apojove, to demonstrate the segmentation of the tour.  Satellite observations (labelled 'M') are prioritised during the $\pm12$ hours surrounding closest approach.  Jupiter atmosphere and auroral science blocks during the $\pm50$ hour perijove windows (red shading, with specific observations in boxes labelled 'A') are interspersed with downlink windows (grey shaded boxes).  Atmospheric monitoring ('D') is interspersed with dedicated magnetospheric and plasma observations ('P') during more distant periods of the orbit.  Magnetospheric observations ride along with remote sensing observations during the perijove encounters.}
\label{orbit}
\end{centering}
\end{figure*}

Given the wide-ranging objectives of JUICE, and the wealth of opportunities provided by the orbital tour, a preliminary architecture for scientific operations and scheduling had to be developed so that requirements on spacecraft resources (telemetry, power, pointing, etc.) could be better understood.  Each scientific discipline, including Jupiter science, evaluated the observing opportunities during the tour.  This included identifying repeated observations of the Jovian atmosphere and auroras, time-critical opportunities for occultations (radio, stellar, and solar), and any unique observational geometries for Jupiter science.  Having identified optimal windows of observations for all disciplines, a preliminary `level-0' segmentation plan for the tour was developed, with specific disciplines being assigned the lead for pointing/operations during different segments.  Jupiter science segments had to be balanced against opportunities for magnetospheric science, satellite encounters (typically within a $\pm12$-hour window around closest approach to a moon), windows for data downlink (typically 8 hours per day), and other segments for navigation, calibration, and distant investigations of the wider Jovian system.  An example is shown in Figure \ref{orbit} for PJ12 in September 2032.  Operations will naturally be more complex during the Jupiter tour, with instruments operating simultaneously (i.e., riding along) during science segments devoted to other disciplines.  Potential disruption of the established observation plan may occur with a time response of a few days only in case of e.g. unexpected events like large asteroid or comet impacts.  Nevertheless, this approach enabled a thorough assessment of JUICE's capabilities to meet the original science requirements.

Proposed Jupiter science investigations mainly fall into the following segment types:

\begin{itemize}
    \item \textbf{Perijove Windows:}  Jupiter science operations will focus on the $\pm50$-hour window surrounding closest approach, with different types of observations planned for dayside, terminator, or nightside encounters.  The highest spatial resolution observations are possible during these windows (particularly during Phase 3), with numerous opportunities for stellar occultations also possible.  At least 50 stellar UV occultations will occur during the tour spanning a wide range of latitudes and local times, as well as repetitions to search for temporal variability. Jupiter science will be interrupted by satellite flybys and windows for downlink, but otherwise these segments will be the top priority for Jupiter science.
    
    \item \textbf{Monitoring Observations:}  Outside of the perijove windows, JUICE is required to provide frequent opportunities to track the evolution of atmospheric phenomena.  Monitoring will likely be organised into campaigns focusing on specific targets (e.g., tracking the changes to a storm or vortex, high-frequency auroral observations, etc.), but preliminary segmentation places a 10-hour monitoring window once every three days (approximately), with significant flexibility in scheduling.
    
    \item \textbf{Phase-Angle Windows:}  When not already covered by perijove or the monitoring windows, JUICE will observe Jupiter within $\pm5$ hours of minimum phase (i.e., dayside), the terminator crossing ($90^\circ$ phase angle), and at maximum phase (i.e., nightside).  These illumination conditions (and phase angles in between) are required to characterise the scattering properties of aerosols, and to provide nightside opportunities for lightning imaging and thermal emission measurements in the absence of scattered sunlight.
    
    \item \textbf{Inclined Windows: } When not already covered by perijove windows, and when the sub-spacecraft latitude exceeds $\pm5^\circ$ during Phase 4, JUICE will observe for $\pm10$ hours either side of the locations of maximum northern and southern sub-spacecraft latitude.  This will enable long-term monitoring of the polar atmosphere and auroras during Phase 4.  Indeed, a special `inclined aurora' segment is also planned for when the phase angle exceeds $160^\circ$.  These inclined windows will be planned on a case by case basis, as overlap with magnetospheric science segments is highly likely during Phase 4.
    
    \item \textbf{Occultation Windows:}  These time-critical events are scheduled independent of the perijove windows, but sometimes occur within them.  A window is reserved for radio occultation ingress and egress, as the spacecraft moves behind the planet as seen by Earth.  Similarly, a $\pm1$ hour window is reserved for solar occultations, at least five of which are planned during the tour.  Stellar occultations are extremely numerous, with optimal stellar types selected by the UV and IR instrument teams, and a subset of these (at least fifty) will be scheduled throughout the tour.
    
\end{itemize}

These generic Jupiter science segments are driven by the tour geometry, and although monitoring of atmospheric and auroral variability is a high priority, the detailed science operations plans will remove any unnecessary redundancies.  Furthermore, new segments that cover multi-instrument campaigns for specific phenomena (Section \ref{synergies}) will be built into the plan.  A generic Jupiter orbit is shown in Figure \ref{orbit}, highlighting a proposed sequence of activities.  This `level-zero' plan for segmentation of the JUICE orbit will be the basis of a rigorous activity plan that ensures proposed observations remain within the resource envelope (data volume, power, etc.) for JUICE.

\begin{figure*}[ht]
\begin{centering}
\centerline{\includegraphics[angle=0,width=0.9\textwidth]{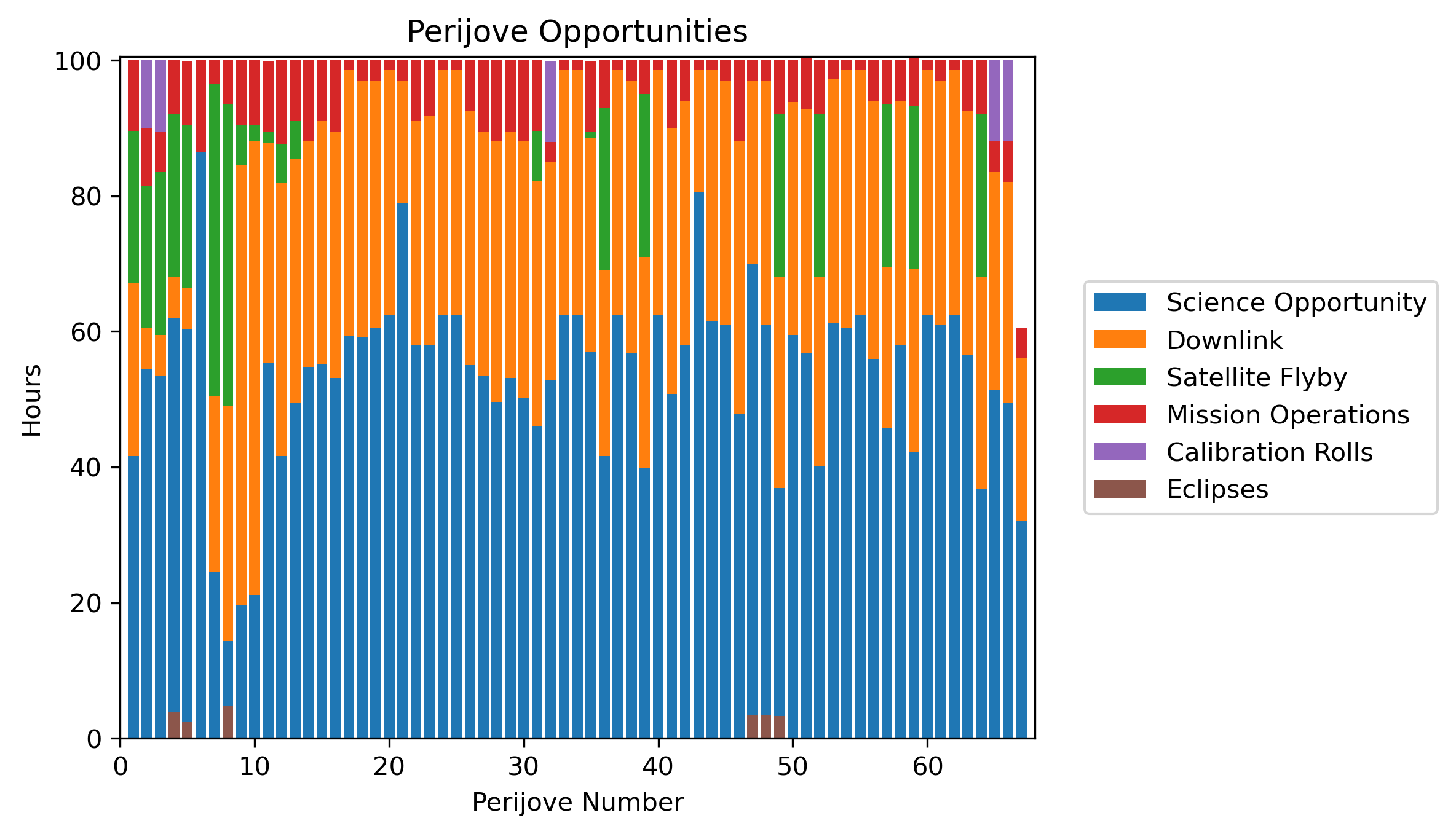}}
\caption{For each perijove, the $\pm50$-hour perijove windows are subdivided into sub-segments - time available for science observations, time devoted to satellite encounters, telemetry downlink, mission operations (such as navigation images and wheel off-loading), calibration of the magnetometer, and time removed due to Jupiter eclipses. }
\label{metrics}
\end{centering}
\end{figure*}

\begin{figure*}[ht]
\begin{centering}
\centerline{\includegraphics[angle=0,width=1.4\textwidth]{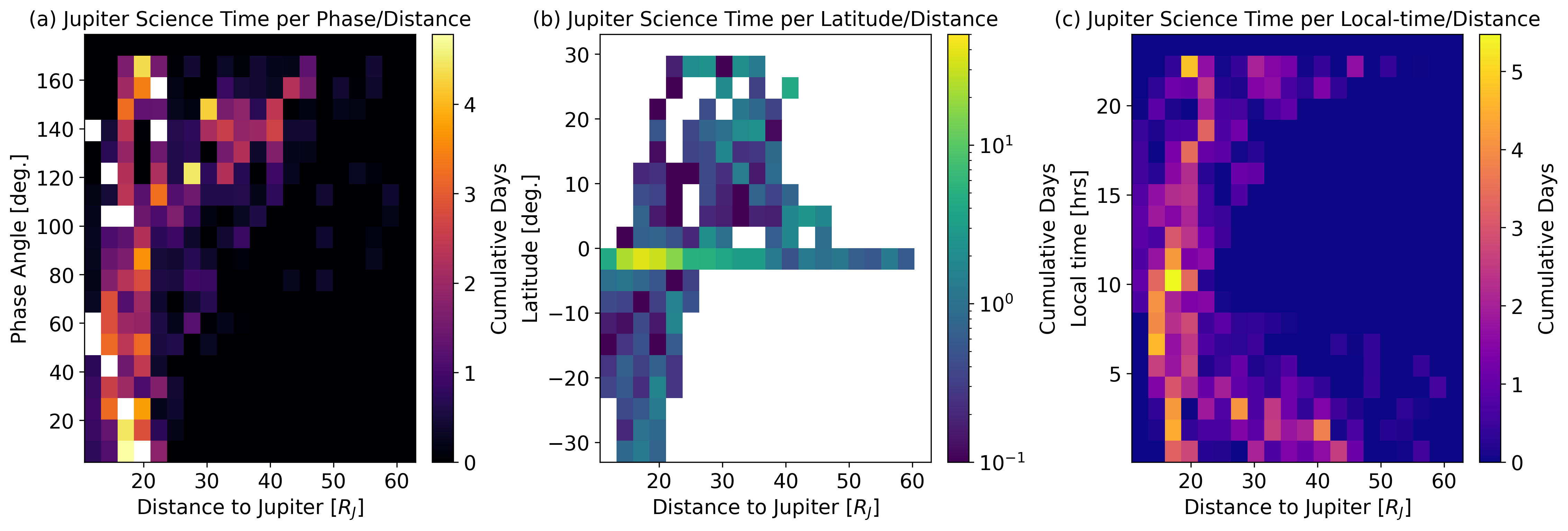}}
\caption{The cumulative time in days spent in different locations within the jovian system, showing the phase-angle versus distance to Jupiter in (a); the sub-spacecraft latitude versus distance in (b); and the local time versus distance in (c).  Note the logarithmic colour range in (b) to display time spent in the high-inclination phases.  Times are only counted if they fall within the Jupiter working group segments.}
\label{metrics2}
\end{centering}
\end{figure*}

To aid in the assessment of different tours, Figure \ref{metrics} shows the amount of time available to Jupiter remote sensing observations during each of the $\pm50$-hour perijove windows.  Once the satellite encounters, spacecraft navigation images, wheel off-loading, and downlink windows are all removed, this leaves approximately 50\% of the available time for science.  During phase 4, we envisage sharing of perijoves with magnetospheric science, further reducing the time available for remote sensing.  However, Jupiter observations are more likely to be data-volume limited rather than time-limited, and as Figures \ref{metrics} and \ref{metrics2} reveal, there are plentiful opportunities for observations within the `level-0' segmentation from (i) close proximity to Jupiter during perijoves; (ii) a range of phase angles from dayside to nightside; and (iii) a range of local times.  The JUICE tour therefore fulfils the requirements listed at the start of this Section.

\section{JUICE Instrumentation}
\label{instruments}

Based on the tour phases and high-level segmentation described in Section \ref{opportunities}, the instrument teams developed a number of observing modes that can be used in an interchangeable fashion, serving as the building blocks to develop the operations plan.  In this section, we provide brief overviews of the instruments and their capabilities for Jupiter science (summarised in Table \ref{tab:instruments}), describing how they will operate in the Jovian system.

\begin{table}[ht]
\centering
\resizebox{1.3\textwidth}{!}{\begin{tabular}{|l|l|l|l|l|l|}
\hline
Instrument  & Name & Spectral Range  & Spatial Properties  & Section & Reference \\
\hline

\textbf{UVS} & Ultraviolet Imaging  & 50-204 nm, & 7.3\degre$\times$0.1\degre slit plus  &  \ref{uvs} & \citet{23gladstone} \\
 & Spectrograph & spectral resolution $<$1\,nm & additional 0.2\degre$\times$0.2\degre  box &  &  \\

\hline

\textbf{MAJIS} & Moons And Jupiter  & 0.49-2.36\,$\mu$m, FWHM of 3.7-5.0 nm;  & 400 pixel slit of IFOV 150\,$\mu$rad; &  \ref{majis} & \citet{23poulet} \\
 & Imaging Spectrometer & 2.27-5.56\,$\mu$m; FWHM of 9.2-10.5 nm & 3D cubes from slews/scans  &  &  \\

\hline

\textbf{SWI} & Sub-millimetre  & 530-625\,GHz (479-565\,$\mu$m) and & Scanning $\pm$72$^\circ$ along-track and &  \ref{swi} & \citet{23hartogh} \\
 & Wave Instrument & 1067-1275\,GHz (235-281\,$\mu$m), $R\sim10^7$ & $\pm$4.3$^\circ$ cross-track, 30 cm antenna &  &  \\
\hline

\textbf{JANUS} & Jovis, Amorum ac  & 340-1080 nm with 13 filters & FOV of 1.29\degre $\times$ 1.72\degre,  &  \ref{janus} & \citet{23palumbo} \\
 & Natorum Undique Scrutator &  & IFOV of 15 $\mu$rad/pixel &  & \\

\hline

\textbf{3GM} & Gravity \& Geophysics of  & X band (7.2-8.4\,GHz) and & $\sim6$ km resolution for  &  \ref{3gm} & \citet{23iess} \\
 & Jupiter and Galilean Moons & Ka band (32.5-34.5\,GHz) & stratospheric profiles & &  \\

\hline

\textbf{RPWI} & Radio \& Plasma  & Radio emissions 80 kHz to 45\,MHz & - &  \ref{rpwi} & \citet{23wahlund} \\
 & Wave Investigation &  &  &  & \\

\hline

\textbf{PRIDE} & Planetary Radio   & VLBI network at X-  & - &  \ref{pride} & \citet{23gurvits} \\
 & Interferometry and & and Ka-band &  &  &  \\
 & Doppler Experiment &  &  &  &  \\

\hline
\end{tabular}}
\caption{JUICE Jupiter Instrument Summary, demonstrating spectral ranges, resolutions, and parameters for spatial coverage.  Note that only instruments and experiments with specific Jupiter science contributions are listed here, JMAG, PEP, RIME and GALA are omitted.}
\label{tab:instruments}
\end{table}


\subsection{UVS}
\label{uvs}


\subsubsection{UVS Description}
\begin{figure*}[ht]
\begin{centering}
\centerline{\includegraphics[angle=0,scale=1]{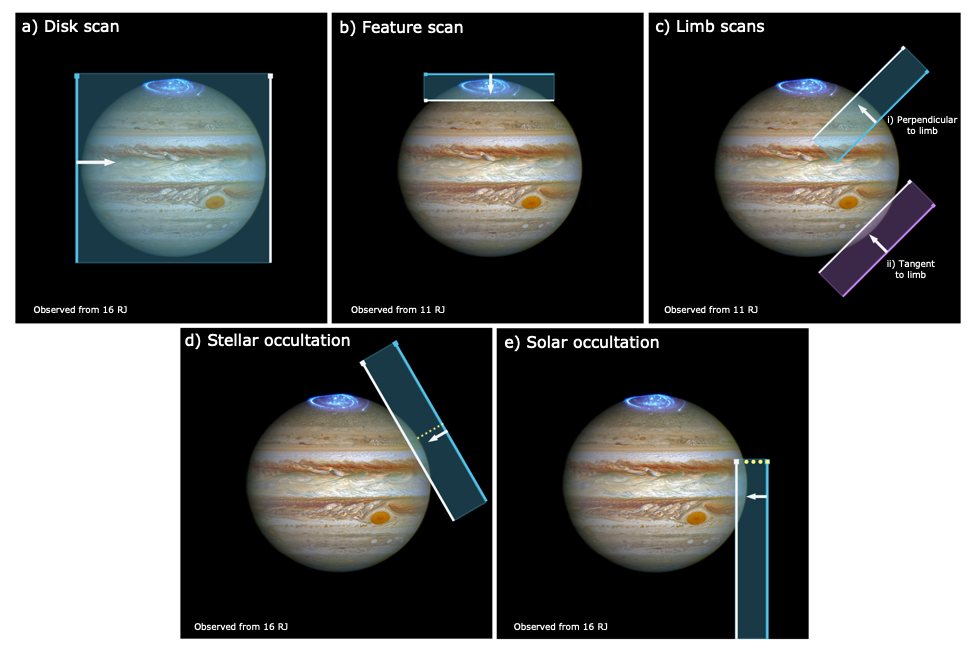}}
\caption{Examples of JUICE UVS observation techniques at Jupiter. In each example the UVS slit is scaled to its projected size when observing from the distances shown in the bottom left of each panel. The start and end position of the slit is shown, with arrows indicating the direction of slit motion.}
\label{UVS_obs_modes}
\end{centering}
\end{figure*}

\begin{figure*}[ht]
\begin{centering}
\centerline{\includegraphics[angle=0,width=\textwidth]{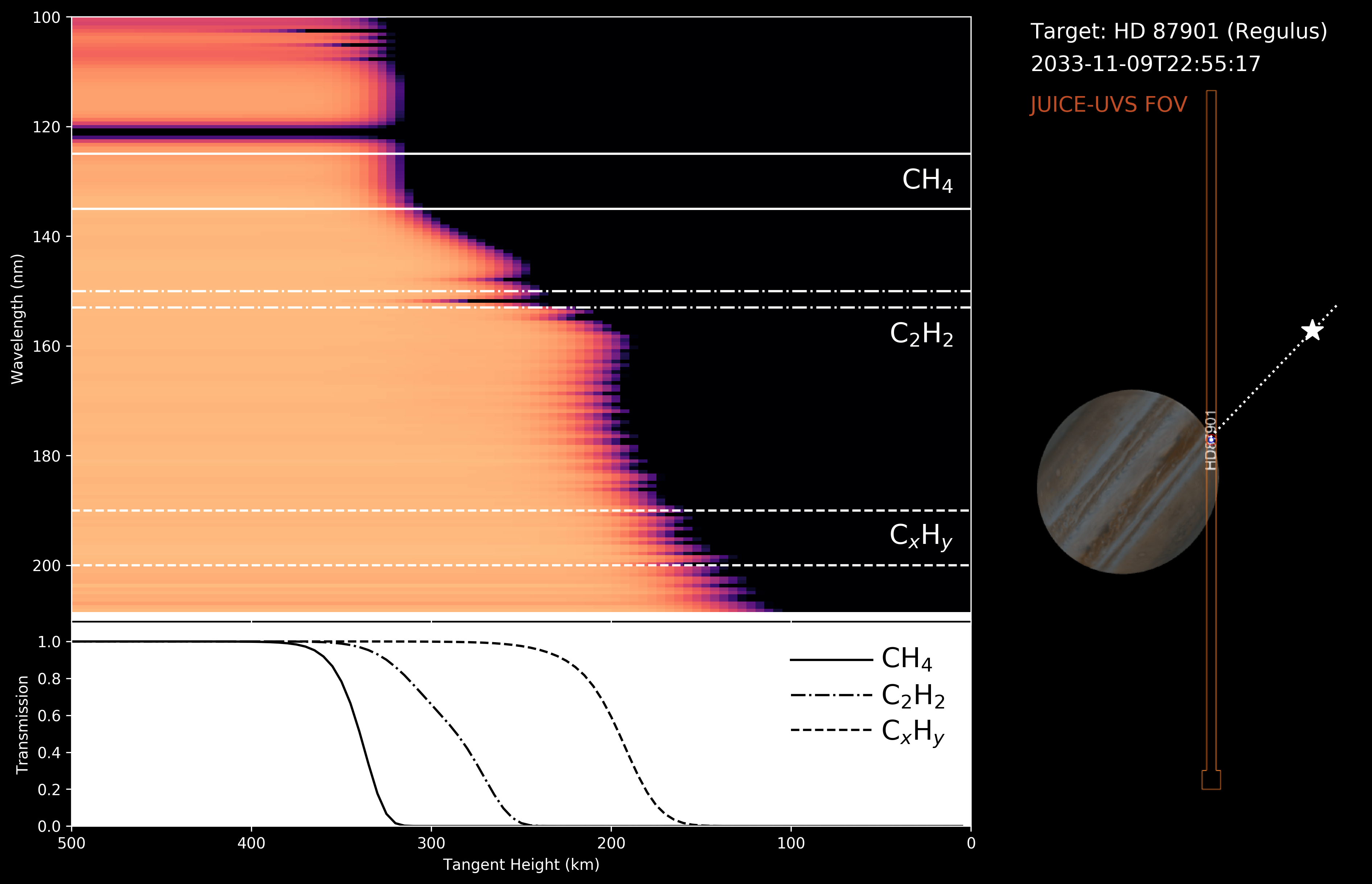}}
\caption{Simulation of a Jupiter occultation of bright star Regulus from JUICE UVS.  JUICE would hold Regulus fixed in the center of the UVS slit starting a few minutes before ingress until all of its light is blocked by Jupiter.  The panels on the left show the ingress occultation light curves spectral resolved (above) and binned over wavelength and labeled by dominant absorber (below). Starting on the left with no light blocked and moving to the right we observe the absorption of the shortest wavelengths first due to H$_2$ and CH$_4$ and then followed at longer wavelengths by absorptions by C$_2$H$_2$, C$_2$H$_6$ and a mix of higher order hydrocarbons (C$_\mathrm{x}$H$_\mathrm{y}$).}
\label{UVS_stellar_occ}
\end{centering}
\end{figure*}

UVS \citep[the Ultraviolet Imaging Spectrograph,][]{23gladstone} observes photons in the 50-204\,nm range of the far-UV (the spectrum is shown in Fig. \ref{spectra}). The UVS field of view is defined by its slit, which comprises a 7.3\degre$\times$0.1\degre  rectangular section with an additional 0.2\degre$\times$0.2\degre  box at one end for solar occultation observations (see Figure \ref{UVS_obs_modes}e), giving a total slit length of 7.5\degre. Spectral images of larger regions are built up by scanning across the target in the direction perpendicular to the long axis of the slit. Observations use a series of `ports', including the main 4\,cm $\times$ 4\,cm airglow port (AP) for auroral observations, stellar occultations (Figure \ref{UVS_stellar_occ}), and nadir view; a 1\,cm $\times$ 1\,cm high- resolution port (HP) that allows observations of bright targets with improved spatial resolution at the expense of a 16$\times$ reduction in the light collected; and a solar port (SP) offset from the main instrument aperture, which uses a 0.25\,mm diameter pinhole and additional pick-off mirror to reduce the solar flux to a suitable level for UVS solar occultation observations. 

The spatial resolution of UVS varies with wavelength and with position along the slit, with the best resolution achieved near the center of the slit (the on-axis position). The AP resolution is $<$0.3\degre  at all wavelengths and along-slit positions and is $<$0.16\degre  at most field positions for wavelengths in the range 70 - 130\,nm \citep{Davis2020}. The HP resolution ranges from 0.038\degre to 0.12\degre; again, the highest spatial resolution is achieved for wavelengths near 70--130\,nm. The point source spectral resolution is $<$1\,nm at all wavelengths when observing on-axis light \citep{Davis2020}.

\begin{figure*}[ht]
\begin{centering}
\centerline{\includegraphics[angle=0,width=\textwidth]{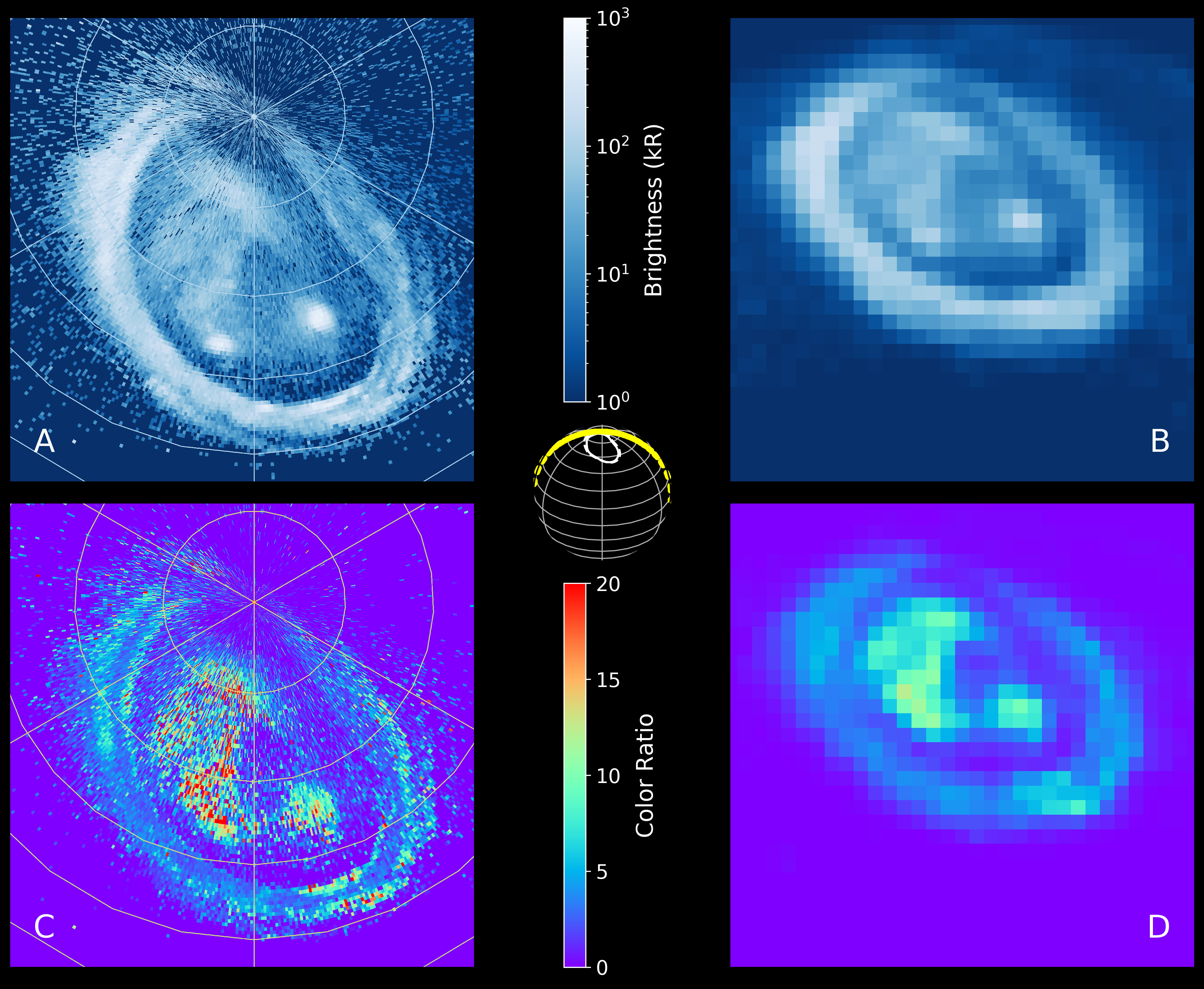}}
\caption{Juno UVS maps from Perijove 3 display the brightness (A) and color ratio (C) of the emissions from Juno's polar orbit vantage point taken (from supp. in \citealt{Greathouse2021}).  The wire frame model in the center shows the view of Jupiter from JUICE during the high inclination phase and the orientation of the aurora and terminator for the time of the UVS observations (aurora on the night side).  Even from a much greater distance, JUICE UVS simulations of the same observation from Juno now displayed in B and D will allow for detailed study of the auroral night side emissions unobservable by Earth-orbiting observatories.}
\label{UVS_aurora_zoom}
\end{centering}
\end{figure*}

UVS aims to study how Jupiter's upper atmosphere interacts with the lower atmosphere below and the ionosphere and magnetosphere above. Unlike the spinning Juno spacecraft, JUICE will enable UVS to perform `point and stare' observations, revealing fainter auroral features, and allowing synoptic-scale imaging of the polar auroras. However, as the minimum JUICE-Jupiter distance is considerably larger than the Juno perijove distances the spatial resolution of JUICE UVS auroral images will be degraded relative to Juno UVS images (see comparison in Figure \ref{UVS_aurora_zoom}). JUICE UVS will also perform stellar occultation measurements, which are not possible with Juno UVS. These observations use inertial pointing to observe UV-bright stars as they are occulted by Jupiter, providing measurements of atmospheric structure, composition, and variability. 

\subsubsection{UVS Operations}

During the $\pm$50-hour period around each JUICE perijove, UVS will use the spacecraft motion to perform a series of scans of Jupiter's disk using both the AP and HP ports (with the choice of port depending on distance and illumination) to map the aurora and airglow (Figure \ref{UVS_obs_modes}a). High-resolution feature scans of smaller areas will be performed when key regions of interest are visible (Figure \ref{UVS_obs_modes}b), and scans along Jupiter’s limb will be used to study the structure and variability of the upper atmosphere (Figure \ref{UVS_obs_modes}c). In both feature scans and limb scans, the region of interest is placed in the center of the slit to achieve the best possible spatial resolution.  UVS does not possess an internal scan mechanism, but instead relies on spacecraft slews to perform these observations.  The orientation of the UVS slit during scans is not strongly constrained, provided that the scan motion is in the direction perpendicular to the slit as indicated by the arrows in Figure \ref{UVS_obs_modes}. Disk scans may be performed North-South, East-West or at any intermediate pointing, while limb scans at different orientations provide different information: scans with the slit pointed perpendicular to the limb probe the vertical structure at high resolution while scans at a tangent to the limb provide larger instantaneous latitude coverage.

UVS also aims to observe $\sim$4 stellar occultations per perijove (Figure \ref{UVS_obs_modes}d), targeting a range of longitudes, latitudes, and local times, including repeat measurements of the same regions to assess temporal variability. Occultations in the auroral regions are a high priority, as are events where two UV bright stars may be observed simultaneously at different positions along the UVS slit. Some nightside perijoves include opportunities to observe solar occultations (Figure \ref{UVS_obs_modes}e); UVS aims to observe a minimum of five of these events, which are particularly important for investigating atmospheric absorption at the short end of the UVS bandpass, where stellar flux is typically very low. 

UVS will perform scans of both the dayside and nightside of Jupiter, with the dayside observations providing the best signal-to-noise ratio for airglow observations and the nightside observations allowing auroral emissions to be devoid of reflected sunlight background. Juno UVS observations of Jupiter’s nightside proved unexpectedly interesting, detecting bright flashes associated with Transient Luminous Events \citep{20giles_tle} and a bolide \citep{21giles_impact}. JUICE UVS will likely not achieve the required spatial resolution to detect similar events, but full disk nightside scans will facilitate searches for tropical nightglow arcs that might be detectable along/near the dip equator, similar to the infrared H$_3^+$ feature detected by \citep{Stallard2018}.

Outside of the perijove periods, UVS will continue to regularly (1--2 times per week) monitor Jupiter's airglow, aligning the slit along Jupiter’s North-South axis and staring for $\sim$10 hours to build up maps over a full Jupiter rotation period. Additional stellar occultation measurements will also be performed whenever suitable opportunities are identified (approximately 10-20\% of the planned UV stellar occultations currently fall outside of the perijove windows), since the JUICE-Jupiter distance is not an important constraint for these observations, but the quality is improved with slower angular velocities of the star as seen by JUICE.

\subsection{MAJIS}
\label{majis}

\subsubsection{MAJIS Description}



\begin{figure*}[ht]
\begin{centering}
\centerline{\includegraphics[angle=0,width=\textwidth]{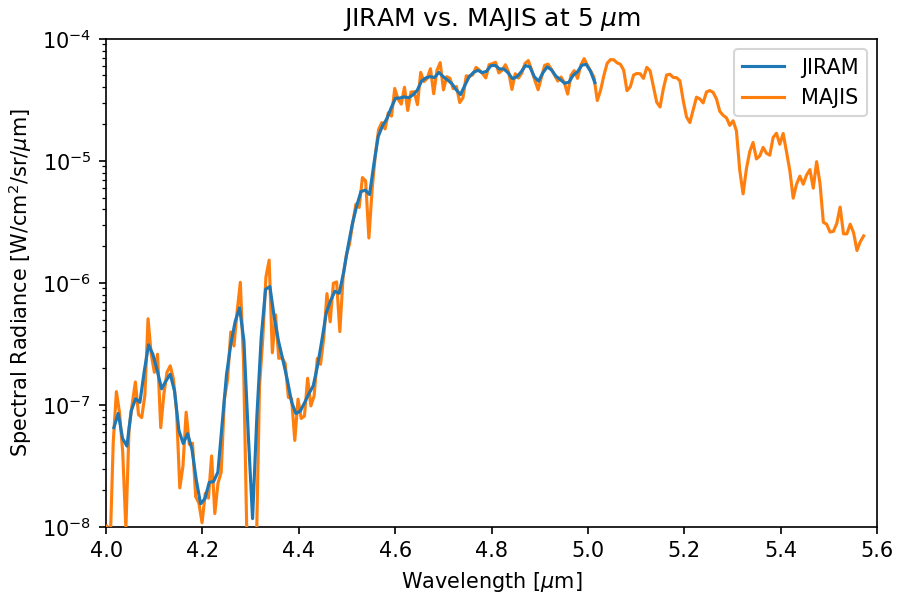}}
\caption{Simulated comparison of Jupiter 5-$\mu$m spectra from Juno/JIRAM (blue) and JUICE/MAJIS (orange), based on thermal emission calculations by \citet{10grassi}.  This shows the higher spectral resolution of MAJIS (9.2-10.5 nm) compared to JIRAM (15 nm) in this spectral range, and reveals the longward extension of MAJIS to 5.56 $\mu$m.  The 5-$\mu$m `window' is sculpted by PH$_3$ absorption on the short-wave side (near 4.3 $\mu$m) and NH$_3$ absorption on the long-wave side (near 5.3 $\mu$m), with numerous contributions from minor species (AsH$_3$, CO, GeH$_4$, CH$_3$D and H$_2$O) as described in the main text. }
\label{jiram_vs_majis}
\end{centering}
\end{figure*}

MAJIS \citep[Moons And Jupiter Imaging Spectrometer,][]{23poulet} is the visible and near-infrared imaging spectrometer for JUICE. It performs global imaging spectroscopy from 0.49 to 5.56\,$\mu$m across all local times. The MAJIS spectral range is divided in two separate channels.  The VIS-NIR channel covers the 0.49-2.36\,$\mu$m region with a spectral sampling of $\sim4$\,nm and a FWHM of 3.7-5.0 nm, depending on wavelength.  The IR channel covers the 2.27-5.56\,$\mu$m region with a spectral sampling of about $\sim6$\,nm and a FWHM of 9.2-10.5 nm, depending on wavelength.  A typical Jupiter spectrum in the infrared is shown in Fig. \ref{spectra}, with a comparison between simulated spectra in the 5-$\mu$m region from Juno/JIRAM and JUICE/MAJIS shown in Fig. \ref{jiram_vs_majis}.  In typical operating conditions, MAJIS observes a line of 400 spatially-contiguous pixels (a `slit') and a spectrum is recorded simultaneously for each pixel, in the entire spectral range. The instantaneous field-of-view (IFOV) of an individual pixel is 150\,$\mu$rad, providing therefore a spatial resolution of about 150 km over the Jupiter’s disk at the typical distance of perijove passages ($\sim10^6$\,km, see Figure \ref{resolution}).

The design of the instrument enables a large number of operative options, conceived to optimise science return. Notably, the size of MAJIS cubes can be reduced by masking (i.e., selective removal) along both the spatial and the spectral dimensions, in order to reduce the downlink burden. On the other hand, spectral regions of special interest (e.g., absorption lines) can be observed with a 2$\times$ spectral oversampling.

Usually, during Jupiter observations, a mosaic of spatially-contiguous slits (a `cube') is created by repointing the instrument between consecutive acquisitions. This technique eventually produces a 3D hyperspectral image of the observed scene.  The spatial repointing required between different slits can be achieved either by means of a slow spacecraft slew or, more commonly, by the motion of MAJIS' own internal pointing mirror. This device has one degree of freedom and can shift the slit position in the sky parallel to its longer axis.

Specific techniques (summing and comparison of several individual observations for a single slit) have been developed to reduce the adverse effects by impinging energetic particles (mostly electrons) upon the instrument and to improve, at the same time, the signal-to-noise (SNR) ratio. In the case of the brightest infrared hot spots observed between the Equatorial Zone and the North Equatorial Belt \citep{17grassi}, MAJIS SNR is expected to exceed 1000 at 4.7\,$\mu$m.

The radiance observed by MAJIS in a single cube is determined by a range of phenomena, and therefore allows one to address a variety of scientific objectives:
\begin{itemize}
    \item Properties, distribution and variability of hazes and clouds on Jupiter offers key insights on the vertical and horizontal dynamics of the troposphere. Indeed, dynamics drives the distribution of condensable species (H$_2$O, H$_2$S, NH$_3$) that forms clouds and the transport of gases that, by photodissociation and through complex chemical patterns, produces the tropospheric and stratospheric hazes. The signal measured by MAJIS between 0.4 and 3.2\,$\mu$m is dominated by the scattering of solar radiation by aerosols and, in lesser extent, by atmospheric gases (Rayleigh scattering). The large variability of gas opacity (mostly due to methane) within this spectral range determines different effective sounding altitudes at different wavelengths and, in turn, offers a method to probe the vertical structure of clouds and hazes \citep{10sromovsky,Braude2020,Perez-Hoyos2020,Grassi2021,Anguiano-Arteaga2021}. These observations will be instrumental in addressing objectives R1-J-6 and R1-J-10 mentioned in Section \ref{science_case}. Maps presented in \citet{Grassi2021} on the basis of JIRAM data demonstrated the high spatial variability of cloud properties - as retrievable from IR-spectroscopy – at least down to scales of few hundreds of kilometers (see their Figures 8.c and 8d). On the other hand, latitudinal mean profiles are clearly detected (see their Figures 8a and 8b), revealing contrasts in aerosols over several thousands of kilometres. In MAJIS observations, it is therefore important to pursue - at the same time - the highest spatial resolution and largest spatial coverage made possible by available data volume. Notably, the observations presented in the Juno JIRAM study \citep{Grassi2021} remains both extremely sparse in space as well as limited to a single perijove passage. The latter issue justifies a specific MAJIS strategy aimed to characterise the time variability of aerosol properties at least down to a frequency of a few days.
    
    \item Distribution and variability of minor atmospheric components is another strong experimental constraint to dynamic models. Measurement of disequilibrium species (PH$_3$, GeH$_4$, AsH$_3$) is of special interest in assessing the nature of vertical motions (Section \ref{dynamics}). The signal measured by MAJIS at $\lambda > $ 4\,$\mu$m is mostly driven by the thermal emission of the atmosphere, as shown in Fig. \ref{jiram_vs_majis}. This spectral range hosts vibro-rotational bands of several compounds such as water, ammonia, phosphine, arsine and germane. In cloud-free regions, the mixing ratios of these molecules can be retrieved down to 5-6\,bars \citep{98irwin,10grassi,20grassi}. These observations will be instrumental in addressing objectives R1-J-8 and R1-J-9 mentioned in Section \ref{science_case}. Spatial and temporal desiderata for the investigation of minor species are similar to the ones implied by aerosol studies, given the intertwined nature of underlying phenomena. Moreover, \citet{20grassi} separately demonstrated the variability of minor species mixing ratios at least down to scales of few hundreds of kilometers (see their Figures 7.c and 8.c) as well as consistent spatial patterns over features with dimensions of several thousands of kilometres, such as the hot-spots and the warm ring around the Great Red Spot.

    \item Physical conditions in the upper stratosphere and thermosphere of Jupiter are strongly coupled with the magnetosphere of the planet, which provides a substantial energy deposition at these altitudes, especially at polar latitudes. The spectral region between 3.2 and 4.0\,$\mu$m hosts a very opaque methane band, that precludes photons to emerge below the approximate level of 30\,mbar \citep{10sromovsky} without being absorbed. Here, auroral emission from H$_3^+$ (ultimately, a by-product of the impinging of electrons over the upper atmosphere, \citealt{89drossart}, \citealt{Dinelli2017}) or non-LTE emission by methane \citep{Moriconi2017} can be investigated, providing a method to map phenomena occurring in the magnetosphere \citep{Mura2018,Moirano2021}. These observations will be instrumental in addressing objectives R1-J-4 and R1-J-9.5 mentioned in Section \ref{science_case}. While MAJIS data cannot match the spatial resolution of JIRAM auroral observations (that reach in most favourable cases values close to 10\,km), they will allow one to explore feature variability at time scales of few hours. This is a frequency regime not probed by fast JIRAM passages over the poles and therefore particularly important to characterise interactions with rapid magnetospheric phenomena. 
\end{itemize}

A particular strength of MAJIS, unlike previous imaging spectrometers, is access to the full M-band window of thermal emission, sculpted by PH$_3$ on its short-wave side, and NH$_3$ on its long-wave side. The extension to 5.56\,$\mu$m, rather than curtailing near 5.0\,$\mu$m like Juno/JIRAM (Fig. \ref{jiram_vs_majis}), provides significantly stronger constraints on NH$_3$ and H$_2$O than previous instruments \citep{10grassi}. Unlike Juno/JIRAM, MAJIS also extends to shorter wavelengths below 2\,$\mu$m, enabling stronger constraints on aerosol properties than possible with JIRAM.

\subsubsection{MAJIS Operations}
MAJIS operations are driven by the following requirements:
\begin{itemize}
   \item To provide continuous coverage of Jupiter's disk, with a time sampling rate as uniform as possible, with periods complementary to those provided by Juno/JIRAM (from timescales from hours to months).
   \item To provide coverage of Jupiter's limbs at each JUICE perijove, with focus on the comparisons of morning vs. dusk and solar vs. antisolar conditions.
   \item To provide coverage of Jupiter's polar regions, with much longer exposure times (5 or more seconds) and specific acquisition modes designed for auroral observations, in the vicinity of each perijove.
\end{itemize}

\begin{figure*}[ht]
\begin{centering}
\centerline{\includegraphics[angle=0,width=\textwidth]{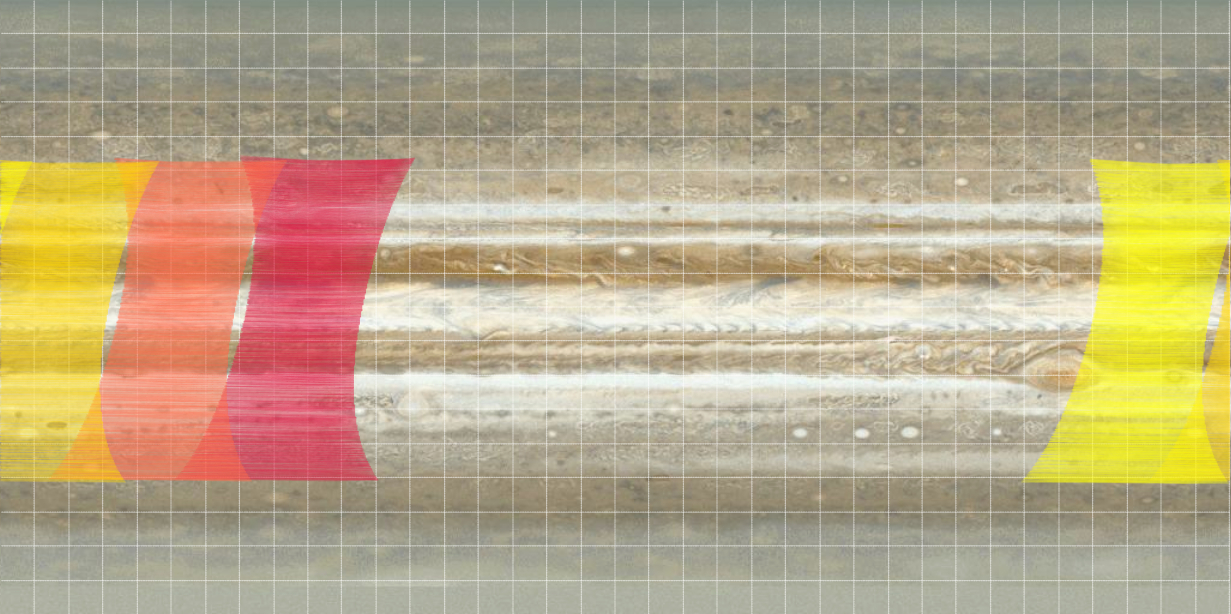}}
\centerline{\includegraphics[angle=0,width=\textwidth]{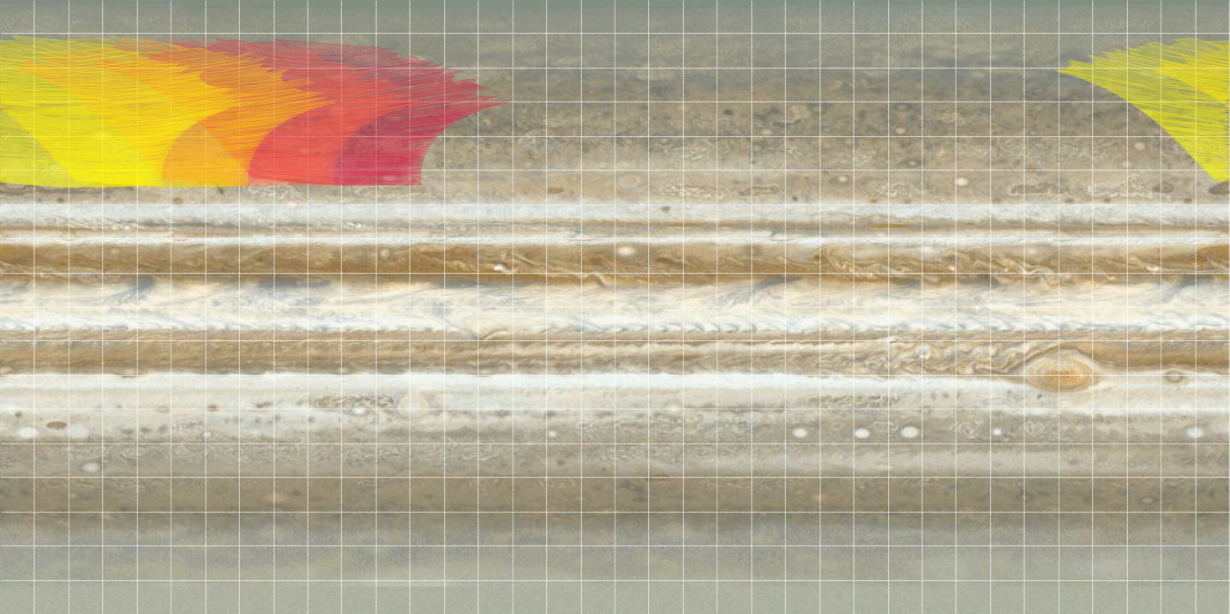}}
\caption{(Top) Expected coverage of Jupiter's disk by MAJIS on 2032-Sep-24, near perijove \#12. Different colours present the coverage of four cubes acquired at time intervals of 1h10m. (Bottom) Expected coverage of Jupiter’s auroras by MAJIS on 2032-Sep-25, a few hours after perijove \#12. Different colours present the coverage of four cubes acquired at time intervals of 46m.}
\label{MAJIS1}
\end{centering}
\end{figure*}

\begin{figure*}[ht]
\begin{centering}
\centerline{\includegraphics[angle=0,width=\textwidth]{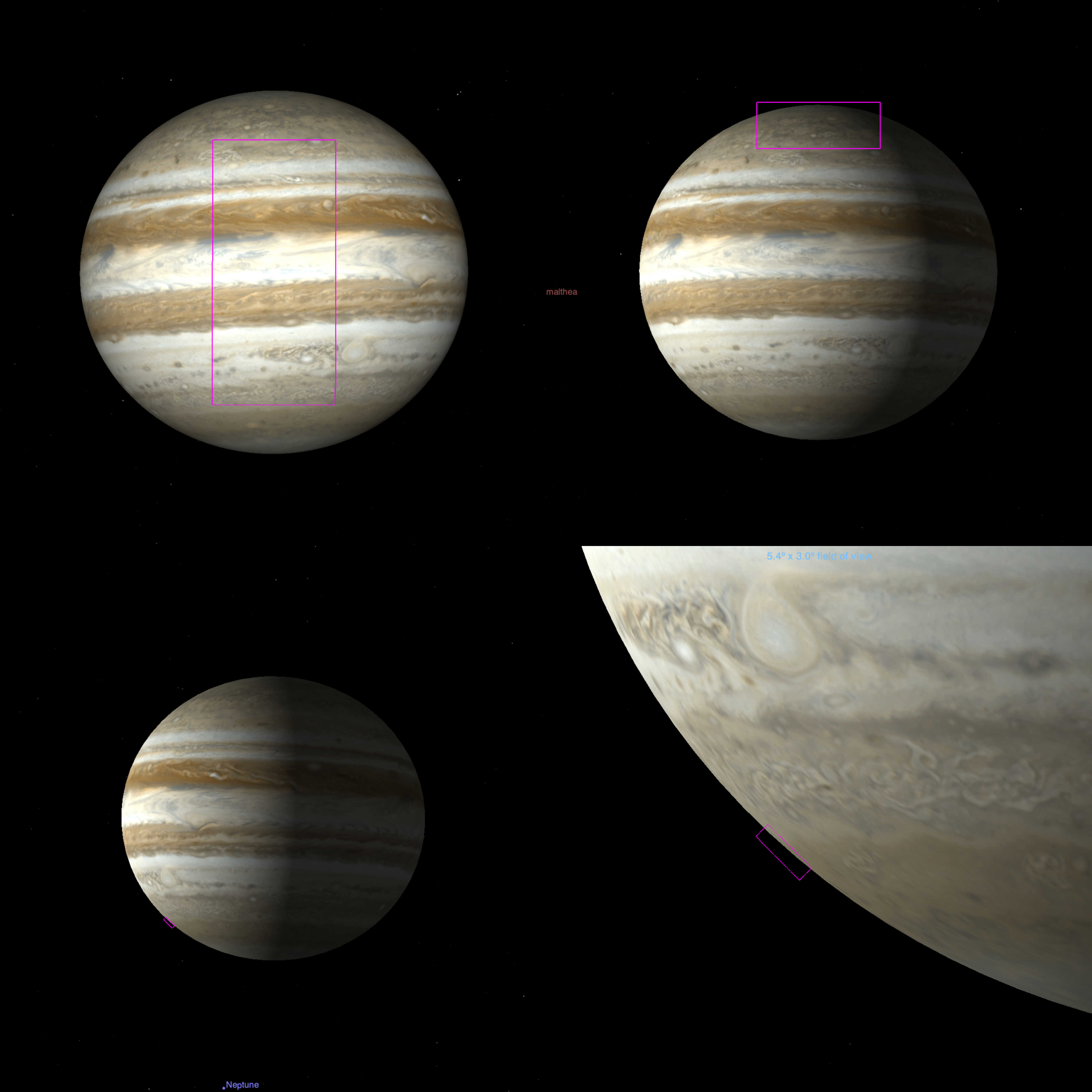}}
\caption{Example fields-of-view of MAJIS for different observing modes. Each panel presents an individual cube from the set of several that are acquired in close temporal sequence:
(top left) A cube from a MAJIS\_JUP\_DISK\_SCAN set; acquisition starts on 2032-Sep-24 at 19:29 UTC, corresponding to the red cube of Fig. \ref{MAJIS1} (top). 
(top right) A cube from a MAJIS\_JUP\_AURORA\_SCAN set; acquisition starts on 2032-Sep-25 at 06:51 UTC, corresponding to the red cube of Fig. \ref{MAJIS1} (bottom). 
(bottom left) A cube from a MAJIS\_JUP\_LIMB\_SCAN set; acquisition starts on 2032-Sep-25 at 15:24 UTC. 
(bottom right) Detail of the same MAJIS\_JUP\_LIMB\_SCAN cube.}
\label{MAJIS2}
\end{centering}
\end{figure*}

These requests are translated in a set of observations types and corresponding rules:
\begin{enumerate}
    \item Observations of Jupiter disk (`MAJIS\_JUP\_DISK\_SCAN') will be performed every three earth days, with small time adjustment performed at each orbit in order to have always an observation around the time of the perijove. Each of these observations will cover the hemisphere of Jupiter visible from JUICE, with a variable number of cubes (from 1 to 4), depending on distance (Figure \ref{MAJIS1}). After the third perijove passage, the MAJIS IFOV will vary between 140 and 680\,km along a typical orbit. The main focus will be toward equatorial regions ($30^\circ$S--$30^\circ$N), that will benefit of better visibility (lower emission angles) for most of the Jupiter tour (i.e., when not in the high-inclination phase). However, more complete latitudinal coverage (implying the acquisition of lines for each cube) or longitudinal coverage (implying the acquisition of more cubes) can be pursued, as far as feasible within data volume constrains imposed by mission resources.
    
    \item Observations of Jupiter's limb (`MAJIS\_JUP\_LIMB\_SCAN') will consist in a series of cubes (typically eight), acquired while keeping the MAJIS slit parallel to the limb, in order to minimise possible straylight from Jupiter's disk (Figure \ref{MAJIS2}). Such a set of cubes will be acquired three times per orbit, at phase angles of 0$^\circ$ and $90^\circ$. Each cube will cover the tangent altitude between 0 and 3000\,km above the nominal 1-bar level. Notably, during these scans, consecutive individual lines will be largely spatially overlapped (up to 90\% in area), in order to achieve a substantial spatial supersampling and to allow, by deconvolution, a reconstruction of the vertical profile of the signal with an effective resolution better than IFOV size. Only the central part (40 pixels) of each line will be transmitted to Earth, to cope with data volume limitations.

    \item Observations of auroras (`MAJIS\_JUP\_AURORAL\_SCAN') will usually consist of a set of four cubes over the northern polar region (north of $50^\circ$N), acquired in proximity of each pericenter passage when sub-spacecraft longitude is closer to 190$^\circ$W, to ensure better visibility of the northern polar oval. Southern polar oval is more symmetric around the pole and will be better observed with during the high inclination phase. Cubes during each perijove will be taken with gaps of approximately 40 minutes, to monitor the rapid variability of auroral structures.
\end{enumerate}

Although the types of observations mentioned above will represent most of the entries in MAJIS Jupiter dataset, they will be nonetheless complemented by other types, to be performed more rarely during the mission. This latter group includes: a global complete mosaic at full spectral coverage (to be acquired just once during the mission, given the massive data volume output), limb observations while in eclipse, stellar occultations and `tracking' observations (repeated observations of the same area at the relatively short period of several minutes).


\subsection{SWI}
\label{swi}

\subsubsection{SWI Description}

The Sub-millimetre Wave Instrument \citep[SWI,][]{23hartogh} is a passively cooled radio telescope with a 29-cm primary mirror that will be sensitive to radiation produced in Jupiter's atmosphere from its upper troposphere to its lower thermosphere. It will be an ideal instrument to constrain the chemistry and dynamics of Jupiter's stratosphere.

\begin{figure*}[ht]
\begin{centering}
\centering
\includegraphics[angle=0,width=\textwidth]{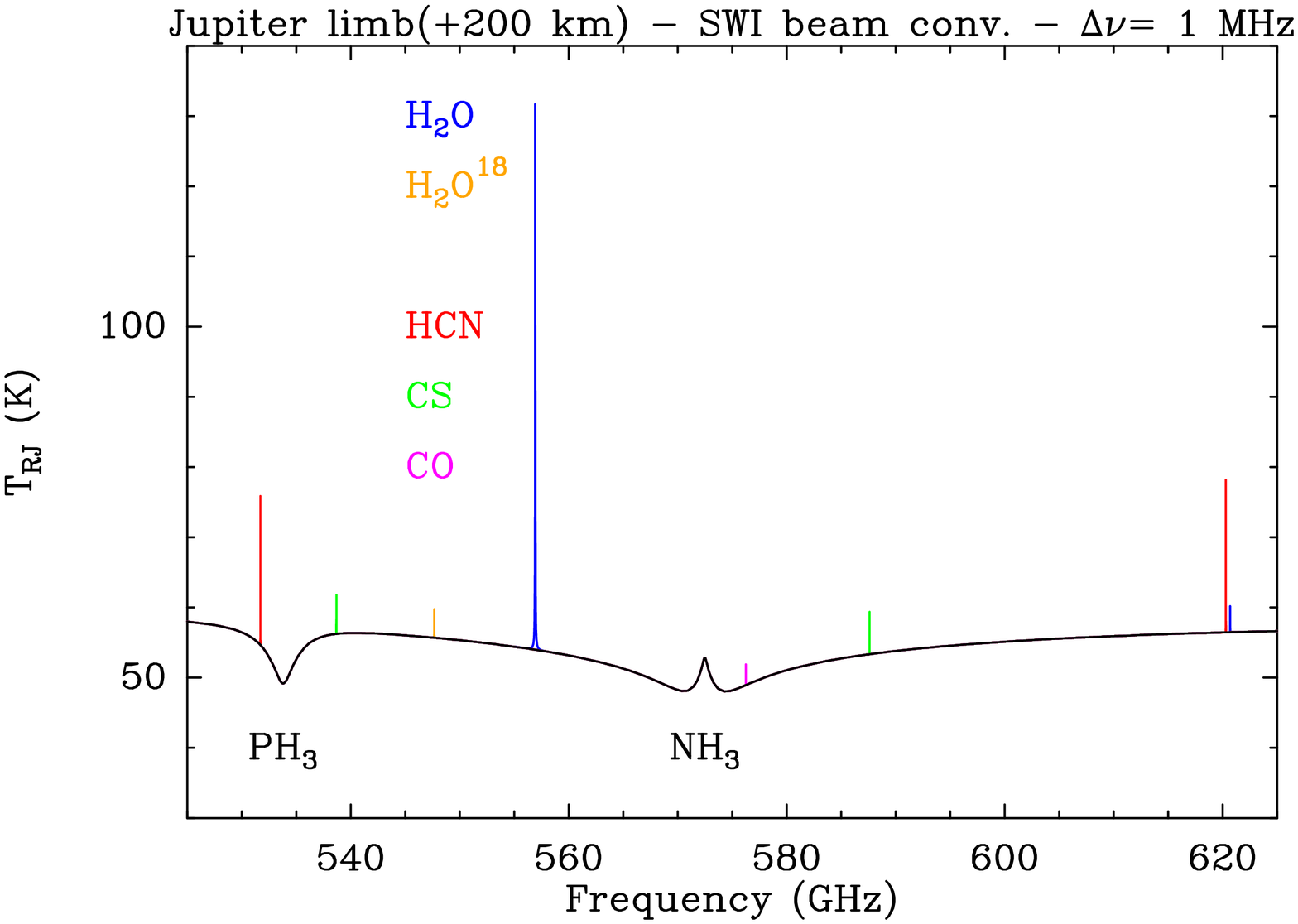}\\
\includegraphics[angle=0,width=\textwidth]{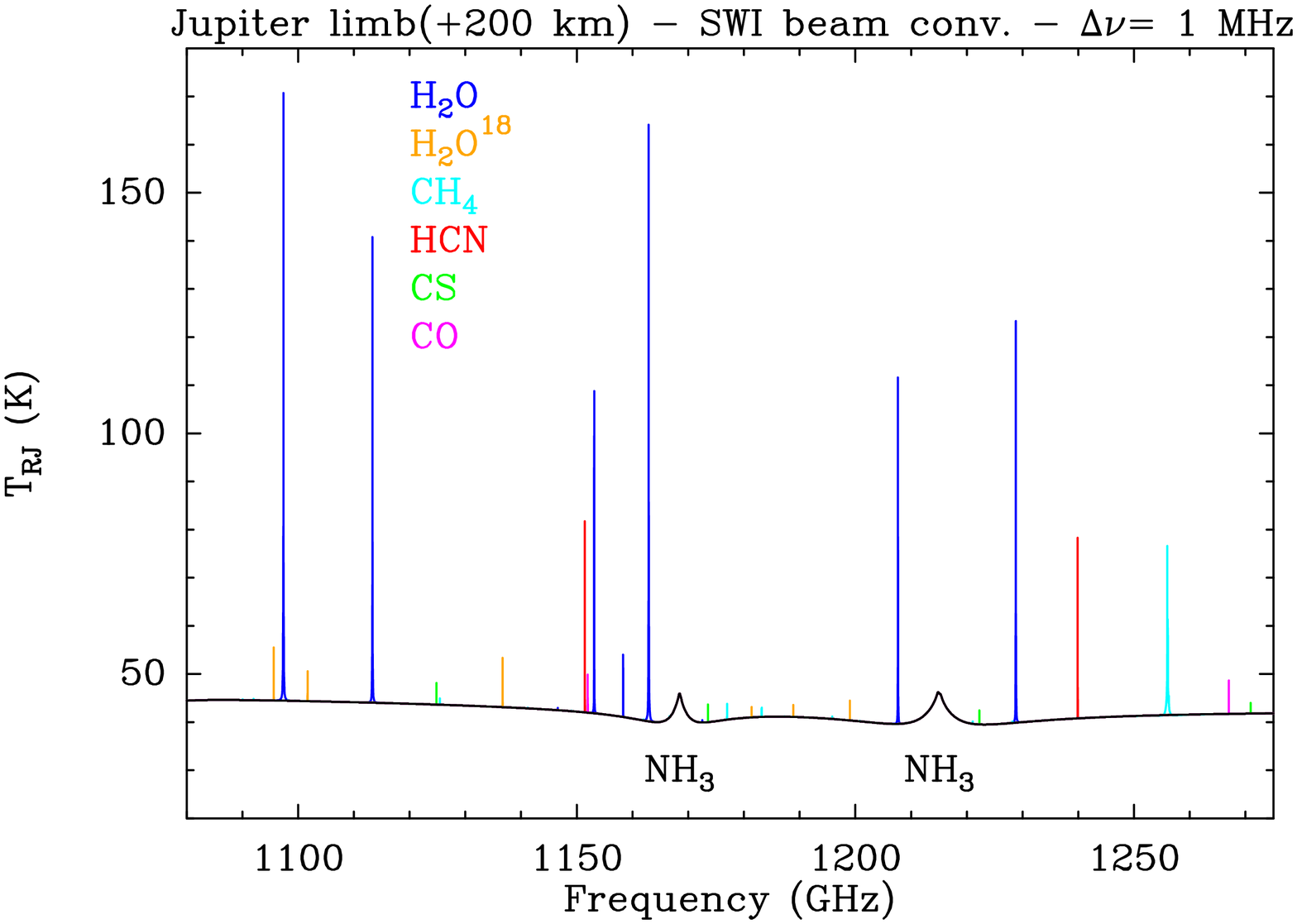}
\caption{Jupiter limb spectra in the 600\,GHz (top) and 1200\,GHz (bottom) bands probed by SWI, with the main observable molecular lines. Spectra are computed at the limb for a tangential height of 200\,km above the 1-bar level and are convolved by the instrument beam and to a spectral resolution of 1\,MHz.}
\label{SWI_bands}
\end{centering}
\end{figure*}

SWI can be tuned to observe in parallel two spectral windows located in two spectral bands, 530-625\,GHz (479-565\,$\mu$m) and 1067-1275\,GHz (235-281\,$\mu$m), with one spatial pixel, and a suite of high resolution spectrometers (up to 10$^7$ resolving power) and continuum channels. These two windows in the sub-millimetre open the possibility to target key neutral and ion species (CH$_4$, H$_2$O, HCN, CO, CS, CH$_3$C$_2$H, etc.) as well as their isotopologues relevant for Jupiter chemistry (see Figure \ref{SWI_bands}). The main limitation of the instrument concerns spatial resolution with a 1\,mrad beam at best (i.e. 1000\,km from about 15\Rj) which will not enable limb scanning. Despite more limited spatial resolution compared to other remote sensing instruments aboard JUICE, SWI will nonetheless get vertically-resolved information on temperature and composition from observations of the pressure broadened lineshapes. The high sensitivity ensured by the low receiver temperatures coupled with the highest resolution spectrometers, i.e. the Chirp Transform Spectrometers (CTS), which will have 1\,GHz bandwidth and 10000 channels of 100\,kHz, will result in very high S/N observations of the lineshapes of H$_2$O and CH$_4$ ($\sim50$ at 600 GHz, and $\sim20$ at 1200 GHz with 200 kHz resolution) in 5 minutes of integration. Observing in the two bands in parallel with the high spectral resolution of the CTS and the high sensitivity of the receivers will enable the simultaneous and independent retrieval of the vertical profiles of temperature (with CH$_4$ lines around 1256\,GHz), abundance of the species observed in the 600\,GHz band, and wind speeds (from the combination of the two lineshapes), from the upper troposphere to the upper stratosphere. SWI will thus improve over previous sub-millimetre observatories like IRAM-30m, ISO, Odin and Herschel for the characterization of the chemistry and general circulation of Jupiter's stratosphere \citep{Moreno2003,Lellouch2002,Cavalie2008c,Cavalie2012,Cavalie2013,20benmahi}. 

Unlike other instruments, SWI will take advantage of its own pointing mechanism that will enable scanning up to $\pm$72$^\circ$ along-track and $\pm$4.3$^\circ$ cross-track away from the attitude of the JUICE spacecraft. This pointing capability has $\sim$30$''$ and $\sim$9$''$ steps along-track and cross-track, respectively, with 30$''$ accuracy. This mechanism will serve several operational goals. SWI will be able to (i) map the whole Jovian disk from any distance $>$15\Rj, (ii) compensate for spacecraft motions when performing scans for other remote sensing instruments to ensure a stable pointing, and (iii) reach any Galilean moon from the equatorial orbit on a daily basis to monitor their emissions and constrain the source of their atmospheres \citep{23tosi}.

\begin{figure*}[ht]
\begin{centering}
\centerline{\includegraphics[angle=0,width=\textwidth]{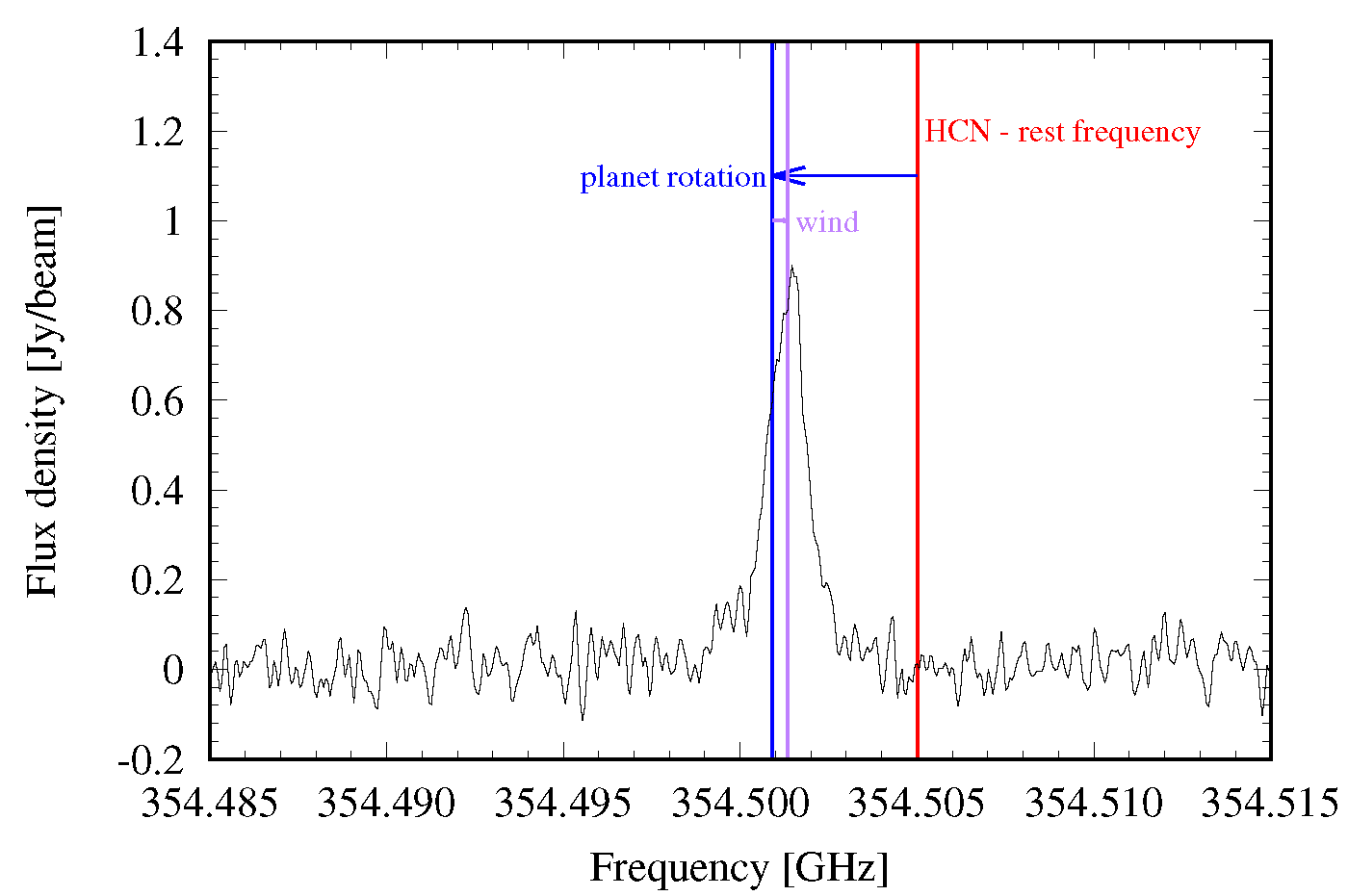}}
\caption{Example of using the Doppler shift of sub-millimetre spectral lines to determine Jupiter’s stratospheric winds \citep[using ALMA data,][]{21cavalie}.  The HCN rest frequency is 354.505 GHz (red line), but the observed line is shifted both by the planetary rotation (blue line) and the strong $\sim$400 m/s westward winds in the south-polar stratosphere (purple line) surrounding regions heated by the jovian auroras.  SWI measurements will use the Doppler shift of several lines to build up a 3D picture of stratospheric winds.}
\label{swi_doppler}
\end{centering}
\end{figure*}

The main Jupiter science goals of SWI concern its stratospheric circulation and chemistry. Constraining the general circulation of Jupiter's stratosphere (e.g. \citealt{Medvedev2013,Guerlet2020}) requires measuring temperature and winds simultaneously and continuously at all latitudes, longitudes and altitudes. This can only be rarely achieved from the ground \citep{21benmahi}, but can be measured precisely via SWI using the Doppler shift of spectral lines away from their rest frequencies (Fig. \ref{swi_doppler}).  The duration of the Jupiter tour will provide SWI with the possibility to build a full 4-dimensional view of Jupiter's stratospheric composition and dynamics, from its QQO-dominated equatorial region to the auroral latitudes, while ALMA can only provide irregularly-spaced snapshots (e.g., \citealt{21cavalie,Cavalie2022b}). A focus will be put on equatorial latitudes to constrain what maintains the QQO \citep{Leovy1991,Orton1991,Li2000,Cosentino2017,20giles_qqo,21benmahi}, and on the auroral regions to understand what controls its chemistry and dynamics \citep{Sinclair2017a,Sinclair2017b,Sinclair2018,Sinclair2019,21cavalie,Cavalie2022b}. Composition maps will help to constrain the stratospheric photochemistry of Jupiter \citep{Moses2005,Hue2018} and spectral scans will be performed in various regions to serendipitously search for new species and isotopologues. SWI will also detect and quantify sources of exogenic species other than SL9, like the Galilean moons, rings and torii \citep{Hartogh2011,19cavalie}, interplanetary dust \citep{Moses2017}, and comet/asteroid impacts. The latter seem to occur regularly in giant planet atmospheres \citep{Bezard2002,Cavalie2010,Cavalie2014,Lellouch2005,Moreno2017,Hueso2010,Hueso2013,Hueso2018,Orton2011}.

\subsubsection{SWI Operations}

\begin{figure*}[ht]
\begin{centering}
\centerline{\includegraphics[angle=0,width=\textwidth]{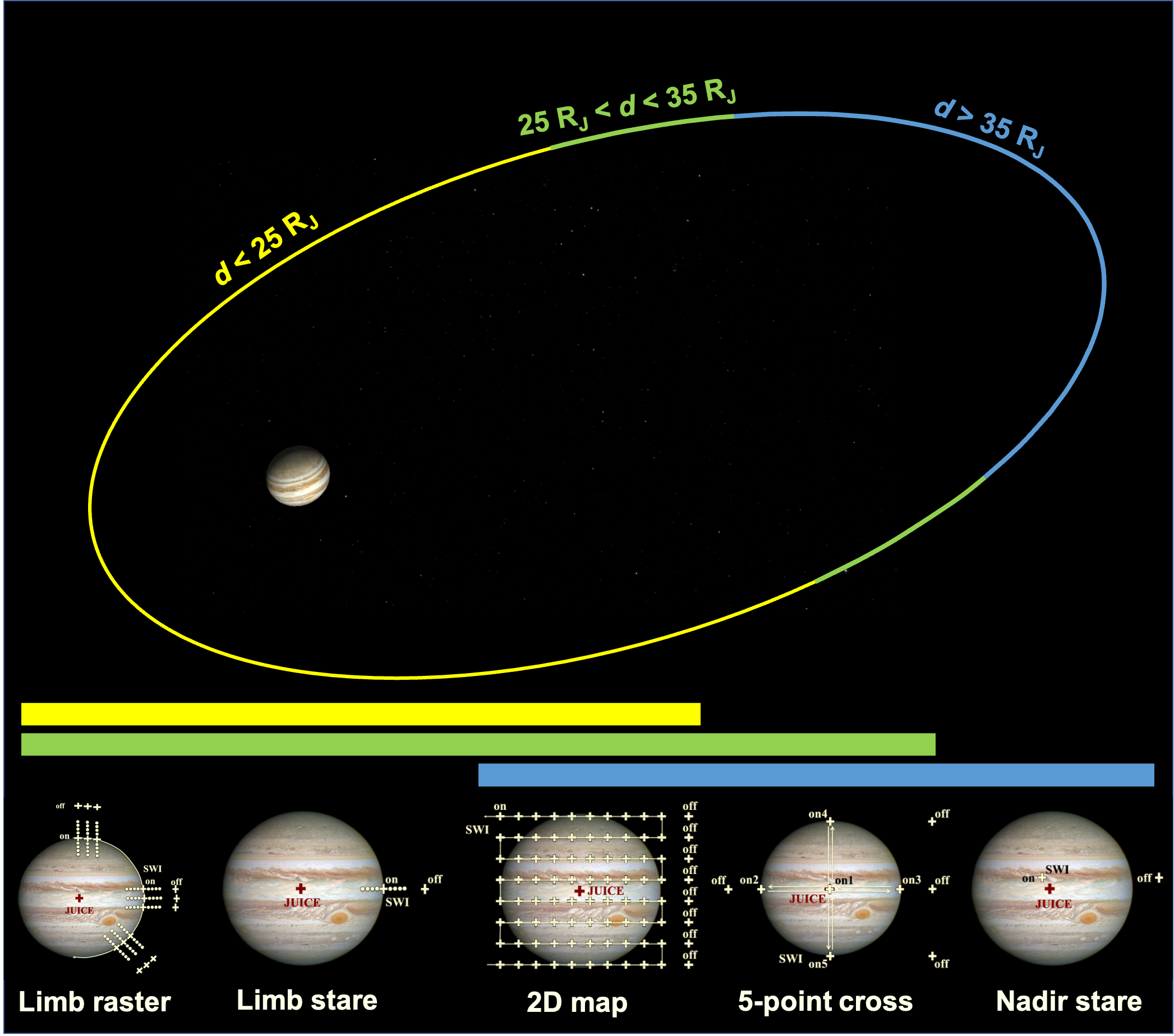}}
\caption{Five main observations modes used by SWI for Jupiter atmospheric investigations. Their use mainly depends on the distance to the planet, as depicted by the color code: yellow, green and blue for modes to be used when JUICE is at distance $d$ $<$25$R_J$, 25$R_J<d<$35$R_J$, and $d>$35$R_J$, respectively. The along-track and cross-track mechanisms of the instrument enable pointing independently from the spacecraft to build various patterns (stares, crosses, scans, and maps). The 2D map mode can be set such that meridional or zonal scans are performed. The limb stares and rasters use a limb-finding procedure at each latitude in which the continuum is recorded over several tens of positions from nadir to sky to reconstruct \textit{a posteriori} the pointing of the long integration for accurate rotation removal in wind measurements.}
\label{swi_fig}
\end{centering}
\end{figure*}

These goals will be achieved from the following observation products: regular 2D wind maps from 10\,$\mu$bar to 400\,mbar with 3\degre~latitudinal resolution; global thermal structure also from 10\,$\mu$bar to 400\,mbar with 3\degre~latitudinal resolution and 20\degre~longitudinal resolution; vertical distribution of a variety of species, tracing atmospheric transport from SL9-derived species distributions; isotopic composition; and search for new species. The observation strategy adopted by SWI mostly depends on the distance to Jupiter (see Figure \ref{swi_fig}). When JUICE is closer than 25\Rj, limb stares and limb rasters will be prioritised to map temperatures (from CH$_4$), abundances and wind speeds with the required latitudinal resolution. Observations from these distances will be complemented by zonal scans to look for thermal waves, 2D maps and meridional scans to measure the latitudinal distributions of high priority species (H$_2$O, HCN, CO). These observations will be clustered around perijoves in the early phases of the tour and during Phase 5. At intermediate distances (between 25\Rj~and 35\Rj), SWI will measure the latitudinal distributions of lower priority species (CS, CH$_3$C$_2$H, isotopologues) with 5\degre~latitudinal resolution and look for new species using limb stares, limb rasters and 2D maps. At distances beyond 35\Rj, SWI will monitor line emissions at lower spatial resolutions from the equatorial region to the auroras, and perform calibration observations, using 2D maps, 5-point crosses and nadir stares.

\subsection{JANUS}
\label{janus}

\subsubsection{JANUS Description}

The JUICE optical camera, JANUS \citep[Jovis, Amorum ac Natorum Undique Scrutator,][]{23palumbo} is a multi-filter camera system with a CMOS detector and thirteen filters spanning the near-UV (340\,nm) to the near-IR (1080\,nm). JANUS addresses a number of the scientific investigations in Section \ref{science_case} and Table \ref{tab:objectives}.  JANUS will investigate the dynamics of the weather layer by providing high-resolution observations of the atmosphere from the cloud layer to the upper hazes in multiple wavelengths. Repeated observations will be used to characterise the atmospheric dynamics and circulation [JA.1] including the role of atmospheric waves at multiple scales [JA.3]. Observations in multiple wavelengths will be used to investigate the vertical cloud structure [JC.1] and the coupling of different clouds with the dynamics [JC.2]. Dayside and night-side observations of convective regions will be used to study the role of moist convection in meteorology [JB.4]. Auroral processes imaged at night [JA.4] will also constitute an important dataset to understand the auroral structure and the impact of auroral processes in forming high altitude aerosol layers in the polar regions. JANUS observations will consist in the combination of mappings of the planet taking advantage of the planet rotation, and high-resolution observations of different targets separated by different time-scales.

The field-of-view of the detector covers a rectangular area over the sky of 1.29\degre $\times$ 1.72\degre, with the largest size of the detector oriented parallel to Jupiter's equator with the nominal spacecraft attitude during Jupiter observations. The detector has 1504 $\times$ 1200 pixels with a pixel IFoV of 15 $\mu$rad that allows images with a 15\,km/pix from a target distance of $10^6$\,km. Images with a Signal to Noise Ratio (SNR) larger than 100 are achievable in most operational conditions foreseen during the Jupiter investigation, with dayside integration times of 1\,s or less even in the filters sensitive to strong methane absorption bands. JANUS observations of the nightside of the planet will be obtained with larger exposure times below an upper operational limit of 112\,s.

Table \ref{tab:JANUS1} provides details about the filters. Each of these filters have spectral transmission curves that are almost rectangular-shaped. The instrument spectral response is also determined by the optical elements and detector spectral response, with a steep decrease at low and high range edges. JANUS filters can be compared with filters from previous missions in Table \ref{tab:JANUS2}.

\begin{table}[ht]
    \caption{JANUS filters}
    \centering
    \begin{tabular}{cccl}
    \hline
Position & Filter id. & Central wavelength / & Note \\
         &            & Bandpass [nm]        &      \\
\hline
F1       & PAN       & 650/500              & Panchromatic (clear) – monochromatic imaging \\
F2       & BLUE      & 450/60               & Blue – aerosol colours \\
F3       & GREEN     & 540/60               & Green,  background for Na – aerosol colours \\
F4       & RED       & 646/60               & Red, background for H$\alpha$ – aerosol colours \\
F5       & CMT medium & 750/20               & Continuum for medium Jovian Methane band \\
F6       & Na         & 590/10               & Sodium D-lines in exospheres \\
F7       & MT strong  & 889/20               & Strong Jovian Methane band \\
F8       & CMT strong & 940/20               & Continuum for strong Jovian Methane band \\
F9       & MT medium  & 727/10               & Medium Jovian Methane band \\
F10      & Violet     & 380/80               & UV slope \\
F11      & NIR 1      & 910/80               & Cloud structure \\
F12      & NIR 2      & 1015/130             & Cloud structure \\
F13      & H$\alpha$  & 656/10               & H$\alpha$-line for aurorae and lightning \\
\hline
    \end{tabular}
    \label{tab:JANUS1}
\end{table}

\begin{landscape}
\begin{table}[ht]
\caption{Comparison of JANUS filters with those on previous missions}
\begin{center}
\begin{tabular}{cccccccccc}
\hline
Wavelength centres [nm] & Galileo SSI (a) & \multicolumn{2}{c}{Voyager} & \multicolumn{2}{c}{Cassini ISS} & \multicolumn{2}{c}{New Horizon} & Juno JunoCam (f) & JUICE
JANUS \\
&& NAC (b) & WAC (b) & NAC (c) & WAC (c) & RALPH (d) & LORRI (e) && \\
\hline
Clear & 611 (440) & 497 (360) & 470 (290) & 611 & 635 & 687 (575) & 600 (500) & - & 650 (500) \\
UV & - & - & - & 258 & - & - & - & - & n.a. \\
UV & - & - & - & 298 & - & - & - & - & n.a. \\
UV & - & 346 & - & 338 & - & - & - & - & n.a. \\
Violet & 404 & 416 & 426 & - & 420 & - & - & - & 380 (80) \\
Blue (2) & - & - & - & 440 & - & - & - & - & - \\
Blue & - & 479 & 476 & 451 & 460 & 475 & - & Bayer Pattern blue & 450 (60) \\
Methane & - & - & 541 & - & - & - & - & - & - \\
Green & 559 & 566 & 560 & 568 & 567 & - & - & Bayer Pattern green & 540 (60)  \\
Sodium-D & - & - & 589 & - & - & - & - & - & 590 (10) \\
Orange & - & 591 & 605 & - & - & - & - & - & - \\
Methane & - & - & 618 & 619 & - & - & - & - & - \\
Continuum 2-lobed & - & - & - & 603, 635 & - & - & - & - & - \\
Red & 671 & - & - & 650 & 648 & 620 & - & Bayer Pattern red & 646 (60) \\
H-alpha & - & - & - & 656 & 656 & - & - & - & 656 (10) \\
Methane & 734 & - & - & 727 & 728 & - & - & - & 727 (10) \\
Continuum / NIR & 756 & - & - & 750 & 752 & - & - & - & 750 (20) \\
NIR (broad) & - & - & - & 752 & 742 & - & - & - & - \\
NIR (broad) & - & - & - & 862 & 853 & 880 & - & - & - \\
Methane & 887 & - & - & 889 & 890 & 885 & - & - & 889 (20) \\
Near Infrared & - & - & - & 930 & 918 & - & - & - & 910 (80) \\
Continuum & - & - & - & 938 & 939 & - & - & - & 940 (20) \\
Near Infrared & 986	- & - & 1002 & 1001 & - & - & - & 1015 (130) \\
\hline
\textbf{NIR (broad)} & - & - & - & - & 1028 & - & - & - & - \\
\hline
\end{tabular}
\end{center}
\textbf{Notes:} Bandwith is given in parentheses for the clear filter (ISS polarization filters and two-filter bandpasses not considered). (a) \citep{Belton1992}, (b) \citep{Smith1977}, (c) \citep{Porco2004}, (d) \citep{Reuter2008}, (e) \citep{Cheng2008}, (f) \citep{Hansen2017}.
\label{tab:JANUS2}
\end{table}
\end{landscape}

\subsubsection{JANUS Operations}

\begin{figure*}[ht]
\begin{centering}
\centerline{\includegraphics[angle=0,width=1.2\textwidth]{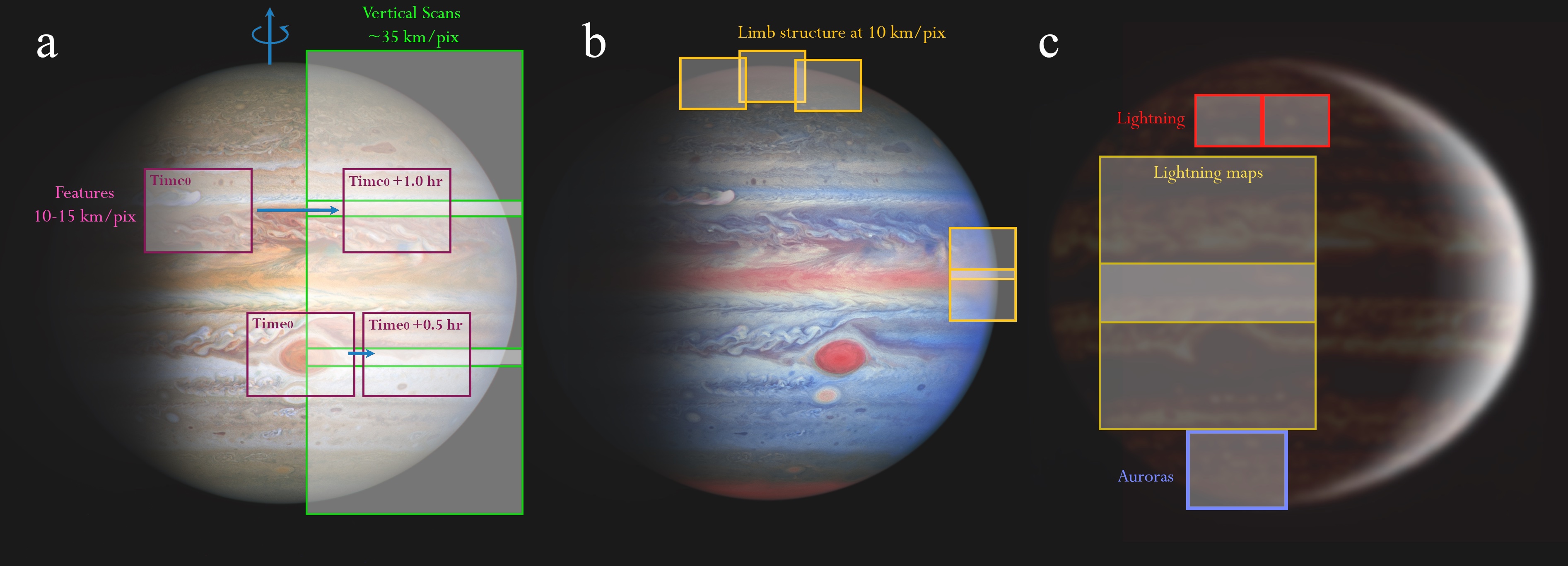}}
\caption{Sketch of different JANUS frames over the planet covering different topics. (a) Dayside observations obtaining vertical scans of the planet as it rotates (green frames), or images of specific features obtained at high spatial resolution with small time differences (purple frames) will be used for dynamics. (b) Limb observations at different wavelengths will complement observations at a variety of phase angles to determine the vertical cloud structure and directly observe haze systems over the limb. (c) Night side observations at high phase angles will map lightning (dark yellow) at moderate resolution but high-resolution images of latitudes with convective activity will also be obtained at high-resolution (red). Auroras observations will also be possible at different spatial and temporal resolutions (blue). These observations will be spread over different phases of the mission.}
\label{fig:JANUS_obs}
\end{centering}
\end{figure*}

Figure \ref{fig:JANUS_obs} shows sketches of some of the operational models of JANUS. JANUS will provide the highest spatial resolution observations over the planet with the capability to map the planet as it rotates and image specific features at very high spatial resolution ($\sim$10\,km/pix). To map the planet JANUS will perform vertical scans of Jupiter (Figure \ref{fig:JANUS_obs}a) in which different frames will cover different latitudinal ranges and different longitudes will be sampled as the planet rotates over its 10-hour rotation period. These maps will be obtained at low and intermediate phase angles ($<$90\degre) over the illuminated side of Jupiter and will attain spatial resolutions of about 15--30\,km/pix. The different opportunities in the orbital tour to obtain full maps of the planet separated by one Jovian rotation (10hr) will result in several determinations of the mean zonal winds and their variability over the course of the orbital tour. These wind fields will have a noise level of less than 1 m/s and will be used to answer fundamental question on the energy inputs of the jets and cloud systems and how is the interaction between the eddies and zonal winds (e.g., \citealt{06salyk}), and on the meridional transport on belts and zones \citep{21duer} at different moments during the mission. Higher resolution observations of particular features will unveil the internal wind field of vortices, cyclonic regions, and convective storms (Figure \ref{fig:montage_vortices}). 

Close to the planet, JANUS will observe specific regions at spatial resolutions down to 10\,km/pix including the limbs (Figure \ref{fig:JANUS_obs}b). Repeated observations of the same regions obtained after a given amount of time will be used to infer dynamics by examining cloud motions on time-scales of 0.5, 2.0 and 10.0\,hr tracking specific features of interest like the Great Red Spot and other regions. Specific observations of the polar regions from sub-spacecraft latitudes of about $\pm$30\degre will be obtained to investigate long-term changes of the polar atmosphere since the Juno observations \citep{18adriani_jiram}.

In addition, by using a combination of filters sensitive to color and cloud altitude via different atmospheric absorption bands, JANUS image will investigate the particle size and optical properties of the clouds, hazes and aerosols, investigating the complex and poorly determined relations between colour distribution and atmospheric dynamics in Jupiter. By observing the limbs at high spatial resolutions over a variety of phase angles over different latitudes JANUS will further investigate the elevated hazes at levels close to the stratopause and higher.

Nightside observations will map the spatial distribution of lightning (Figure \ref{fig:JANUS_obs}c) to complement and extend the results from previous missions \citep{99little,04dyudina,07baines,Becker2020}. Observations of lightning at high spatial resolution will be used to investigate lightning spot sizes to constrain the depth of the lightning source. Dayside observations of the same areas observed on the nightside with time differences of a few hours will also give information about the intensity, depth and vertical transport associated to moist convective storms. Nightside observations of the polar region (Figure \ref{fig:JANUS_obs}c) will also investigate the emissions of the aurora and its vertical structure.

\subsection{3GM}
\label{3gm}

\subsubsection{3GM Description}

The 3GM \citep[Gravity \& Geophysics of Jupiter and Galilean Moons,][]{23iess} is the experiment onboard the JUICE mission responsible for the radio science. This instrument consists of two separate and independently operated units incorporated in the Telemetry Tracking and Command subsystem of the spacecraft: a Ka band Transponder (KaT) and an Ultra Stable Oscillator (USO). The KaT will enable two-way range and Doppler measurements with very high accuracy at Ka band (34.5-32.2\,GHz), measuring the gravity fields of Europa, Ganymede and Callisto \citep{Cappuccio2022}.  Juno measured the gravity field of Jupiter to high accuracy \citep{18iess}, enabling the determination of the zonal wind depth \citep{18kaspi}. This was made possible by Juno's close proximity to Jupiter's cloud tops (a few thousand kilometres altitude). However, JUICE will orbit Jupiter with a proximity of hundreds of thousands of kilometres, rendering Juno-like gravity measurements of Jupiter's deep interior impracticable.  However, the USO will enable radio occultations of Jupiter's atmosphere by providing a highly stable frequency reference for the transmitted signal, establishing the possibility of one-way downlink radio occultation experiments.  Together with X and Ka band transponders, providing various communication links, the USO will be used to perform radio science to retrieve the structure and composition of the neutral atmosphere and ionosphere of both Jupiter and its moons. 

These occultations and bistatic radar radio science experiments permit the retrieval of vertical profiles of density, pressure, temperature and composition in the Jovian neutral atmosphere and electron/ion densities in the Jovian Ionosphere. The wide temporal and spatial variability of the experiments will also enable the study of vertically-propagating waves influencing the heating in the thermosphere, and monitor this variability with time and latitude.

The main principle of radio occultation measurements is the transmission of an electromagnetic signal in the radio spectrum (X and Ka band) between two separated stations (the JUICE spacecraft and an Earth-based ground-station), thereby passing through the atmosphere or ionosphere of a solar system object (Jupiter or its moons). Electromagnetic signals crossing an optical medium experience refraction, which can have two distinct effects on the signal. The first is the increase in the phase speed of the signal as the wave propagates in a neutral medium and the decrease in the phase speed of the signal while propagating in an ionised medium. The second is the bending of the signal towards regions of higher index of refraction, instead of traveling in a straight line from transmission to reception. These effects on the signal result in frequency shifts (Doppler shifts) with respect to a signal that would be traveling in vacuum. This refraction is exploited and studied in radio occultation experiments in order to retrieve the atmospheric properties of planetary atmospheres in the form of profiles varying with depth and latitude (e.g., \citealt{81lindal}, \citealt{Schinder2015}). 

The orbit of JUICE will enable the widespread distribution of these vertical profiles, guaranteeing good spatial coverage.  The vertical temperature profiles will be improved by the instrumental accuracy of the USO that will allow a resolution of about 0.1\,K over 1 km.  The vertical resolution of occultations is also characterised by the Fresnel scale, some 6\,km in the stratosphere, and a few hundred meters in the troposphere (due to refractive defocusing), from the distance of Ganymede. In addition, the attenuation of the signal can be used to deduce the absorptivity of ammonia, since radio occultations (at X band) are expected to be able to probe down to approximately the cloud level at the $\sim$700\,mbar level, whereupon NH$_3$ absorption fully attenuates the signal.

\subsubsection{3GM Operations}

\begin{figure*}[ht]
\begin{centering}
\centerline{\includegraphics[angle=0,width=\textwidth]{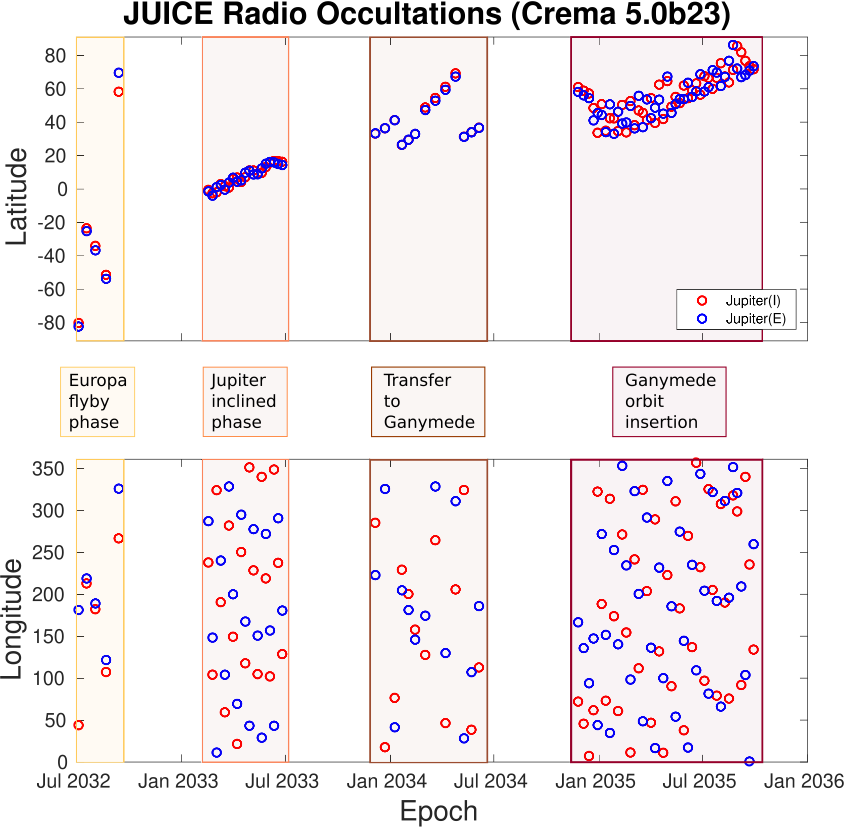}}
\caption{JUICE radio occultation opportunities of Jupiter based on Crema 5.0b23, for both latitudinal coverage (top) and longitudinal coverage (bottom) for Phases 3 through 6 of the mission.  Note the large numbers of occultations while JUICE is in orbit around Ganymede during Phase 6.}
\label{3GM_radio_occultations_fig}
\end{centering}
\end{figure*}

All predicted JUICE radio occultation opportunities of Jupiter are portrayed for the period of July 2032 to January 2036, separated into the different phases of the mission (Figure \ref{3GM_radio_occultations_fig}). The latest scenario of an April 2023 launch campaign (Crema 5.0b23) allows for around 80 radio occultations of the gas giant. The orbit of JUICE allows for not only a large number of radio occultations over the timeframe of the mission, but also for a broad latitudinal coverage. This will enable a better three-dimensional mapping of the upper atmosphere of Jupiter than has been previously possible. Figure \ref{3GM_radio_occultations_fig} shows the wide distribution of the radio occultation possibilities over Jovian latitude (upper) and longitude (lower). Jupiter's northern hemisphere has significantly more coverage due to the timing of the JUICE mission, providing the chance for occultations at approximately the same latitude twice.  Equatorial latitudes will only be observable with occultations during Phase 4 (the inclined phase).  Since the southern hemisphere has fewer opportunities, and all during Phase 3 (the Europa flybys), the portrayed radio occultations are important to include in the science activity plan. Note that future changes in the tour of JUICE can affect the frequency and possibilities of the radio occultation experiments, especially in the southern hemisphere.

The USO on board of JUICE provides a highly stable frequency reference for the transmitted signal in X and Ka bands, leading to the possibility for one-way radio occultation analysis. In one-way mode, the pointing of the spacecraft will be adjusted continuously to the direction that allows the signal transmitted by the spacecraft to be bent in a way that it will be traveling toward the receiving station after traversing Jupiter's atmosphere. The analysis of radio occultations of oblate planetary objects is performed in a different way than for spherically symmetrical ones, due to the fact that the signal in the oblate case refracts three-dimensionally. The ray-tracing method used for the analysis of Jupiter follows \citet{Schinder2015}. This method uses a numerical integration of the equations (Eikonal Equations) describing subsequent optical rays across a barotropic layered atmosphere. 

A few radio occultation experiments of Jupiter have been performed in the past, and can provide benchmarks and constrain more finely the composition of the atmosphere. A series of radio occultations experiments of Jupiter were performed by the Voyager 1 and 2 missions \citep{81lindal, Gupta2022}, with a similar configuration to the JUICE mission. Both Voyager spacecrafts carried a USO on board, which was used to provide a very stable frequency reference for the pure tones in the S and X frequency bands, so that the analysis could be performed in one-way mode. 

The analysis of $\sim20$ upcoming radio occultations by the Juno spacecraft during its extended mission (2023-2025), covering both northern and southern hemispheres, will provide the most recent data for comparison to the JUICE radio science. Juno occultations will be performed in two-way mode since no USO is present on board Juno. In a two-way mode, the highly stable frequency reference is not on the spacecraft but at a transmitting station on the Earth.

\subsection{RPWI}
\label{rpwi}

\subsubsection{RPWI Description}
The JUICE Radio \& Plasma Wave Investigation (RPWI) instrument package and science objectives are described in detail in \citet{23wahlund}. Here we briefly focus on the parts relevant for the Jupiter science goals. The RPWI instrument package will measure the electric- and magnetic-field vectors, as well as thermal plasma properties in a wide frequency range and with high temporal/spatial resolution. The electric field is measured from DC up to 1.4\,MHz by a set of four Langmuir probes. A search coil magnetometer will measure the magnetic field in the frequency range 0.1-20\,kHz, complementing the higher frequency wave components of the J-MAG measurements. Onboard analysis of these measurements, combined with the RPWI electric components, will be used to obtain the polarization and propagation properties of electromagnetic waves in this frequency range. 

Broadband measurements of all these wave field components will be especially useful for analysis of electromagnetic signals emitted from electric discharges in the Jovian atmosphere and waves in different propagation modes linked to Jovian aurora. The near DC electric field measurements, together with the J-MAG magnetic field measurements, will enable us to continuously give values of the $\bm{E}\times\bm{B}$ convection (i.e., wherever JUICE is traversing in the Jovian system). These convection electric fields are mapped, along magnetic field flux tubes, down to Jupiter's atmosphere and ionosphere, there giving rise to electrodynamic momentum and energy exchange primarily within the Jovian thermosphere and ionosphere. Along these flux tubes energy and momentum is also transported by Alfv\'en wave activity, readily monitored by RPWI determining their Poynting flux and dispersive properties. This is especially important on magnetic footprints corresponding to the Jovian auroral regions and the flux tubes connecting to the larger icy moons (Callisto, Ganymede, and Europa). Jovian radio waves are monitored by a special antenna system operating from 80 kHz to 45\,MHz. In addition, the RPWI sensors monitor thermal plasma and $\mu$m-sized dust properties wherever they operate. 

More specifically, the RPWI will contribute to Jupiter science goals in the following ways:
\begin{itemize}
   \item Map the thermal plasma and the electrodynamic (convection) interaction processes along Jovian magnetic footprint fluxtubes corresponding to the icy moons and the Jovian auroral regions.
   \item Characterise Alfv\'en waves, whistler mode waves, plasma waves in the Jovian magnetosphere of importance for the energy and momentum flux down to the Jovian ionosphere/thermosphere.
   \item Characterise radio and whistler wave emissions from lightning at Jupiter. 
   \item Determination of the general dynamic state of the Jovian magnetosphere through monitoring of the Jovian radio emission activity. 
\end{itemize}

\subsubsection{RPWI Operations}
The RPWI sensors will nominally be continuously operating together with J-MAG. The convection electric field (and $\bm{E}\times\bm{B}$), thermal plasma properties, and Jovian plasma and radio wave activity will therefore always be available, with limited impact on the remote sensing plan during the perijove windows in Section \ref{opportunities}. Certain periods will require `burst-mode data` to enable detailed studies near the large icy moons or magnetosphere boundaries, e.g., characterizing the icy moons ionospheres or icy thickness, the icy moons interaction with the Jovian magnetosphere, the energy and momentum processes related to the magnetospheric footprints to Jupiter, etc.  Multidimensional electromagnetic signals from electrical discharges in the Jovian atmosphere will be also captured during the dedicated burst mode periods. The RPWI operations will implement a so-called selective downlink where the RPWI instrument almost always does burst mode measurements, while the telemetry allocation determines how much data RPWI can download at a specific time during a period of a few weeks. 

\subsection{PRIDE}
\label{pride}
The Planetary Radio Interferometry and Doppler Experiment \citep[PRIDE-JUICE,][]{23gurvits} experiment will conduct radio occultation observations, alongside 3GM, using radio telescopes from the European VLBI network (EVN), and other radio telescopes around the world, as receivers in a one-way mode using the X- and, optionally, Ka- bands. The experimental setup is shown in \citet{23gurvits}, whereby the PRIDE telescopes would perform shadow tracking of the spacecraft during the occultation (e.g., \citealt{19bocanegra}). The radio signal would be recorded as it gets refracted by Jupiter's ionosphere and atmosphere in a wideband open-loop configuration, which allows capturing its high dynamic range.  PRIDE has access through the EVN (and other networks) to tens of radio telescopes in Europe, Asia, the Americas, South Africa and Australia, including large antennas such as the 65-m Tianma (China), 64-m Sardinia (Italy) and 100-m Effelsberg (Germany). The use of many antennas, and of large antennas in particular in radio occultation measurements, especially for those geometries with limited SNR \citep{18bocanegra}), would improve the quality of the atmospheric data retrieved and allow sounding deeper in the atmosphere than with antennas of smaller collecting area.  

These observations probe the upper troposphere down to the top-most condensate clouds, with the X-band probing deepest \citep[to approximately 700 mbar,][]{23gurvits}, because the effects of refractive defocussing are larger for the Ka band.  Both the X and Ka bands are needed to disentangle the contributions of NH$_3$ and PH$_3$ absorption to the signal, but these are both independent from the derivation of atmospheric temperature. The absorptivity is derived from the change in amplitude of the signal, whereas the temperature is derived from the refraction caused by the atmosphere, which is measured by the excess Doppler (change in instantaneous frequency) of the observed signal.  Retrievals of NH$_3$ and PH$_3$ abundances from absorptivity, alongside measurements from MAJIS and UVS, could lead to an improved characterisation of NH$_3$ condensation physics and PH$_3$ photochemical destruction processes in Jupiter's atmosphere.

\section{Synergistic Science at Jupiter}
\label{synergies}

\begin{figure*}[ht]
\begin{centering}
\centerline{\includegraphics[angle=0,width=\textwidth]{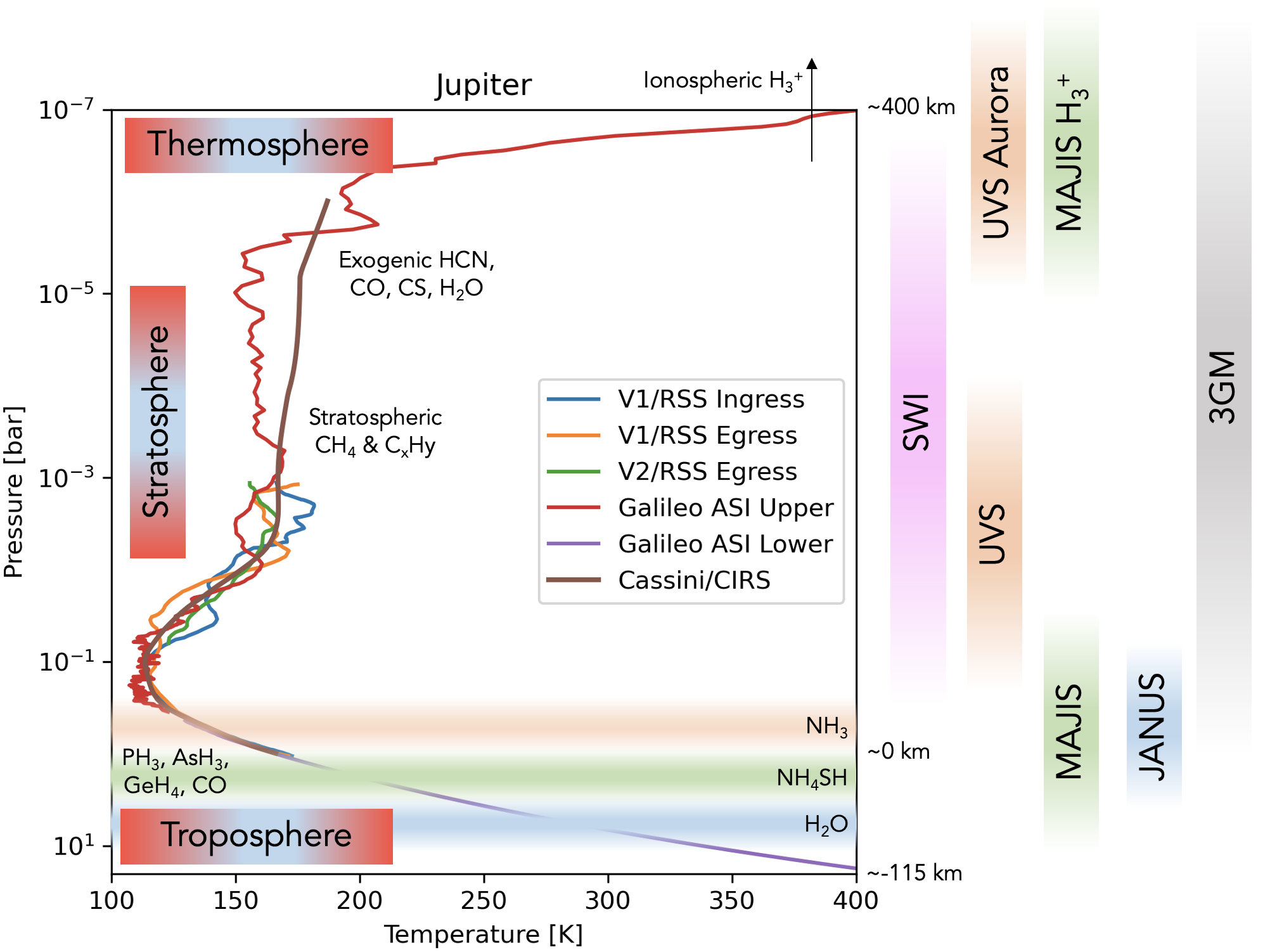}}
\caption{Synergistic observations are possible due to the overlapping vertical sensitivity of each instrument in nadir sounding, compared to Jupiter's thermal structure from Voyager radio occultations \citep{Gupta2022}, the Galileo Probe Atmospheric Structure Instrument \citep{98seiff}, and an average of Cassini/CIRS temperature retrievals \citep{16fletcher_texes}.  Key atmospheric regions, species, and aerosols are labelled where they can be studied via spectroscopy.  At higher altitudes, \textit{in-situ} instruments (J-MAG, RPWI, PEP) will contribute to characterise the energy and dynamics of the thermosphere (and ionosphere).}   
\label{synergies_fig}
\end{centering}
\end{figure*}

The previous sections described the JUICE scientific objectives (Section \ref{science_case}), the requirements on the Jupiter tour and its segmentation (Section \ref{opportunities}), and the capabilities of the payload for Jupiter science (Section \ref{instruments}).  In this Section we provide a subset of examples of how the different instruments might work together synergistically to explore equatorial dynamics, moist convection and lightning, stratospheric chemistry, auroral activity, and vertical wave structures, e.g. by taking advantage from the complementary vertical coverage of the various remote sensing instruments as shown in Fig. \ref{synergies_fig}.

\subsection{Jupiter's Equatorial Circulation and Meteorology}
Jupiter's Equatorial Zone (EZ) and neighbouring belts display a plethora of dynamic activity: from evidence of moist convection and lightning in the belts, to waves, variable winds, large vortices, and a Hadley-like circulation pattern generating extreme contrasts in volatiles \citep[e.g., NH$_3$][]{16depater, 17li}, temperatures and disequilibrium species \citep[e.g.,][]{09fletcher_ph3} between the EZ, NEB and SEB.  The sole \textit{in situ} measurement of a giant planet comes from Galileo Probe observations in a region between the NEB and EZ \citep{98atkinson, 04wong_gal}, and it remains unclear how representative these measurements were of the planetary bulk.  Tropospheric dynamics provide the spectrum of waves responsible for Jupiter's equatorial stratospheric oscillation \citep{91leovy}, which subsequently influences the temperatures, winds and composition of the stratosphere.  All of these processes vary over timescales from hours to months, but systematic multi-wavelength studies have proven challenging.  

JUICE observations are designed to address the need for atmospheric monitoring, both from the perijove windows to the monitoring opportunities throughout Phases 2, 3 and 5.  MAJIS spectral maps of reflected sunlight and thermal emission to map aerosols and gaseous composition will be interspersed with JANUS dayside cloud tracking for tropospheric winds and dynamics, UVS scans for aerosols and stratospheric composition, SWI zonal scans to determine the stratospheric temperature structure, and SWI limb observations to derive the vertical winds associated with the QQO.  UVS, MAJIS and SWI (via the 572\,GHz line) all have sensitivity to Jovian NH$_3$, enabling reconstruction of the gaseous distribution above the condensation clouds to study its temporal variability.  Radio occultations during Phase 4 will provide the vertical $T(p)$ and NH$_3$ absorption for comparison with remote sensing measurements, and RPWI will be able to search for evidence of lightning activity (see below).

\subsection{Moist Convection and Lightning}
As described in Section \ref{convection}, mapping of lightning activity provides a powerful probe of Jupiter's meteorology within the water-cloud regions.  UVS may detect transient luminous events \citep{20giles_tle} that could be associated with lightning; JANUS will use nightside imaging to search for optical flashes (and dayside imaging to find their storm counterparts) using a similar approach to Galileo \citep{00gierasch} and Juno \citep{20becker}.  MAJIS observations of evolving storm plumes could constrain their changing optical depth from bright fresh plumes to darker ovals, following similar techniques for Cassini at Saturn \citep{18sromovsky_storm}.  Lightning whistler measurements will also be performed by the RPWI search coil magnetometer and the Langmuir probes used as electric antennas. Based on the received signals in a range from 10\,Hz to 20\,kHz the LF (low frequency) receiver will produce snapshots a few seconds long, producing spectra of high temporal resolution from which the whistler dispersion can be measured. Measurement of sferics or Jovian dispersed pulses from lightning by RPWI might also be possible using the three RWI antennas and its corresponding HF (high frequency) receiver in the frequency range above 80\,kHz, where each sweep will take a few seconds. Such a receiver mode at high frequencies would work for the detection of Saturn lightning, but previous attempts to use similar frequency-sweeping receivers on a spacecraft at Jupiter have so far failed to detect any Jovian sferics. Similarly, the RPWI LP receiver can use the Langmuir probes in the electric field mode from DC to 1.5\,MHz to produce electric field spectra with low time resolution.  Taken together, these techniques will build a statistical picture of lightning phenomena on Jupiter, relating lightning detection to discrete regions of features.

\subsection{Stratospheric Chemistry}
As described in Section \ref{strat_comp}, the vertical and meridional distributions of Jupiter's stratospheric hydrocarbons remain poorly understood. The high latitude distributions of C$_2$ species was first observed during the Cassini flyby, with C$_2$H$_2$ and C$_2$H$_6$ showing opposite behaviours as a function of latitude \citep{Nixon2007}. Longitudinally resolved observations of the auroral regions unveiled an even more complex situation, with a high degree of variability in all three dimensions for the main C$_2$ species \citep{Sinclair2018} and for more complex hydrocarbons \citep{Sinclair2019}.  Some binary or multi star systems have both UV bright and near-infrared bright stars collocated on the sky.  Leveraging this group of unique targets, UVS and MAJIS will be able to perform simultaneous occultation observations.  These synchronous observations will provide a high fidelity measurement of the thermal and chemical structure structure from the lower thermosphere to the troposphere. JANUS would contribute with contextual astrometry observations. All these observations would be complemented by SWI limb observations of the stratospheric thermal structure from CH$_4$ observations at 1256\,GHz and the vertical profile of CH$_3$C$_2$H. These combined observations of the temperature and hydrocarbon distributions would help to constrain photochemical and transport models of Jupiter's stratosphere \citep{Moses2005,Hue2018}, alongside SWI measurements of the spatial distributions of long-lived H$_2$O, CO, HCN and CS deposited by the SL9 comet impact in 1994 and diffusing with latitude and depth ever since \citep{Moreno2003,Griffith2004,Lellouch2002,06lellouch,Cavalie2013,Cavalie2022b}.

  

\subsection{Auroral Morphology, Chemistry and Dynamics}
Multi-wavelength simultaneous monitoring of the auroral emissions from the UV to the submillimeter will enable better understanding the variability, energetics, and dynamics of the auroras. UVS will perform regular meridional scans of the auroras to monitor the morphology of the auroras. The produced brightness and color ratio maps will make use of CH$_4$, C$_2$H$_6$ and C$_2$H$_2$ distributions retrieved from UV stellar occultation experiments (see Fig. \ref{UVS_stellar_occ}) to capture the flux and energy of the precipitating auroral particles, giving a measure of the auroral energy deposited into the polar region.  MAJIS maps of H$_3^+$ emission give insight into the cooling due to radiation to space of some of the auroral heating via H$_3^+$ emissions \citep{Gerard2023}.  Thermospheric temperatures and H$_3^+$ densities can be inferred from the H$_3^+$ emissions, while the upper stratospheric and lower thermospheric temperatures can be captured by SWI observations of CH$_4$ and HCN lines, revealing how auroral processes contribute to heating and cooling of the neutral atmosphere.  JANUS observations of visible-light emissions will provide context for the auroral morphology and vertical structure of the auroral curtain \citep{99vasavada}.  RPWI will detect electromagnetic waves of various modes linked to auroral phenomena in a broad frequency range, such as Alfv\'en waves, whistler mode waves, and free space mode radio waves. Multicomponent measurements of fluctuating electric and magnetic fields will be used to characterise the wave polarization and propagation properties, including their Poynting flux.


Compositional contrasts between the auroral ovals and their surroundings will be studied with UVS, MAJIS and SWI, and SWI will provide upper stratospheric wind measurements to assess whether auroral winds measured from the ionosphere \citep{Rego1999,Stallard2001,Stallard2003,Johnson2017} down to the stratosphere \citep{21cavalie} play an active role in favouring auroral chemistry interior of the main ovals by confining photochemical products in a region rich in energetic magnetospheric electrons.  UVS and SWI will also look in the moon auroral footprints for exogenic species possibly transported from the Io (and Europa) torus to Jupiter's atmosphere. For instance, UVS is sensitive to SO$_2$ and SWI can look for SO and SO$_2$ lines. These measurements, in combination with MAJIS H$_3^+$ emission detection from the moon footprints, will also try to solve the puzzling structure of the near footprint tail \citep{Mura2018}.

\subsection{Occultations and Waves}
The spatial distribution of radio occultation opportunities was described in Section \ref{3gm} and shown in Figure \ref{3GM_radio_occultations_fig}.  These will probe both the electron density of the ionosphere and the temperature structure of the neutral atmosphere with a high vertical resolution, at a variety of latitudes and local times.  Once the occultation is complete and JUICE has rotated back to Jupiter, remote sensing instruments will be able to provide contextual measurements for the same locations:  for example, MAJIS observations of aerosols and NH$_3$, SWI measurements of temperatures and winds; and JANUS and UVS observations of any layering observed in the aerosol field on the Jovian limb. Independently of the radio occultations, stellar occultations observed by UVS, MAJIS and JANUS will be coordinated to ensure multi-spectral context for any vertical variability.  This would allow vertically-propagating waves observed during the radio/stellar occultations to be tied to thermal, aerosol, and chemical layering observed on the planetary limb, and its variability with time. 









\section{JUICE Science in Context}
\label{support}

\subsection{Complementary Earth-Based Observations}

JUICE will be orbiting the Jupiter system for the first half of the 2030s, exploring the atmosphere and auroras from vantage points that can only be achieved from an orbiting spacecraft.  It will be joined by NASA's Europa Clipper mission, dedicated to the exploration of Europa but carrying exciting additional capabilities for Jupiter science.  However, just as with numerous previous planetary missions (including Cassini and Juno), Earth-based observatories, both professional and amateur, will be able to provide significant spatial, temporal, and spectral support for these missions.  In terms of \textbf{spatial context,} imaging and spectroscopy from ground- and space-based telescopes provide global, moderate-resolution views of Jupiter to supplement the close-in high-resolution views of the spacecraft, aiding in registration of observations and providing broader context (e.g., by characterising environmental conditions across a broad belt or zone to support JUICE observations of an embedded thunderstorm).  

Given limitations on system resources (e.g., data volume), competition between scientific disciplines, and the finite duration of the Jupiter tour, Earth-based monitoring can also provide \textbf{temporal context.}  As storms, vortices, and the planetary bands evolve and shift over timescales ranging from days to years, ground-based records can be used to track meteorological features.  Lucky-imaging techniques employed by amateur astronomers \citep{14mousis_proam} produce high-quality Jupiter imaging on a near nightly basis.  By stacking only the sharpest frames, observers can reduce the blurring effects of atmospheric seeing to create excellent images, which are then shared with the community via repositories such as the Planetary Virtual Observatory \citep[PVOL,][]{18hueso} and the Association of Lunar and Planetary Observers (ALPO-Japan\footnote{http://alpo-j.sakura.ne.jp/indexE.htm}).  Discrete storm features are tracked from night to night to reconstruct zonal winds, wave patterns, and drift of active domains.  During Juno's prime mission, such drift charts proved essential for targeting JunoCam visible-light images.  Asteroid and cometary impacts caught by amateur imaging \citep{Hueso2010,Hueso2013,Hueso2018} can also provide opportunities for JUICE to observe impact-driven alterations of atmospheric composition and temperature.  With advances expected in amateur capabilities in the coming decade, and machine-learning approaches to tracking atmospheric features, such long-term atmospheric monitoring will help to connect the Jovian phenomena of the 2030s to the record of observations spanning back decades \citep{95rogers}.  Furthermore, ground-based infrared (e.g., 3-4 $\mu$m H$_3^+$) observations and space-based UV (e.g., HST, if still operational in the 2030s) observations can track the fluctuating auroral emissions over short timescales to complement the JUICE observations.

But perhaps the most important contribution from Earth-based observations is access to wider wavelength ranges to provide \textbf{spectral context}, at potentially higher spectral resolutions, than is possible from the JUICE spacecraft.  For example, JUICE does not have instrumentation spanning Jupiter X-ray emission \citep[e.g.,][]{17dunn,21yao}, so cannot access the hard X-ray bremsstrahlung emission, pulsed/flared soft X-ray emissions, and the dim flickering aurora observed by the likes of Chandra and XMM Newton.  These facilities (or their successors) should be able to observe Jupiter in tandem with JUICE to connect the X-ray emission to the remote sensing discussed in Section \ref{instruments}.  There is also a significant infrared gap in JUICE's capabilities, between the long-wave cutoff of MAJIS at 5.5 $\mu$m, and the 250 $\mu$m channel of SWI.  This spans the mid- and far-infrared spectrum previously studied by Cassini/CIRS \citep{04flasar_jup}, which provides a means of mapping tropospheric and stratospheric temperatures via a host of absorption and emission features, respectively.  The MIRI instrument on JWST, which is expected to be operating well into the 2030s, provides spectroscopic mapping in the 5-11 $\mu$m range without saturation, but the fields of view are too small (3-7" in size) to view the entirety of Jupiter without complex mosaicking \citep{16norwood}.  Thus the only imaging capabilities for the mid- and far-infrared are likely to come from ground-based telescopes with 3- and 8-m diameter primary mirrors, which will hopefully still be available to the planetary community in the 2030s (at the time of writing, only a small number of ageing mid-infrared instruments are still operational). 

At even longer wavelengths, JUICE observations could be supported in the millimetre, centimetre, and metre ranges by facilities like ALMA \citep{21cavalie}, the next-generation Very Large Array \citep[ngVLA,][]{18depater}, and the Square Kilometre Array \citep[SKA,][]{04butler}.  In the continuum bands, these facilities are sensitive to temperatures, ammonia, and possibly PH$_3$ at depths below the cloud-forming region described in Section \ref{science_case}.  Joint campaigns between JUICE and these ground-based observatories could help connect Jupiter's dynamic weather layer to what is happening at great depth. At decimeter to decameter wavelengths, Jupiter is one of the most prominent celestial radio sources. LOFAR has revealed the first low frequency resolved images of the radiation belts of Jupiter \citep{16girard}; the long wavelength range makes the observations sensitive to the lower energy end of the Jovian radiation belt electrons. Io-induced and non-Io decametric emission have also been observed \citep{19turner}. LOFAR's unprecedented long baselines and spectra-temporal resolution also provide the opportunity to image the dynamics of charge bunches causing Jovian decametric emission and test physical models of the associated plasma instabilities \citep{04zarka}. Jupiter is also likely to become an object of studies with prospective spaceborne radio telescopes at wavelengths longer than 20-m at the hitherto unreachable (to ground-based radio telescopes) ultra-low frequency part of the electromagnetic spectrum \citep{20bentum}.

JUICE operations in the 2030s will be in an era of major new facilities for astronomy.  The European Extremely Large Telescope (ELT), a 39-m diameter observatory based in the Atacama desert, will provide high-resolution observations in the visible, near-infrared, and mid-infrared out to approximately 14 $\mu$m.  The primary science targets for the first-light instruments require small fields of view, creating a substantial mosaicking challenge for the large disc of Jupiter (but excellent for mapping the Galilean satellites), but could nevertheless observe atmospheric phenomena at the same time as JUICE.  The Vera C. Rubin Observatory, previously known as the Large Synoptic Survey Telescope (LSST), will begin operating in the 2020s, increasing the catalogue of small objects throughout the solar system.  Within that extensive new survey, potential objects on collision courses with the Jovian system could be identified early, and then JUICE observations adjusted to investigate the aftermath.  The JUICE tour will therefore retain some flexibility in planning, to take advantage of potentially unique and unforeseen opportunities during the mission, as highlighted in Section \ref{opportunities}.

\subsection{JUICE and Exoplanets}
\label{exoplanets}
Finally, the JUICE mission will be operating in an era when our exploration of extrasolar planets moves from a phase of detection into an era of spectroscopic characterisation, via techniques such as transit spectroscopy and direct imaging.  The discovery of the first hot Jupiter exoplanet \citep{Mayor1995} fundamentally modified our vision of the Solar System planets. Firstly, planetary formation models have been challenged by the discovery of the Jupiter-mass planets closer to their star than Mercury, with the introduction of migration processes as a natural step following planet formation \citep{Morbidelli2020}. Secondly, the mass distribution  of the $\sim$5000 exoplanets discovered to date exhibits a very different pattern than that in the Solar System, with a peak in the mass range of 5--10 Earth masses, corresponding to the ``intermediate planets'', mini-Neptunes or super-Earths, which are not found in our Solar System. 

JUICE's exploration in the 2030s will allow us to compare Jupiter's atmospheric phenomena and composition with the wider collection of extrasolar giants, as the closest member of a class of astrophysical object.  The wider context and temperature range offered by exoplanet spectroscopic surveys could reveal whether Jupiter is a more or less \textit{typical giant planet} in composition and structure, or a more rare and special product of our Solar System.  Similarly, the growing census of exo-planetary system architectures may also lead to broader understanding of how Jupiter's formation and migration shaped the structure of our own Solar System.  The climate, meteorology, energetics and variability of Jupiter explored by JUICE will serve as the archetype for hydrogen-dominated giant planets and Brown Dwarfs.  

ESA's ARIEL mission, to be launched in 2029 \citep{Tinetti2018}, is expected to be contemporaneous to JUICE. ARIEL is devoted to a statistical study of the composition of $\sim$1000 exoplanets from warm super-Earths to hot Jupiters.   Interaction between the science teams of these two missions could therefore improve the science return of both, by addressing astrophysical questions about planetary origins and environments in a much broader context.

\section{Summary}
\label{summary}

The exploration of Jupiter as an archetype for giant planets, both in our Solar system and beyond, has been one of the two primary scientific objectives of ESA's JUICE mission since its original inception.  Since that time, NASA's Juno mission has revealed a wealth of new insights into the interior, atmosphere, and magnetosphere of the gas giant, prompting us to revisit the JUICE scientific requirements in this article.  Section \ref{science_case} reviewed the current status of Jupiter exploration, from the dynamic weather layer with its belts, zones, vortices, and convective storms; to the chemistry and circulation of the middle atmosphere; the global composition as a window onto planetary origins; and the energetics and circulation of the ionosphere, thermosphere, and auroras.  JUICE will explore how these different layers are interconnected and coupled, both to the circulations of the deep interior, and to the processes at work in Jupiter's magnetosphere.  These themes are used to justify the JUICE Jupiter science requirements in Table \ref{tab:objectives}.  

Section \ref{opportunities} then described how the 3.5-year Jupiter tour, with its equatorial and inclined phases bringing JUICE within 700,000 km of Jupiter at the closest perijoves, provides the observational opportunities needed to achieve the science goals.  JUICE's long-term monitoring of atmospheric and auroral phenomena, its global perspective from a variety of phase angles (dayside to nightside), and its near-equatorial vantage point all complement Juno's close-in regional views of Jupiter from its highly-inclined orbit, its 40-to-50-day separation of perijoves, and its near-terminator ($90^\circ$ phase angle) viewing geometry.  The high-inclination phase enables excellent spatial and temporal coverage of Jupiter's polar atmosphere and auroras, some 5-10 years after the Juno mission.

Section \ref{instruments} then described the subset of instruments that are needed to achieve closure with respect to the JUICE science requirements:  namely the remote sensing instruments (JANUS, MAJIS, UVS and SWI), radio occultations (3GM) and radio and plasma wave measurements (RPWI).  These instruments offer flexible opportunities and modes, working within the spacecraft resource envelope of power and data volume, and can operate both independently and synergistically (Section \ref{synergies} to explore phenomena across a broad spectral range.  UVS covers the 50-204\,nm range of the far-UV (similar to the 68-210\,nm range of Juno/UVS); JANUS provides multi-wavelength imaging from 340 to 1080\,nm (complementing the R, G, B and CH$_4$ bands on Juno/JunoCam); and MAJIS provides visible and near-infrared spectroscopy from 0.49 to 5.56\,$\mu$m (extending the 2.0-5.0\,$\mu$m coverage of Juno/JIRAM).  SWI provides a unique capability to access the temperatures, winds, and composition high in Jupiter's stratosphere, using sub-millimetre spectroscopy in two channels, 530-625\,GHz (479-565\,$\mu$m) and 1080-1275\,GHz (235-277\,$\mu$m).  JUICE does not have capabilities in the mid-infrared (5.5-30\,$\mu$m), nor in the microwave ($>1$ cm, like Juno/MWR) or X-ray.  JUICE observations will therefore be supported by ground- and space-based observatories (including JWST) wherever possible (Section \ref{support}).  Finally, 3GM radio occultations will provide electron and neutral temperature profiles at a range of latitudes, longitudes, and local times, complementing those being acquired by Juno during its extended mission.

JUICE launched in April 2023, and the scientific questions will no doubt evolve as JUICE cruises towards Jupiter in the 2020s, informed by discoveries made by Juno, by ground- and space-based observatories, by theoretical modelling, by laboratory investigations, and by the ongoing characterisation of giant exoplanets.  Most importantly, Jupiter has the ability to surprise us, with unforeseen connections, unexpected events, and new puzzles.  The JUICE spacecraft, scientific payload, and orbital tour have been designed to maximise our capability to explore the unexpected, and to provide our best four-dimensional characterisation of this archetypal giant planet.

\begin{acknowledgements}
We wish to express our gratitude to the teams of scientists, engineers, and mission planners that have made JUICE a reality, far beyond those scientists listed here as co-authors.  Without them, the scientific potential expressed in this article would remain unrealised.  
L.N. Fletcher is a JUICE Interdisciplinary Scientist supported by a European Research Council Consolidator Grant (under the European Union's Horizon 2020 research and innovation programme, grant agreement No 723890) at the University of Leicester.  
T. Cavali\'e acknowledges support from CNES and the Programme National de Plan\'etologie (PNP) of CNRS/INSU. Fletcher and Cavali\'e have jointly chaired the `JUICE Jupiter Working Group' since 2015.
D. Grassi, P. Palumbo, G. Piccioni, F. Altieri and A. Mura were supported by Italian Space Agency ``Accordo ASI-INAF n. 2018-25-HH.0, Attivit\`a scientifiche per JUICE fase C/D''.  The Italian contribution to MAJIS has been coordinated and funded by the Italian Space Agency (contract 2021-18-I.0), and supported by the ASI-INAF agreement number 2023-6-HH.0.
R. Hueso and A. S\'anchez-Lavega were supported by grant PID2019-109467GB-I00 funded by MCIN/AEI/10.13039/501100011033/ and Grupos Gobierno Vasco IT- 1742-22. 
L. M. Lara was supported by project PGC2018-099425-B-I00 (MCI/AEI/FEDER, UE). 
Y. Kaspi, E. Galanti and M. Smirnova were supported by the Israeli Space Agency. 
T. K. Greathouse, P. M. Molyneux, G.R. Gladstone and K. D. Retherford acknowledge NASA funding supporting the UVS team for ESA's JUICE mission.
M. Galand was supported by the UK Space Agency [grant number ST/W001071/1].
The JUICE/RPWI efforts are supported by the Swedish National Space Agency (SNSA).
I. Kolmasova and O. Santolik were supported from the Czech contribution to the ESA PRODEX programme and from the M\v{S}MT grant LUAUS23152.
P. G. J. Irwin acknowledges the support of the UK Science and Technology Facilities Council (ST/S000461/1).
A. Coustenis was supported by the Centre National d'\'Etudes Spatiales (CNES) in France.
G. H. Jones acknowledges the support of the UK Science and Technology Facilities Council, as well as to the International Space Science Institute, ISSI, Bern, for support as a Visiting Scientist.
We wish to express our gratitude to two anonymous referees, and to T. M. Bocanegra-Baham\'{o}n, for their help in improving this manuscript.
\end{acknowledgements}

\section*{Conflict of Interest Statement}

The authors declare that they have no competing financial or non-financial interests to declare that are relevant to the content of this article.

\bibliographystyle{aps-nameyear}      

\bibliography{biblio}   

\end{document}